%% file: master.tex
\documentclass{jfm}

\usepackage{graphicx}
\usepackage{natbib}
\usepackage{amsmath}
\usepackage{psfrag}
\usepackage{epsfig}
\usepackage{xcolor,import}
\usepackage{pstool}
\usepackage{placeins}
\usepackage{amssymb}
\usepackage{mathabx}
\usepackage{graphicx,xcolor}        
\usepackage{subfig}                             
\usepackage{wasysym}        
\usepackage{floatrow}
\usepackage{dashrule}
\usepackage{tikz}
\usepackage{booktabs}

\usetikzlibrary{arrows,decorations.markings}
\usetikzlibrary{calc}
\ifCUPmtlplainloaded \else
  \checkfont{eurm10}
  \iffontfound
    \IfFileExists{upmath.sty}
      {\typeout{^^JFound AMS Euler Roman fonts on the system,
                   using the 'upmath' package.^^J}%
       \usepackage{upmath}}
      {\typeout{^^JFound AMS Euler Roman fonts on the system, but you
                   dont seem to have the}%
       \typeout{'upmath' package installed. JFM.cls can take advantage
                 of these fonts,^^Jif you use 'upmath' package.^^J}%
      }
  \else
  \fi
\fi

\ifCUPmtlplainloaded \else
  \checkfont{msam10}
  \iffontfound
    \IfFileExists{amssymb.sty}
      {\typeout{^^JFound AMS Symbol fonts on the system, using the
                'amssymb' package.^^J}%
       \usepackage{amssymb}%

      }{}
  \fi
\fi

\ifCUPmtlplainloaded \else
  \IfFileExists{amsbsy.sty}
    {\typeout{^^JFound the 'amsbsy' package on the system, using it.^^J}%
     \usepackage{amsbsy}}
    {}
\fi

\newcommand\Rey{\mbox{\textit{Re}}}  

\newsavebox{\astrutbox}
\sbox{\astrutbox}{\rule[-5pt]{0pt}{20pt}}

\tikzset{%
  >=latex,
  inner sep=0pt,%
  outer sep=2pt,%
  mark coordinate/.style={inner sep=0pt,outer sep=0pt,minimum size=3pt,
    fill=black,circle}%
}
\usepackage{amsmath}
\usetikzlibrary{arrows}
\usepackage{pgfplots}
\usetikzlibrary{calc,fadings,decorations.pathreplacing}
\usepackage{commath}

\title[]{Lumley Decomposition of the Turbulent Round Jet Far-field. Part 1 - Kinematics}

\author[Azur Hod\v zi\' c, Knud Erik Meyer, Clara M. Velte and William K. George]%
{Azur Hod\v zi\' c$^1$
  \thanks{Email address for correspondence: azuhod@mek.dtu.dk}, Knud Erik Meyer$^1$,\\
 Clara M. Velte$^1$, and William K. George$^2$}

\shortauthor{A. Hod\v zi\' c, K. E. Meyer, C. M. Velte and W. K. George}

\affiliation{$^1$Department of Mechanical Engineering, Technical University of Denmark,
2800, Kgs. Lyngby, Denmark\\[\affilskip]
$^2$Department of Aeronautics, Imperial College London, South Kensington Campus, London SW7 2AZ, UK}

\pubyear{2010}
\volume{650}
\pagerange{119--126}
\date{?; revised ?; accepted ?. - To be entered by editorial office}
\begin{document}

\maketitle

\begin{abstract}
The current work presents a tensor formulation of the Lumley Decomposition (LD), introduced in its original form by \cite{lumley1967structure}, allowing decompositions of turbulent flow fields in curvilinear coordinates. The LD in this form is shown to enable semi-analytical decompositions of self-similar turbulent flows in general coordinate systems. The decomposition is applied to the far-field region of the fully developed turbulent axi-symmetric jet, which is expressed in stretched spherical coordinates in order to exploit the self-similar nature of the flow while ensuring the self-adjointness of the LD integral. From the LD integral it is deduced that the optimal eigenfunctions in the streamwise direction are stretched amplitude-decaying Fourier modes (SADFM). The SADFM are obtained from the LD integral upon the introduction of a streamwise-decaying weight function in the vector space definition. The wavelength of the Fourier modes is linearly increasing in the streamwise direction with an amplitude which decays with the $-3/2$ power of distance from the virtual origin. The streamwise evolution of the SADFM resembles reversed wave shoaling known from surface waves. The energy- and cross-spectra obtained from these SADFM exhibit a $-5/3$- and a $-7/3$-slope region, respectively, as would be expected for regular Fourier modes in homogeneous and constant shear flows. The approach introduced in this work can be extended to other flows which admit to equilibrium similarity, such that a Fourier-based decomposition along inhomogeneous flow directions can be performed.
\end{abstract}
\begin{keywords}
\end{keywords}
\section{Introduction}
The self-similarity of the far-field region of the turbulent axi-symmetric jet has been studied extensively over the years, herein the scaling laws reflecting the spatial development of the flow field. The self-similarity of the jet has on the other hand rarely been combined with more advanced analysis techniques, such as Fourier analysis and the Lumley Decomposition. In order to facilitate the analysis of the turbulent jet far-field the current work incorporates elements from tensor analysis together with the Lumley Decomposition (LD) introduced in \cite{lumley1967structure}, in order to identify the exact form of the energy-optimized eigenfunctions for the self-similar region of the turbulent jet.

The formulation of the LD in tensor form demonstrates the possibility of extending the approach introduced in \cite{lumley1967structure} for homogeneous and stationary flow directions, to flow directions along which the flow is not statistically homogeneous but for which equilibrium similarity theory (\cite{George1989}) holds (see \cite{george201750} for an extensive review of how these ideas fit together). This entails that the statistics develop downstream according to specific scaling laws, which for the case of the jet are usually the jet half-width and the centerline velocity, \cite{Ewing2007}. Effectively, this represents a relaxation of the homogeneity requirements for a Fourier-based decomposition in flow directions characterized by equilibrium similarity. In particular, the success of \cite{Ewing2007} in describing the single and two-point statistics of the far jet by similarity coordinates along the centerline implies that a Fourier-based LD of the jet should be possible. The work of \cite{Ewing2007} showed using experiments in the same jet described below that the single and two-point similarity results were valid descriptions of the flow. The work of \cite{Wanstrom2009} and \cite{Hodzic2018b} used full-field PIV measurements to carry out a full-three dimensional decomposition of the entire fully-developed jet. This paper and its companion paper (\cite{Hodzic2019_part2}) describe in detail those results along with a new theoretical understanding of them. 

The purpose of the present work is twofold. The first major focus is to demonstrate for a turbulent flow for which equilibrium similarity holds, how a proper tensor formulation of the LD along inhomogeneous flow directions can be performed. This has the advantage of yielding semi-analytical basis functions as opposed to the strictly numerical ones obtained by evaluating the correlation tensor directly. The deduction of the Fourier-based decomposition in the streamwise direction of the jet far-field results from the appropriate coordinate transformations combined with a specific choice of weight function in the definition of the inner product space, $L^2_w$. From this it is possible to analyze the self-similar region of the turbulent jet in terms of a combination of stretched amplitude decaying Fourier modes and numerical modes obtained from the LD in the transverse direction. 

The deduction of the Fourier-based eigenfunctions from the LD starts from the physical field, as opposed to the scaled field. This is crucial in order to reveal the physical form of the basis functions which the energy spectrum is based upon. The derivation provided in the current work thereby yields a direct relation between the energy spectrum along the streamwise direction in the jet far-field and the turbulent scales - a gap currently missing in literature. 

The second major focus of the paper lies in the analysis of the modal building blocks of the flow in order to gain a deeper understanding of the dynamics of the turbulent jet far-field. This is performed by analyzing the energy density spectra of the derived stretched amplitude-decaying modes and analyzing the modal components along the inhomogeneous transverse direction. The semi-analytical form of the deduced basis functions reinforce the coupling of the analysis to the physics of the flow.

The current work is structured as follows. Initially the governing equations are introduced in tensor form along with the tensor formulation of the Lumley decomposition. From this formulation the semi-analytical form of the modes in the streamwise and azimuthal direction is obtained, with the expansion of the field being restricted to a weighted $L^2$-space, denoted by $L^2_w$. Having identified the basis functions in $L^2_w$ these are then expressed in $L^2$, which can be argued to be a more intuitive space as the inner product weight is unity across the entire domain. The results related to the decomposition of the flow are then discussed. 
\section{Governing equations\label{sec:governing_equations}}
The invariant velocity vector, $\overline{V}\in\mathbb{R}^3$, is presented as a linear combination of a contravariant and a covariant tensor, e.g. $V^i\overline{z}_i$, where repeated indices imply the Einstein summation convention. In this work the tensor notation from \cite{Grinfeld2013} is applied. For an incompressible Newtonian fluid with constant material properties the Navier-Stokes equations with no body forces have the following contravariant form in curvilinear coordinates
\begin{equation}
\frac{\partial V^{i}}{\partial t}+V^{j}\nabla_{j}V^{i} = -\frac{1}{\rho}\nabla^{i}P+\nu\nabla^{j}\nabla_{j}V^{i}\label{eq:NS-curvilinear},
\end{equation}
where, $\nabla_j$, denotes the covariant derivative. Applying the Reynolds decomposition on, $V^i$, and ensemble averaging (denoted by angle brackets $\left\langle\cdot\right\rangle$) the Reynolds Averaged Navier-Stokes (RANS) equations are obtained
\begin{equation}
\frac{\partial \left\langle V^i\right\rangle}{\partial t}+\left\langle V^j\right\rangle\nabla_j\left\langle V^i\right\rangle=-\frac{1}{\rho}\nabla^i\left\langle P\right\rangle+\nabla_j\left(\nu\nabla^j\left\langle V^i\right\rangle-\left\langle v^iv^j\right\rangle\right)\label{eq:RANS},
\end{equation}
where, ${\left\langle V^i\right\rangle}$, and, $v^i$, are the mean and fluctuating parts of the contravariant velocity field and, $\left\langle v^iv^j\right\rangle$, is the contravariant density-normalized Reynolds stress tensor. The turbulence kinetic energy production budget takes the following form 
\begin{eqnarray}
\underbrace{\frac{D K_t}{D t}}_{I}+\underbrace{\left\langle v_iv^j\right\rangle\nabla_j\left\langle V^i\right\rangle}_{II}+\underbrace{\frac{1}{2}\nabla_j\left\langle v_iv^iv^j\right\rangle}_{III} &=& -\underbrace{\frac{1}{\rho}\nabla_i\left\langle v^ip\right\rangle}_{IV}+\label{eq:turb_energy_terms}\\
&+&\underbrace{\nabla_j\left\langle v_i\tau^{ij}\right\rangle}_{V}-\underbrace{\left\langle \tau^{ij}\nabla_jv_i\right\rangle}_{VI},\nonumber
\end{eqnarray}
where the material derivative, the ensemble averaged turbulence kinetic energy, the deviatoric stress tensor and the contravariant strain tensor are respectively defined as
\begin{equation}
\frac{D}{Dt}=\frac{\partial}{\partial t}+\left\langle V^j\right\rangle\nabla_j\hspace{0.2cm},\hspace{0.2cm}K_t=\frac{1}{2}\left\langle v^iv_i\right\rangle\hspace{0.2cm},\hspace{0.2cm}\tau^{ij}=2\nu s^{ij}\hspace{0.2cm},\hspace{0.2cm}s^{ij}=\frac{1}{2}\left(\nabla^jv^i+\nabla^iv^j\right).
\end{equation}
It is noted that Appendix \ref{app:Transport_equations} contains the derivation of the energy equations in curvilinear coordinates for laminar, mean, and turbulent flow. In \eqref{eq:turb_energy_terms} the term, $I$, is the material derivative of the turbulence kinetic energy, $II$, represents the turbulence energy production by mean shear, $III$, is the diffusion due to velocity fluctuations, $IV$, is the diffusion due to pressure fluctuations, $V$, is the viscous transport and, $VI$, is the viscous dissipation.

The divergence of the (constant density) contravariant velocity field takes on the following form in curvilinear coordinates
\begin{equation}
\nabla_{i}V^{i} = \frac{1}{\sqrt{Z}}\frac{\partial}{\partial z^{i}}\left(\sqrt{Z}V^{i}\right),\label{eq:continuity}
\end{equation}
where $\sqrt{Z}$ is the volume element of the coordinate system, and $Z$ is the determinant of the covariant metric tensor of the coordinate system.
\section{The Lumley decomposition}
The solution to the LD integral is a basis, which is represented by the eigenfunctions of the flow. The eigenfunctions in themselves can therefore be considered as a coordinate basis. The eigenfunctions are characterized by a maximization of statistical parallelism with respect to an $L^2_w$-inner product. The eigenfunctions are quantified by the eigenvalues. The vector space choice for the projection is defined through the definition of the inner product. In the formulation of the LD integral the function space is restricted to the $L^2_w$-space for vector-valued functions which is defined as
\begin{equation}
L^2_w\left(\Omega,\mathbb{C}^3\right):=\left\lbrace \overline{\Phi}:\Omega\rightarrow\mathbb{C}^3 \rvert \int_\Omega \lVert \overline{\Phi} \rVert^2wd\mu <\infty\right\rbrace,\label{eq:L2-function_space}
\end{equation}
where $\Omega\subseteq\mathbb{R}^3\times t$. The weighted $L^2$-inner product, $(\cdot,\cdot)_w$, of two vector-valued functions, $\overline{\Phi},\overline{\psi}\in\mathbb{C}^3$, with a positive weight function, $w:\mathbb{R}\mapsto\mathbb{R}_{>0}$, is defined in terms of the inner product
\begin{equation}
\left(\cdot,\cdot\right):\mathbb{C}^3\times\mathbb{C}^3\mapsto \mathbb{C},
\end{equation}
such that this inner product is the scalar product between two complex vector-valued functions
\begin{equation}
\left(\overline{\Phi},\overline{\Psi}\right) = \overline{\Phi}\cdot\overline{\Psi}^*,
\end{equation}
where the asterisk denotes the complex conjugate and the corresponding norm is given by
\begin{equation}
\lVert \overline{\Phi} \rVert = \left(\overline{\Phi},\overline{\Phi}\right)^{\frac{1}{2}}.
\end{equation}
The $L_w^2$-inner product between two vector-valued functions $\overline{\Phi}$ and $\overline{\psi}$ is then defined as the composite of the scalar product and the weighted inner product defined by the following
\begin{equation}
(\overline{\Phi},\overline{\Psi})_w=\int_\Omega\left( \overline{\Phi},\overline{\Psi}\right) wd\mu,\label{eq:inner_product}
\end{equation}
where $w$ is the imposed weight function. The variable $\mu$ here collectively designates the differential volume for both space and time, $\mu=\left\{z^i,t\right\}$ and the corresponding weighted norm is defined as
\begin{equation}
\|\overline{\Phi}\|_w=(\overline{\Phi},\overline{\Phi})_w^\frac{1}{2}.
\end{equation}
The weighted LD integral then takes the form, \cite{Holmes2012}
\begin{equation}
\overline{R}\,\overline{\Phi}=\left\langle\overline{v}\left(\overline{\Phi},\overline{v}\right)_w\right\rangle=\lambda\overline{\Phi},\label{eq:LD_invariant}
\end{equation}
where $\overline{R}$ is self-adjoint with respect to the $L^2_w$-inner product and $\left\langle\cdot\right\rangle$ designates an averaging operation. 
\subsection{The LD in tensor form}
Let $\overline{\Phi}^\alpha\in L^2\left(\Omega,\mathbb{C}^3\right)$ be the $\alpha$'th eigenfunction of the LD. By decomposing $\overline{v}$ and $\overline{\Phi}^{\alpha}$ in terms of their covariant basis vectors,~${\overline{v}=v^i\overline{z}_i}$ and~${\overline{\Phi}^\alpha=\varphi^{i\alpha}\overline{z}_i}$, the inner product in curvilinear coordinates is expressed by
\begin{equation}
\left(\overline{\Phi},\overline{v}\right)_w=\int_\Omega v^{*}_{\hat{i}}\varphi^{\hat{i}}\hat{w}d\hat{\mu},
\end{equation}
where the hat over the index is used to indicate the coordinate that is being integrated over. Applying the same decomposition of the velocity vector and the eigenfunctions to \eqref{eq:LD_invariant} we obtain the following tensor formulation of the LD
\begin{eqnarray}
\int_\Omega R^i_{\cdot\hat{i}}\varphi^{\hat{i}}\hat{w}\sqrt{\widehat{Z}}d\mu^{\widehat{4}}=\lambda \varphi^i,
\label{eq:LD_arithmetic}
\end{eqnarray}
which is valid in curvilinear coordinates, where the two-point covariant-contravariant correlation tensor is defined as $R^i_{\cdot\hat{i}}=\left\langle v^i v^*_{\hat{i}}\right\rangle$, and, $(\cdot)$, is used as an index placeholder to indicate the ordering of the indices in the second-order mixed tensor. Here $\widehat{Z}$ is the determinant of the covariant metric tensor evaluated at the integration point, and $d\mu^{\hat{4}}$ represents the four differential elements of the coordinate system (space and time). The form of the LD given in \eqref{eq:LD_arithmetic} is most convenient for implementation purposes and this is the form which will be used for the deduction of the Fourier-based decomposition of the jet in section \ref{sec:LD_in_SSC}. We note that $\sqrt{\widehat{Z}}$ is equal to the Jacobian for the purpose of this work, as the stretched spherical coordinates (introduced in section \ref{sec:LD_in_SSC}) are expressed in terms of Cartesian coordinates. For a coordinate system defined in terms of a non-affine coordinate system the volume element would not correspond to the Jacobian and $\sqrt{\widehat{Z}}$ should be used as stated in \eqref{eq:LD_arithmetic}.

Extending the tensor notation to elements of $L_w^2\left(\Omega,\mathbb{C}^3\right)$, the inner product between bases can be expressed as
\begin{equation}
\delta^\alpha_\beta=\left(\overline{\Phi}^\alpha,\overline{\Phi}_\beta\right)_w,\label{eq:orthogonality_eigenfunctions}
\end{equation}
where Greek indices denote the mode numbers. Since the $L^2_w$-bases are orthonormal the placement of the Greek indices in the sub- and superscript is exclusively chosen in order to support the Einstein summation convention and thereby the tensor notation. It is therefore worth noting that lowering the Greek index in \eqref{eq:orthogonality_eigenfunctions} yields the metric tensor in Fourier space. Since this tensor is an identity operator due to the orthonormality of the eigenfunctions with respect to the inner product in \eqref{eq:inner_product}, it is likewise denoted by a delta
\begin{equation}
\delta_{\alpha\beta}=\left(\overline{\Phi}_\alpha,\overline{\Phi}_\beta\right)_w.
\end{equation}
Analogous to the decomposition of the vector field with respect to the covariant basis, we can decompose $\overline{v}$ in terms of the eigenfunctions
\begin{equation}
\overline{v} = v^\alpha\overline{\Phi}_\alpha,\label{eq:decomposition_basis}
\end{equation}
where the vector coefficients are obtained by a projection of the instantaneous field onto the new basis. This is achieved by taking the inner product of \eqref{eq:decomposition_basis} and $\overline{\Phi}^\beta$ to yield
\begin{equation}
v^\alpha=\left(\overline{v},\overline{\Phi}^\alpha\right)_{w}.\label{eq:LD_coefficients}
\end{equation}
The indices are $\alpha=1,2,...,N$, where $N$ is the dimensionality of the new space. Observe, that we have chosen to denote the coefficients in \eqref{eq:decomposition_basis} in a very similar manner to the contravariant velocity components, namely with the letter $v$. The difference between the components in the two spaces is designated by the Greek indices which will consistently be used throughout the text for eigenfunctions and eigenvalues. The $v^\alpha$ are therefore analogous to the velocity components but with respect to the basis functions $\overline{\Phi}_\alpha$. 

The total kinetic energy of the field can be obtained from the eigenfunctions using the $L_w^2$-inner product of the velocity field with itself. By decomposing the velocity components as \eqref{eq:decomposition_basis} we obtain
\begin{equation}
\left(\overline{v},\overline{v}\right)_w=v^\alpha v^*_\alpha.\label{eq:instantaneous_kinetic_energy}
\end{equation}
Ensemble averaging \eqref{eq:instantaneous_kinetic_energy}, yields the eigenvalues related to the decomposition of the field 
\begin{equation}
\lambda=\left\langle v^\alpha v_\alpha^*\right\rangle.\label{eq:eigenvalues_modal_coefficients}
\end{equation}

In the following, the expression \eqref{eq:LD_arithmetic} will be used to deduce the eigenfunctions in SSC. This will lead to a discussion of the role of the weight function in the definition of the inner product.
\subsection{Evaluation of transformation objects for SSC}
\begin{figure}[t]
\includegraphics[scale=1]{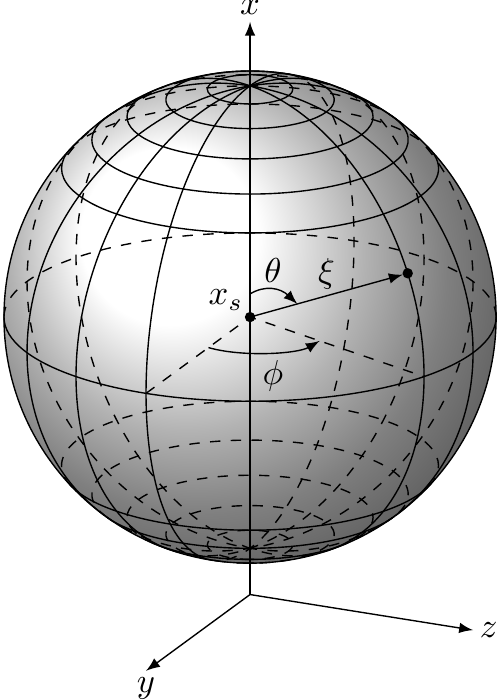}\label{fig:var_phi_1_real}
\caption{Sketch of the stretched spherical coordinate system, $\left(\xi,\theta,\phi\right)$ where $x_s$ is the start of the self-similar region and corresponds to $\xi=0$. The main streamwise direction is along the $\xi$-coordinate where the centerline of the flow is defined by $\theta=0$. The center of the nozzle lip is located at $x=y=z=0$ in the sketch.\label{fig:SSC}}
\end{figure}

The stretched spherical coordinates (SSC) are a set of orthogonal coordinates which contain many of the essential features of the similarity coordinates applied in \cite{Ewing2007}. They produce a homogeneous flow field in the streamwise direction if the field is scaled by the local centerline velocity and the logarithmic stretching "neutralizes" the spatial scale growth. Most importantly the SSC have the additional advantage of directly yielding a Hermitian symmetric cross-correlation matrix for the LD - a direct result of their orthogonality - which ensures that the eigenvalues of the matrix are real and the eigenvectors are orthogonal. The relation between Cartesian coordinates, $z^{i'}$, where $z^{1'} = x$, $z^{2'} = y$, $z^{3'} = z$ and the SSC, $z^{i}$, where $z^{1} = \xi$, $z^{2} = \theta$, $z^{3} = \phi$, can be written as follows
\begin{eqnarray}
\xi\left(x,y,z\right) &=& \ln\left(\frac{1}{C}\sqrt{\left(x-x_0\right)^2+y^2+z^2}\right),\\
\theta\left(x,y,z\right) &=& \arccos\left(\left(x-x_0\right)\left(\left(x-x_0\right)^2+y^2+z^2\right)^{-\frac{1}{2}}\right),\\
\phi\left(x,y,z\right) &=& \arctan\frac{z}{y},
\end{eqnarray}
and the inverse relations
\begin{eqnarray}
x(\xi,\theta) &=& Ce^{\xi}\cos\theta+x_0,\label{eq:x_SSC}\\
y(\xi,\theta,\phi) &=& Ce^{\xi}\sin\theta\cos\phi,\label{eq:y_SSC}\\
z(\xi,\theta,\phi) &=& Ce^{\xi}\sin\theta\sin\phi\label{eq:z_SSC}.
\end{eqnarray}
In the above $C=x_s-x_0$ is the distance from the virtual origin, $x_0$, to the start of the self-similar region of the flow and similarly. Figure \ref{fig:SSC} shows a sketch of the SSC along with the Cartesian coordinates. The center of the nozzle is then located at the Cartesian origo ($x=z=y=0$), and $\theta=0$ defines the centerline of the jet. Note that $\xi=0$ corresponds to the start of the self-similar region. The placement of the $\xi$-coordinate origo at this point turns out to be crucial for the orthogonality criterion of the SADFM in the self-similar region of the flow. This, however, will be discussed in more detail later.
\FloatBarrier
The Jacobian object, $J^{i'}_i=\partial z^{i'}/\partial z^i$, relating the Cartesian and SSC takes on the following form,
\begin{equation}
J^{i'}_{i} = \begin{Bmatrix}
C{{e}^{\xi}}\cos\theta &-C{{ e}^{\xi}}\sin\theta &0\\
C{{ e}^{\xi}}\sin\theta\cos\phi &C{{ e}^{\xi}}\cos\theta\cos\phi &-C{{ e}^{\xi}}\sin\theta 
\sin\phi\\
C{{ e}^{\xi}}\sin\theta\sin\phi &C{{ e}^{\xi}}\cos\theta \sin\phi &C{{ e}^{\xi}}\sin\theta\cos\phi
\end{Bmatrix},\label{eq:Jacobian_sim_to_cart}
\end{equation}
where the upper and lower indices correspond to the row- and column numbers, while the inverse transformation $J^i_{i'}=\partial z^i/\partial z^{i'}$ evaluates to
\begin{equation}
J^{i}_{i'} = \begin{Bmatrix}
{\left(Ce^{\xi}\right)^{-1}\cos\theta}&{\left(Ce^{\xi}\right)^{-1} \sin\theta \cos\phi }&{\left(Ce^{\xi}\right)^{-1}\sin\theta\sin\phi}\\
-{\left(Ce^{\xi}\right)^{-1}\sin\theta}&{\left(Ce^{\xi}\right)^{-1}\cos\theta 
\cos\phi}&{\left(Ce^{\xi}\right)^{-1}\cos\theta\sin\phi}\\
0&-{\left(Ce^{\xi}\right)^{-1}\sin\phi\sin^{-1}\theta
}&{\left(Ce^{\xi}\right)^{-1}\cos\phi\sin^{-1}\theta}
\label{eq:Jacobian_cart_to_sim}
\end{Bmatrix}.
\end{equation}
These yield the following covariant metric tensor for the SSC
\begin{equation}
z_{ij} = \begin{Bmatrix}
\left(Ce^{\xi}\right)^{2} & 0 & 0\\
0 & \left(Ce^{\xi}\right)^{2} & 0\\
0 & 0 & \left(Ce^{\xi}\sin\theta\right)^{2}
\end{Bmatrix},\label{eq:covariant_metric_tensor_SSC}
\end{equation}
as well as the contravariant counterpart
\begin{equation}
z^{ij} = \begin{Bmatrix}
\left(Ce^{\xi}\right)^{-2} & 0 & 0\\
0 & \left(Ce^{\xi}\right)^{-2} & 0\\
0 & 0 & \left(Ce^{\xi}\sin\theta\right)^{-2}
\end{Bmatrix}.\label{eq:contravariant_metric_tensor_SSC}
\end{equation}
The \textit{orthogonality} of the SSC is confirmed by the diagonality of the metric tensors, \eqref{eq:covariant_metric_tensor_SSC} and \eqref{eq:contravariant_metric_tensor_SSC}, a property that generally reduces the number of cross-coupling terms in the formulation of the equations of motion. This, for instance, is contrasted by the similarity coordinates used in \cite{Ewing2007} which are not orthogonal, consequently increasing the number of terms appearing in the governing equations. 

The square root of the determinant of the covariant metric tensor is then
\begin{equation}
\sqrt{Z} =  \left(Ce^{\xi}\right)^3\sin\theta,\label{eq:volume_element_SSC}
\end{equation}
and the non-zero elements of the Christoffel-symbols evaluate to
\begin{eqnarray}
\Gamma_{11}^{1} &=& 1 \hspace{0.2cm},\hspace{0.2cm} \Gamma_{22}^{1} = -1\hspace{0.2cm},\hspace{0.2cm} \Gamma_{33}^{1} = -\sin^2\theta ,\\
\Gamma_{12}^{2} &=& 1\hspace{0.2cm},\hspace{0.2cm} \Gamma_{33}^{2} = -\sin\theta\cos\theta ,\\
\Gamma_{13}^{3} &=& 1 \hspace{0.2cm},\hspace{0.2cm} \Gamma_{23}^{3} = \frac{\cos\theta}{\sin\theta}.
\end{eqnarray}
These represent a total of ten non-zero elements due to the symmetry of the second order Christoffel symbol in its lower indices. The physical velocity components (\cite{Truesdell1953}) in SSC are then expressed in terms of Cartesian velocity components
\begin{eqnarray}
V^\xi &=& V^x\cos\theta+V^y\sin\theta\cos\phi+V^z\sin\theta\sin\phi, \label{eq:physical_velocity_xi_SSC}\\
V^\theta &=& V^y\cos\theta\cos\phi-V^x\sin\theta+V^z\cos\theta\sin\phi, \label{eq:physical_velocity_theta_SSC}\\
V^\phi &=& V^z\cos\phi-V^y\sin\phi. \label{eq:physical_velocity_phi_SSC}
\end{eqnarray}
which are the SSC velocity components with respect to SSC basis vectors of unit length. Note that the physical velocity components are denoted in the superscript by the Greek coordinate symbols $\xi,\theta$ and $\phi$ in correspondence to the coordinate axes of the SSC.
\subsection{The LD in SSC\label{sec:LD_in_SSC}}
A new definition of the characteristic velocity in SSC is introduced, which depends on the absolute distance from $x=x_0$. From the centerline velocity, $U_c=BM_0^{\frac{1}{2}}/(x-x_0)$, \cite{Hussein1994}, we can rewrite the contravariant centerline velocity as 
\begin{equation}
\widetilde{U}_c = \frac{BM^{\frac{1}{2}}_0}{\left(Ce^{\xi}\right)^2},\label{eq:Uc_wiggle}
\end{equation}
where it is noted that $\xi$ is the radial coordinate in figure \ref{fig:SSC}. The velocity components and the metric tensor can then be written as a product of a $\xi$-invariant part and $\widetilde{U}_c$ from \eqref{eq:Uc_wiggle} in the following manner along with the weight function
\begin{eqnarray}
v^i &=& \tilde{v}^i\widetilde{U}_c,\\
z_{\hat{i}\hat{j}} &= & \tilde{z}_{\hat{i}\hat{j}}C^2e^{2\hat{\xi}},\\
\widehat{w} &=& e^{-\hat{\xi}}.\label{eq:weight_function}
\end{eqnarray}
From the following definitions of coordinate differences
\begin{eqnarray}
\zeta &=&\widehat{\xi}-\xi,\label{eq:zeta}\\
\Theta &=& \widehat{\phi}-\phi,\\
\tau &=& \widehat{t}-t,
\end{eqnarray}
the LD integral in SSC takes on the following form, (see Appendix \ref{app:Derivation_of_POD_scaled} for the derivation)
\begin{eqnarray}
\int_{\Omega} \widetilde{R}^j_{\cdot\hat{j}}\tilde{\varphi}^{\hat{j}}_\alpha\sin\widehat{\theta} d\mu^{\widehat{4}} = \tilde{\lambda} \tilde{\varphi}^{j}_\alpha,
\label{eq:LD_innermost_integral_SSC}
\end{eqnarray}
where
\begin{eqnarray}
\widetilde{R}^j_{\cdot\hat{j}}&=&\tilde{z}_{\hat{j}\hat{k}}\left\langle \tilde{v}^j\tilde{v}^{\hat{k}}\right\rangle,\label{eq:correlation_tensor_wiggle}\\
\tilde{\varphi}^{\hat{j}}_\alpha&=&e^{2\hat{\xi}}\varphi^{\hat{j}}_\alpha,\\
\tilde{\varphi}^{j}_\alpha &=&e^{2\hat{\xi}}\varphi^{j}_\alpha\label{eq:modes_scaled_SSC},\\
\tilde{\lambda}&=&\lambda/\left(CB^2M_0\right).
\end{eqnarray}
Since the correlation tensor in \eqref{eq:correlation_tensor_wiggle} is invariant with respect to the separation in \eqref{eq:zeta} the eigenfunctions $\tilde{\varphi}^{j\alpha}$ can be decomposed with respect to Fourier modes in $t$ and in the $\xi$-, and $\phi$-directions, \cite{lumley1967structure}, \cite{Holmes2012}. The following form is then obtained
\begin{equation}
\tilde{\varphi}^{j}_\alpha=\psi^{j}_\alpha(\omega,\kappa,\theta,m)e^{i\left(t\omega+\kappa\xi+m\phi\right)}\label{eq:eigenfunction_decomposed},
\end{equation}
where
\begin{equation}
\psi^{j}_\alpha=\frac{\widetilde{\psi}^{j}_\alpha}{\sqrt{C^5\Vol\sin{\theta}}}\label{eq:psi_scaled},
\end{equation}
represents the contravariant eigenfunctions along the $\theta$-coordinate, where~${\Vol=2\pi TL_\xi}$. From \eqref{eq:LD_innermost_integral_SSC}, \eqref{eq:eigenfunction_decomposed} and \eqref{eq:psi_scaled}  the following expression for the LD integral is obtained in SSC
\begin{equation}
\int\widetilde{\Phi}^j_{\cdot\hat{j}}\widetilde{\psi}^{\hat{j}}_\alpha d\hat{\theta}=\tilde{\lambda}\widetilde{\psi}^{j}_\alpha,\label{eq:LD_SSC}
\end{equation}
where 
\begin{equation}
\widetilde{\Phi}^j_{\cdot\hat{j}}=\int_0^{2\pi}\int_{-\infty}^\infty\int_{-\infty}^\infty\widetilde{R}^j_{\cdot\hat{j}} e^{-i\left(\tau\omega+\kappa\zeta+m\Theta\right)}\sqrt{\sin\theta\sin\hat{\theta}}d\tau d\zeta d\Theta.\label{eq:Omega_SSC}
\end{equation}
The formulation of the LD in SSC allows for the solution of the integral eigenvalue problem using standard matrix operations. One should note that the two-point correlation in SSC, unlike the case of similarity coordinates, is taken along a curved line (circle segment), in the $\theta$-direction. By combining \eqref{eq:modes_scaled_SSC}, \eqref{eq:eigenfunction_decomposed}, and \eqref{eq:psi_scaled} the contravariant components, $\varphi^{j}_\alpha$, are then
\begin{subequations}
\begin{eqnarray}
\varphi^{1}_\alpha &=& \widetilde{\psi}^{1}_\alpha\frac{e^{i\left(t\omega+\kappa\xi+m\phi\right)-2\xi}}{\sqrt{C^5\Vol\sin\theta}}\label{eq:varphi_1},\\
\varphi^{2}_\alpha &=& \widetilde{\psi}^{2}_\alpha\frac{e^{i\left(t\omega+\kappa\xi+m\phi\right)-2\xi}}{\sqrt{C^5\Vol\sin\theta}},\\
\varphi^{3}_\alpha &=& \widetilde{\psi}^{3}_\alpha\frac{e^{i\left(t\omega+\kappa\xi+m\phi\right)-2\xi}}{\sqrt{C^5\Vol\sin\theta}}\label{eq:varphi_3}.
\end{eqnarray}
\end{subequations}
In the case of Fourier-based eigenfunctions such as \eqref{eq:varphi_1}-\eqref{eq:varphi_3} the form \eqref{eq:decomposition_basis} should be understood in terms of an integration/sum in Fourier space in addition to the summation over $\alpha$. Otherwise, at face value the right hand side of \eqref{eq:decomposition_basis} produces a spectrum. The \textit{physical components} (corresponding to the components of $\overline{\Phi}^\alpha$ with respect to a set of basis vectors of unit length, \cite{Truesdell1953}) of the eigenfunctions take the form
\begin{subequations}
\begin{eqnarray}
\varphi^{\xi}_\alpha &=& \psi^{\xi}_\alpha\frac{e^{i\left(t\omega+\kappa\xi+m\phi\right)-\xi}}{\sqrt{C^3\Vol\sin\theta}}\label{eq:varphi_xi},\\
\varphi^{\theta}_\alpha &=& \psi^{\theta}_\alpha\frac{e^{i\left(t\omega+\kappa\xi+m\phi\right)-\xi}}{\sqrt{C^3\Vol\sin\theta}},\\
\varphi^{\phi}_\alpha &=& \psi^{\phi}_\alpha\frac{\sqrt{\sin\theta}e^{i\left(t\omega+\kappa\xi+m\phi\right)-\xi}}{\sqrt{C^3\Vol}}\label{eq:varphi_phi}.
\end{eqnarray}
\end{subequations}
Note that the eigenfunction components, $\psi^{\xi}_\alpha=\widetilde{\psi}^1_\alpha$, $\psi^{\theta}_\alpha=\widetilde{\psi}^2_\alpha$, and $\psi^{\phi}_\alpha=\widetilde{\psi}^3_\alpha$ are non-dimensional. From \eqref{eq:varphi_xi}-\eqref{eq:varphi_phi} it is seen that the streamwise components of the LD basis functions consist of stretched amplitude decaying Fourier modes (SADFM). Note that this was achieved by introducing a specific inner product weight in the inner product definition in SSC (Appendix \ref{app:Derivation_of_POD_scaled}). This means that the basis \ref{eq:varphi_xi}-\ref{eq:varphi_phi} spans the space $L_w^2(\Omega,\mathbb{C}^3)$, and not $L^2(\Omega,\mathbb{C}^3)$ which is the corresponding space with weight unity. Orthogonality of the SADFM with respect to the $L^2_w\left(\Omega,\mathbb{C}^3\right)$-inner product, \eqref{eq:inner_product}, is demonstrated in~{Appendix~\ref{app:Orthogonality_of_eigenfunctions}}. The additional $e^{\xi}$-term appearing in the LD integral, which motivated the introduction of a 'neutralizing' weight function, appears due to the logarithmic stretching of the coordinate system in the $\xi$-direction and can be compared to the radial coordinate appearing in the Jacobian in the transformation from Cartesian to polar coordinates. The stretching in the latter case is naturally a result of the linearly increasing displacement between adjacent azimuthal coordinates with increasing radial distance. The consequence of the weighted inner product in the LD integral for the jet far-field, however, is that energy is filtered away with downstream distance and redistributed across eigenvalues, as shown in Appendix \ref{app:effect_of_w}. 

In order to obtain a more intuitive form of the modes which is orthogonal with respect to the uniformly weighted inner product
\begin{equation}
\delta^\alpha_\beta=\left(\overline{\chi}^\alpha,\overline{\chi}_\beta\right)_1,
\end{equation}
where $\left(\cdot,\cdot\right)_1$ is the corresponding inner product for $L^2_w\left(\Omega,\mathbb{C}^3\right)$ with weight $w=1$, the following transformation of the eigenfunctions is applied
\begin{equation}
\overline{\chi}^\alpha = e^{-\xi/2}\overline{\Phi}^\alpha\, ,\,\overline{\chi}_\beta = e^{-\xi/2}\overline{\Phi}_\beta\label{eq:chi_eigenfunctions}.
\end{equation}
Since the basis functions in \eqref{eq:chi_eigenfunctions} are orthogonal with respect to $\left(\cdot,\cdot\right)_2$ it means that they span the vector space $L^2\left(\Omega,\mathbb{C}^3\right)$. From \eqref{eq:varphi_xi}-\eqref{eq:varphi_phi} and \eqref{eq:chi_eigenfunctions} the physical components of \eqref{eq:chi_eigenfunctions} are obtained
\begin{subequations}
\label{eq:chi_equations}
\begin{eqnarray}
\chi^{\xi}_\alpha &=& \psi^{\xi}_\alpha\frac{e^{i\left(t\omega+\kappa\xi+m\phi\right)}}{\left(Ce^{\xi}\right)^{\frac{3}{2}}\sqrt{\Vol\sin\theta}},\label{eq:chi_xi}\\
\chi^{\theta}_\alpha &=& \psi^{\theta}_\alpha\frac{e^{i\left(t\omega+\kappa\xi+m\phi\right)}}{\left(Ce^{\xi}\right)^{\frac{3}{2}}\sqrt{\Vol\sin\theta}}\label{eq:chi_theta},\\
\chi^{\phi}_\alpha &=& \psi^{\phi}_\alpha\frac{e^{i\left(t\omega+\kappa\xi+m\phi\right)}\sqrt{\sin\theta}}{\sqrt{\Vol}\left(Ce^{\xi}\right)^{\frac{3}{2}}}\label{eq:chi_phi}.
\end{eqnarray}
\end{subequations}
\begin{center}
\begin{figure}[t]
\subfloat[]{\includegraphics[scale=0.45]{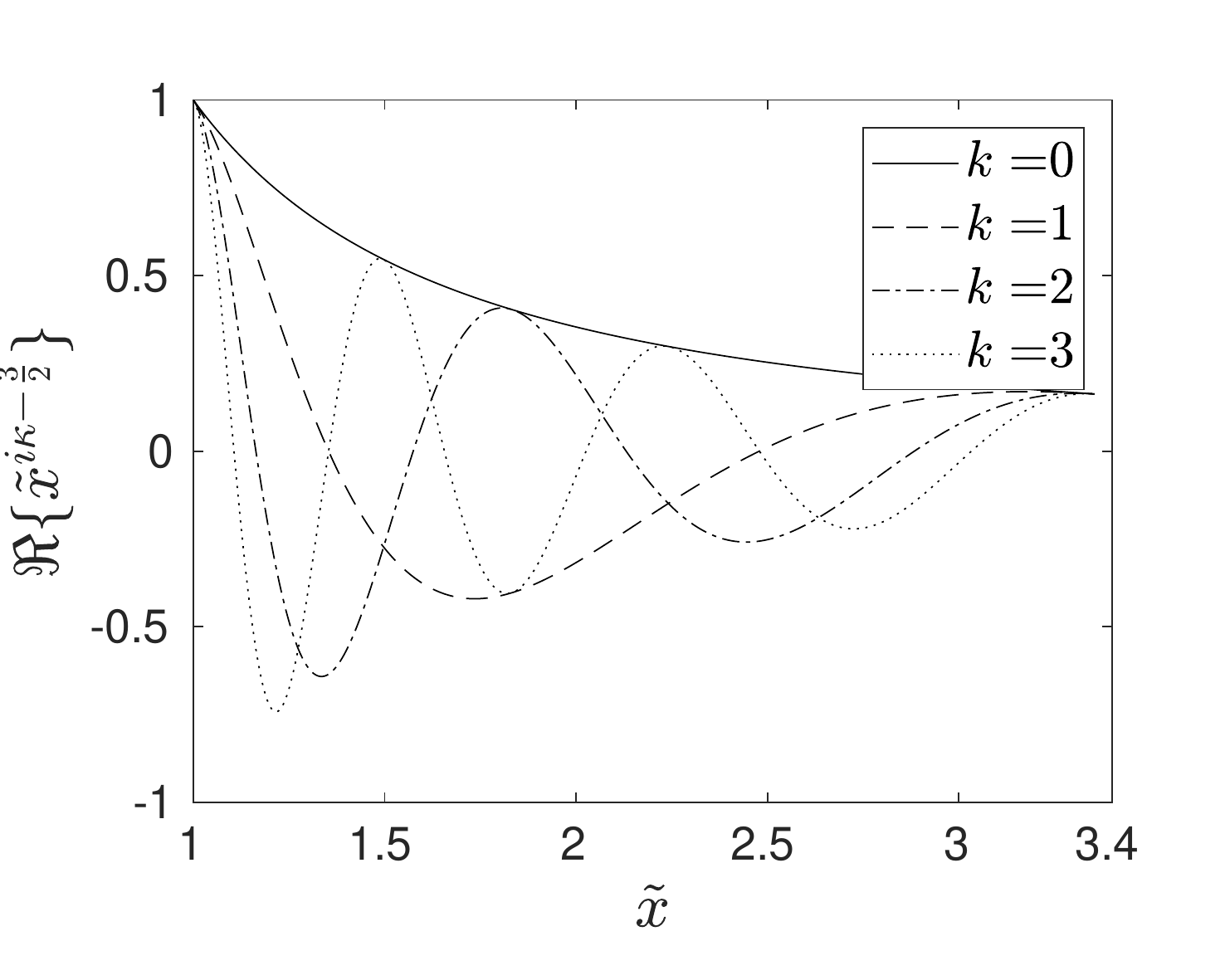}\label{fig:var_phi_1_real}}
\subfloat[]{\includegraphics[scale=0.45]{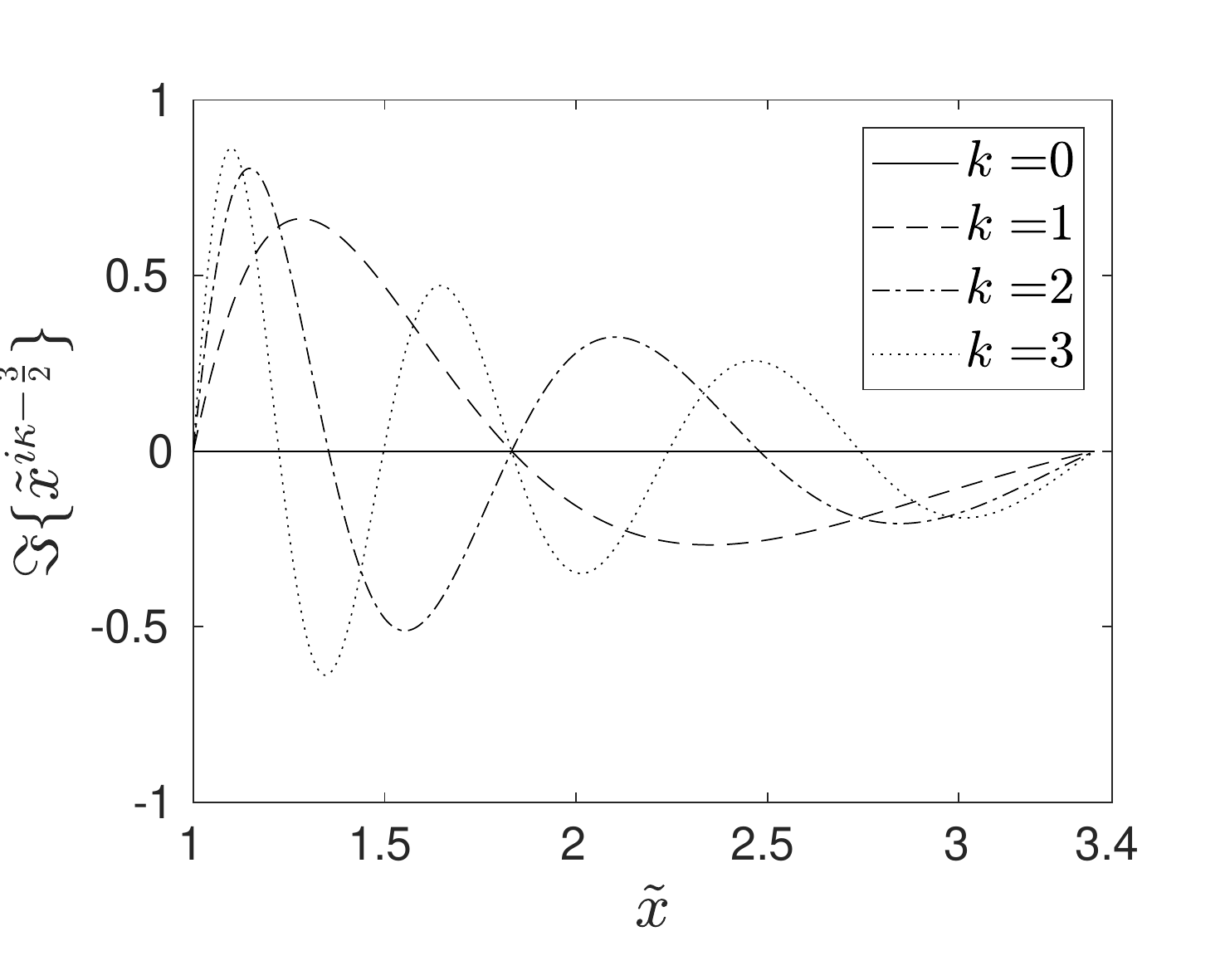}\label{fig:var_phi_1_imag}}\\
\subfloat[]{\includegraphics[scale=0.45]{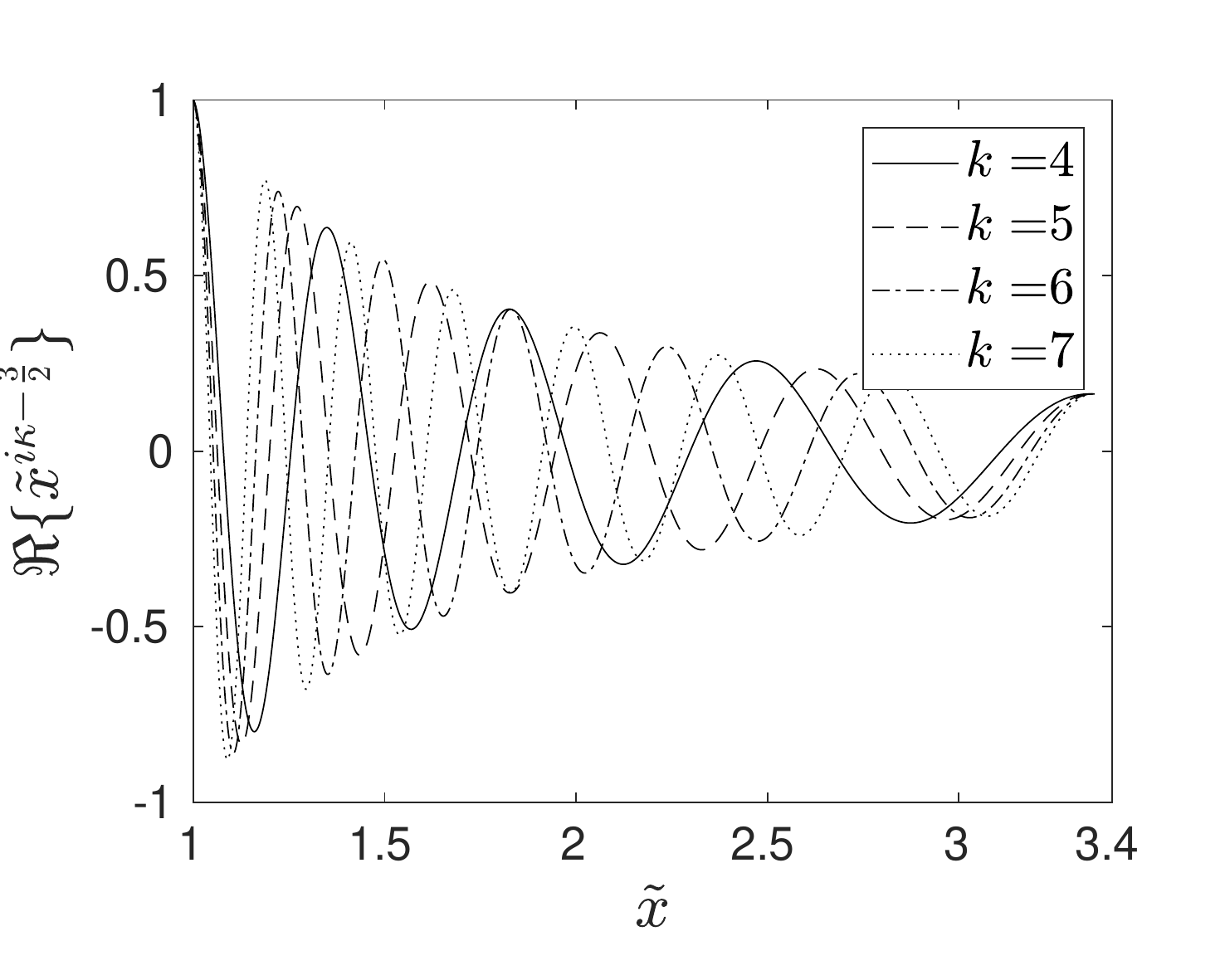}\label{fig:var_phi_2_real}}
\subfloat[]{\includegraphics[scale=0.45]{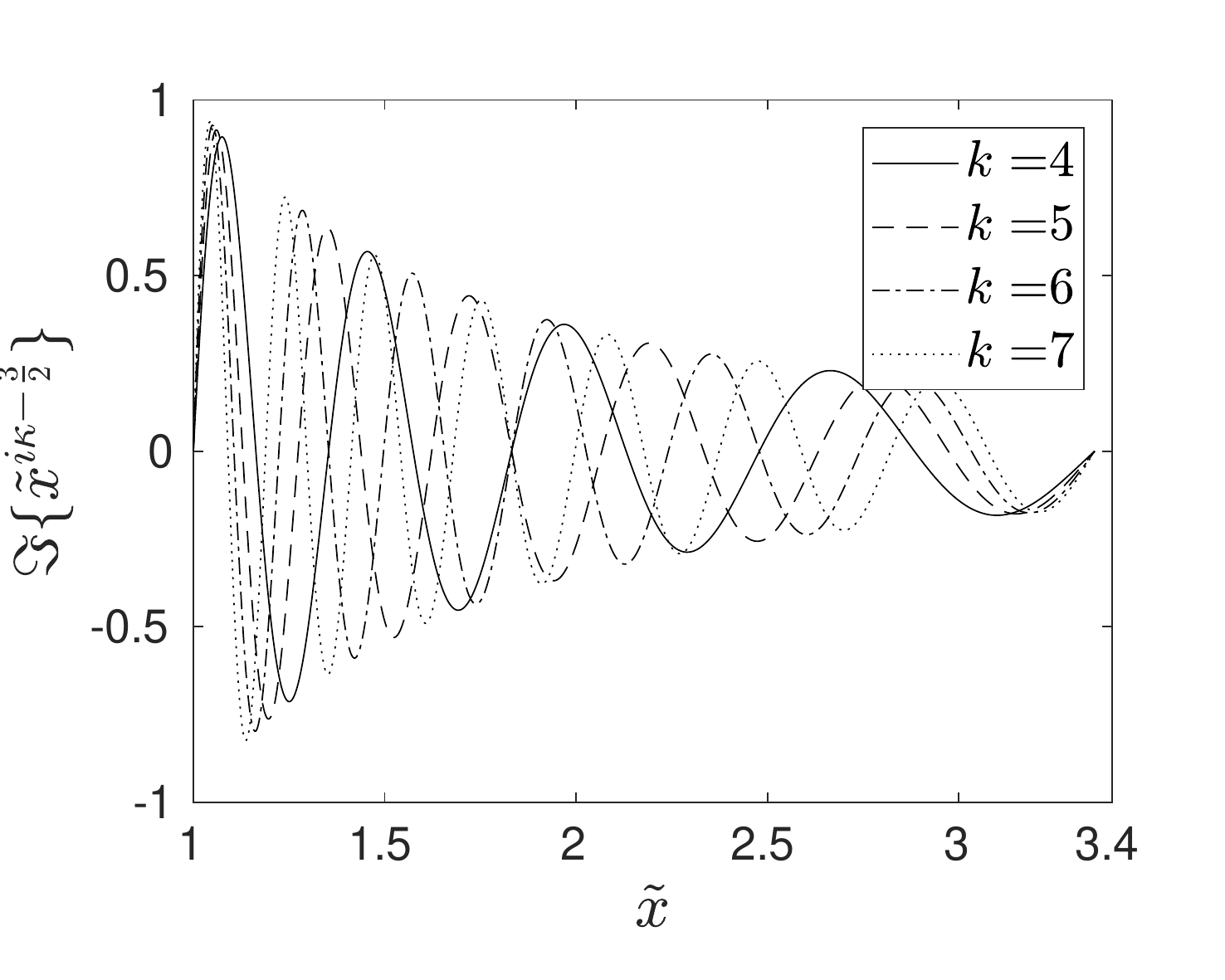}\label{fig:var_phi_2_imag}}
\caption{Figure (a) and (b) show the real- and imaginary parts of $\tilde{x}^{ik-\frac{3}{2}}$, where $\tilde{x}=\left(x-x_0\right)/C$ for $k={0,1,2,3}$, and Figures (c) and (d) show the real- and imaginary parts, for $k={4,5,6,7}$. Here $\kappa=2\pi k/L_\xi$, and $L_\xi$ is the length of the domain. As the amplitudes decay the wavelength increases with distance from the nozzle in the self-similar region of the jet.\label{fig:example_of_mode}}
\end{figure}
\end{center}
\noindent
The streamwise spatial evolution of the modes can then be expressed in terms of the distance from the nozzle, $\tilde{x}=(x-x_0)/C$, by $e^{i\kappa\xi-\frac{3}{2}\xi} = \tilde{x}^{i\kappa-\frac{3}{2}}$, from which we can write the latter expression in terms of the power series
\begin{equation}
\tilde{x}^{i\kappa-\frac{3}{2}} = \tilde{x}^{-\frac{3}{2}}\sum_{n=0}^\infty\frac{i^n}{n!}\left(\sum_{m=0}^\infty\frac{\left(\tilde{x}-1\right)^m}{m\tilde{x}^m}\right)^n.
\end{equation}

These results show that the amplitudes of the basis spanning the field in the self-similar region of the axi-symmetric jet decay with a power of $-3/2$ with distance from the virtual origin, \eqref{eq:chi_xi}-\eqref{eq:chi_phi}. The $-3/2$-power appears due to the orthogonality criterion imposed on the functions in three spatial dimensions (see Appendix \ref{app:Formulation_energy_spectra} for the basis in one-dimensional space). Orthogonality is therefore only satisfied if the power $-3/2$ appears in the eigenfunction definition. From a physical perspective, the $-3/2$-power decay describes the scattering of the wave energy with distance from the nozzle, and is therefore related to the logarithmic growth of the scales with downstream distance from the jet. An example of these waves is depicted in figure \ref{fig:example_of_mode}, where the real and imaginary parts of the spatial evolution of $\tilde{x}^{i\kappa-\frac{3}{2}}$ are shown as a function of the distance from the nozzle for various $\kappa$. Figures \ref{fig:var_phi_1_real} and \ref{fig:var_phi_2_real} show the real parts for $\kappa=0,1,2,..,7$ whilst Figures \ref{fig:var_phi_1_imag} and \ref{fig:var_phi_2_imag} show the corresponding imaginary parts. Note that this is the physical form of the modes with respect to an unstretched streamwise coordinate. It is noted that these streamwise eigenfunctions resemble water waves on a decline as the wavelength is gradually increasing in physical space, while the amplitude is decaying with increasing distance. Polynomials of decaying amplitudes are regularly found in literature appearing as eigenfunctions of certain operators in cylindrical or spherical coordinates. Interestingly the functions, $\tilde{x}^{i\kappa}$, appear as eigenfunctions to the Cauchy-Euler (CE) equation (which appears as the radial component of the Laplace equation in spherical coordinates). The full significance of the coupling between the eigenfunctions \ref{eq:chi_xi}-\ref{eq:chi_phi} and the CE equation is not clear at this point. 

The dependence of the eigenfunctions on the operator $\overline{R}$ in \eqref{eq:LD_invariant} goes beyond merely the correlation tensor. It depends on the choice of averaging operator as well as the inner product definition - the latter depending on the choice of domain. The deduction of \eqref{eq:chi_xi}-\eqref{eq:chi_phi} was based on $\Omega$ being a subset of all space and time. However, choosing the domain to be one-dimensional, say $\Omega = [a,b]$, would change the necessary form of $w$ in deducing a Fourier-based decomposition of the field based on the arguments in \cite{lumley1967structure} (see Appendix \ref{app:Formulation_energy_spectra}). Arguing, however, that the domain choice should encapsulate the entire evolution of the flow (in space and time), the eigenfunctions \eqref{eq:chi_xi}-\eqref{eq:chi_phi} deduced for $\Omega\subset \mathbb{R}^3\times t$ can be viewed as the ones that capture the physics of the flow to the highest degree. The units of $\rho\lambda$ in the case when $\Omega\subseteq\mathbb{R}^3\times t$ are $\mathrm{J\cdot s}$, which are the units of \textit{action}. By reducing the domain to consist of space exclusively ($\Omega\subseteq\mathbb{R}^3$) the units of the density multiplied eigenvalues become those of energy, $\mathrm{J}$. The choice of $\Omega$ therefore determines the units of both eigenvalues and of eigenfunctions, which is relevant to keep in mind in cases where experimental limitations reduce the dimensionality of the measured field. In the current work and in \cite{Hodzic2019_part2} the domain is chosen to be $\Omega\subseteq\mathbb{R}^3$ and thereby representing exclusively the spatial domain. This implies that the volume element used when integrating over the spatial domain remains to be defined as \eqref{eq:volume_element_SSC}.

The current decomposition of the far-field turbulence of the axi-symmetric jet into Fourier modes was based on the extension of the procedure presented in \cite{lumley1967structure}, where the turbulent field was assumed homogeneous in Cartesian or cylindrical coordinate systems. The prerequisite for applying a Fourier-based decomposition of the jet far-field was that the weight function in \eqref{eq:weight_function} was introduced. The above procedure demonstrates how a deduction of eigenfunctions from the LD is performed for which the LD modes are not regular Fourier modes, by making use of the inner product weight function. Analogous procedures can be applied to any velocity field along directions for which the two-point correlation tensor can be shown to be independent of position. The weight function can then be used as a tool in the search for analytical solutions to the LD integral as long as it is positive definite everywhere in the domain. 

\section{Numerical implementation of LD in SSC}
In the current section it is illustrated how the matrix implementation of the LD in SSC should be performed. In the following the implementation will be performed with respect to scaled contravariant velocity components consistent with the preceding sections. In order to formulate the integral eigenvalue problem in SSC, \eqref{eq:Omega_SSC}, in matrix form we must first formulate the Jacobian operator and the metric tensor in matrix form. The first- and second index of the tensors refer to rows and columns, respectively, where the upper index of the Jacobian designates the row number. Initially, the contravariant velocity components are obtained from the Cartesian velocity components using standard linear algebraic methods. Hereafter, the implementation of the LD is defined for the full four-dimensional decomposition. 
The general matrix formulation of the analytical form of the LD can be written in terms of the covariant metric tensors and the contravariant two-point correlation matrices. Note, however, that in general coordinates Hermitian symmetry is not guaranteed due to the non-zero off diagonal elements of the metric tensors in cases of non-orthogonal coordinates. In SSC, on the other hand, the matrix formulation of the problem reduces to the following form
\begin{equation}
\begin{bmatrix}
\tilde{z}_{\hat{1}\hat{1}}\widetilde{\Phi}^{1\hat{1}} & \tilde{z}_{\hat{2}\hat{2}}\widetilde{\Phi}^{1\hat{2}} & \tilde{z}_{\hat{3}\hat{3}}\widetilde{\Phi}^{1\hat{3}}\\
\tilde{z}_{\hat{1}\hat{1}}\widetilde{\Phi}^{2\hat{1}} & \tilde{z}_{\hat{2}\hat{2}}\widetilde{\Phi}^{2\hat{2}} & \tilde{z}_{\hat{3}\hat{3}}\widetilde{\Phi}^{2\hat{3}}\\
\tilde{z}_{\hat{1}\hat{1}}\widetilde{\Phi}^{3\hat{1}} & \tilde{z}_{\hat{2}\hat{2}}\widetilde{\Phi}^{3\hat{2}} & \tilde{z}_{\hat{3}\hat{3}}\widetilde{\Phi}^{3\hat{3}}
\end{bmatrix} \begin{bmatrix}
\widetilde{\psi}^{\hat{1}}\\
\widetilde{\psi}^{\hat{2}}\\
\widetilde{\psi}^{\hat{3}}
\end{bmatrix} = \lambda \begin{bmatrix}
\widetilde{\psi}^{{1}}\\
\widetilde{\psi}^{{2}}\\
\widetilde{\psi}^{{3}}
\end{bmatrix},
\label{eq:LD_matrix_SSC}
\end{equation}
where each element in the matrix in \eqref{eq:LD_matrix_SSC} represents a submatrix. In the current case where only two velocity components are measured \eqref{eq:LD_matrix_SSC} takes on the following form 
\begin{equation}
\begin{bmatrix}
\tilde{z}_{\hat{1}\hat{1}}\widetilde{\Phi}^{1\hat{1}} & \tilde{z}_{\hat{2}\hat{2}}\widetilde{\Phi}^{1\hat{2}}\\
\tilde{z}_{\hat{1}\hat{1}}\widetilde{\Phi}^{2\hat{1}} & \tilde{z}_{\hat{2}\hat{2}}\widetilde{\Phi}^{2\hat{2}}\\
\end{bmatrix} \begin{bmatrix}
\widetilde{\psi}^{\hat{1}}\\
\widetilde{\psi}^{\hat{2}}\\
\end{bmatrix} = \lambda \begin{bmatrix}
\widetilde{\psi}^{{1}}\\
\widetilde{\psi}^{{2}}
\end{bmatrix},
\label{eq:LD_matrix_SSC_2D}
\end{equation}
where the elements of \eqref{eq:LD_matrix_SSC_2D} can be expressed as cross-correlations of Fourier coefficients across the span of the jet
\begin{equation}
\tilde{z}_{\hat{i}\hat{j}}\widetilde{\Phi}^{m\hat{j}}=\begin{bmatrix}
\tilde{z}_{\hat{i}\hat{j}}\widetilde{\Phi}^{m\hat{j}}(k,\theta_1,\theta_1) & \cdots & \tilde{z}_{\hat{i}\hat{j}}\widetilde{\Phi}^{m\hat{j}}(k,\theta_1,\theta_n)\\
\vdots & \ddots & \vdots\\
\tilde{z}_{\hat{i}\hat{j}}\widetilde{\Phi}^{m\hat{j}}(k,\theta_n,\theta_1) & \cdots & \tilde{z}_{\hat{i}\hat{j}}\widetilde{\Phi}^{m\hat{j}}(k,\theta_n,\theta_n)
\end{bmatrix}.
\end{equation}
The eigenvalue problem is then solved numerically for each $\kappa=2\pi k/L_\xi$, using a total of $572$ grid points in the $\xi$-direction. The half-spectrum thus consists of $286$ points and the maximum value for $\kappa$ is obtained at $k=285$, corresponding to $\kappa_{\text{max}}=1479.4$.

After obtaining the eigenfunctions $\widetilde{\psi}^{j}_\alpha$ these are transformed back to physical space using \eqref{eq:chi_equations}. Defining the domain to be, ${\Omega := \Omega_\xi\times\Omega_\theta\times\Omega_\phi}$, where ${\Omega_\xi := [\xi_1:\xi_2]}$, ${\Omega_\theta := [\theta_1:\theta_2]}$ and ${\Omega_\phi := [\phi_1:\phi_2[}$. The streamwise limits of the domain are then $\xi_1=0$, and ${\xi_2=1.21}$, such that the length of the domain is ${L_\xi=\xi_2-\xi_1 = 1.21}$. The $\theta$-limits are $\theta_1=0$, and ${\theta_2=0.23\,\mathrm{rad}}$, while the azimuthal limits are ${\phi_1=0}$, and ${\phi_2=2\pi\,\mathrm{rad}}$. This is thus the definition of $\Omega$ on which the ${L^2_w\left(\Omega,\mathbb{C}^3\right)}$-space is defined.
\section{Experimental procedure\label{sec:experimental_procedure}}\noindent
Two datasets, $E_1$ and $E_2$, from \cite{Hodzic2018b} and \cite{Wanstrom2009}, respectively, were obtained independently. The experiments were performed in tents with dimensions\\
 ${2.5\times3.0\times10.0\,\mathrm{m^3}}$, in order to minimize the induced backflow due to confinement, \cite{Hussein1994}. The tent was sealed off during measurements and was seeded with tracer particles generated by an in-house seeding generator equipped with an atomizing nozzle producing glycerine droplets of approximately ${2-3\,\mathrm{\mu m}}$. A fan was driving the flow and was placed inside the tent in order to ensure that air inside the jet-box had a similar concentration of seeding particles to the bulk of air inside the tent. The fan supplied air through the back-side of a jet-box with inner dimensions $58.5\times58.5\times59\,\mathrm{cm^3}$, where this particular jet-box was used in~\cite{Gamard2004} in order to generate their~${D=1\mathrm{cm}}$ jet. It was equipped with baffles and screens in order to break down any transient structures, and thus provide a constant pressure field across the nozzle inlet. The axi-symmetrical nozzle designs were based on fifth-order polynomials in order to create a smooth contraction from $32,\mathrm{mm}$ to $D=10\,\mathrm{mm}$ at the outlet for both experiments. Before commencing the data acquisition the fan was left running for approximately one hour in order to ensure that any transient effects have passed, as well as to ensure that the particle concentration had approached a homogeneous distribution inside the tent.
\begin{figure}%
\centering
\def\svgwidth{0.7\textwidth} 
\input{experimentalsetup_pdf}
\caption{Sketch of the experimental setup.\label{fig:experimental_setup}}
\end{figure}
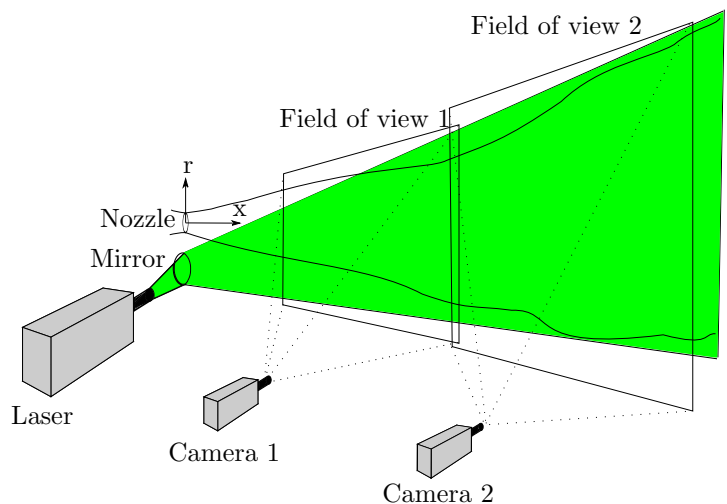
\subsection{Data acquisition}\noindent
The data were acquired using an experimental setup shown in figure~\ref{fig:experimental_setup}. For $E_1$ the 2C-PIV system consisted of two $16\,\mathrm{MPix}$ ($4872\times3248\,\mathrm{pix}$) Dantec FlowSenseEO cameras, with a pixel pitch of $7.4\,\mathrm{\mu m}$ using $60\, \mathrm{mm}$ Nikon lenses with an aperture of $f^{\#}\,2.8$. The particles were illuminated by a $200\,\mathrm{mJ}$ ND:YAG $532\,\mathrm{nm}$ laser equipped with dual cavities. Both experiments produced a similar laser sheet thickness of approximately $2\,\mathrm{mm}$. For $E_2$ the 2C-PIV system consisted of two $4\,\mathrm{MPix}$ ($2048\times2048\,\mathrm{pix}$) Dantec HiSense cameras, with a pixel pitch of $7.4\,\mathrm{\mu m}$ using $50\, \mathrm{mm}$, and $60\, \mathrm{mm}$ Nikon lenses, for the up- and downstream cameras, with an aperture of $f^{\#}\,2.8$. The particles were illuminated by a $120\,\mathrm{mJ}$ YAG $532\,\mathrm{nm}$ laser with dual cavities.

In order to reduce the distorting effect of windowing (spectral leakage) on the stream-wise energy spectra, the cameras needed to cover a sufficient extent of the jet in the stream-wise direction. This effect was modeled in \cite{Wanstrom2009} by assuming a correlation function for homogeneous turbulence of the shape $\exp(-|\Xi|/I_\xi)c$, where $\Xi$ is the spatial separation in the homogeneous direction, $\xi$, between two interrogation points, $c$ is the variance of the stream-wise velocity component and $I_\xi$ is the integral length scale in the $\xi$-direction. Since the energy spectrum of the correlation functions shows noticeable distortions when a rectangular window is applied, for $|\Xi|<10 I_\xi$, the conservative window sizes used in these experiments were~${\approx 1.25}$ and ~${L_\xi \approx 1.1}$ and, corresponding to a window size of $75D$ and $70D$ for $E_1$ and $E_2$ in physical space. The time between pulses, $\Delta t$, was carefully optimized in correspondence with the dynamic range of the field by analyzing measurements for various $\Delta t$. This was done to ensure that the captured dynamic range was as large as possible, by reducing particle loss at the beginning of the field-of-view (FOV) of camera 1 and reducing the peak-locking bias at the end of the FOV of camera 2. $\Delta t = 150\,\mathrm{\mu s}$ and $300\,\mathrm{\mu s}$ were found to be optimal for $E_1$ and $E_2$, respectively.

The minimum sampling rate in order to ensure uncorrelated fields, was estimated from preliminary measurements of the downstream flow field. For measurements to be uncorrelated at any given point in the measuring space, the minimum time between samples was required to be two integral time scales, corresponding to $0.12\,\mathrm{s}$. This was estimated using Taylor's hypothesis along the centerline from the range $x/D=\left[70:100\right]$ downstream of the nozzle, applying a convection velocity corresponding to the local mean velocity at one jet half-width (corresponding to $50\%$ of the local centerline velocity). A sampling rate of $1\,\mathrm{Hz}$ was used in the experiments and the data was acquired in a single sitting in order to obtain $11\,000$ and $10\,841$ uncorrelated realizations for $E_1$ and $E_2$, respectively. 
\subsection{Data processing}\noindent
The velocity fields were converted from particle images using the Dantec software DynamicSudio v4.0 with a correlation-based interrogation scheme (adaptive correlation). This uses a multi-grid processing of the data with outlier detection, resulting in a final grid of ${32\times 32\,\mathrm{pix}}$ interrogation areas with $50\%$ overlap, corresponding to a physical interrogation window size in $E_1$ of ${\Delta^2 = \left(1.7\,\mathrm{mm}\right)^2}$ and ${\left(2.6\,\mathrm{mm}\right)^2}$ for cameras 1 and 2. For $E_2$ the interrogation windows were ${\Delta^2 = \left(2.5\,\mathrm{mm}\right)^2}$ and ${\left(3.0\,\mathrm{mm}\right)^2}$ for cameras 1 and 2, respectively. Window shifting with moving averages was applied in order to increase the dynamic range of particle displacement estimates in both experiments.  
\FloatBarrier
\subsection{Validation of results}\label{sec:data_acquisition_and_validation}
\begin{figure}[h]
\centering
\subfloat[]{\includegraphics[width=0.5\linewidth]{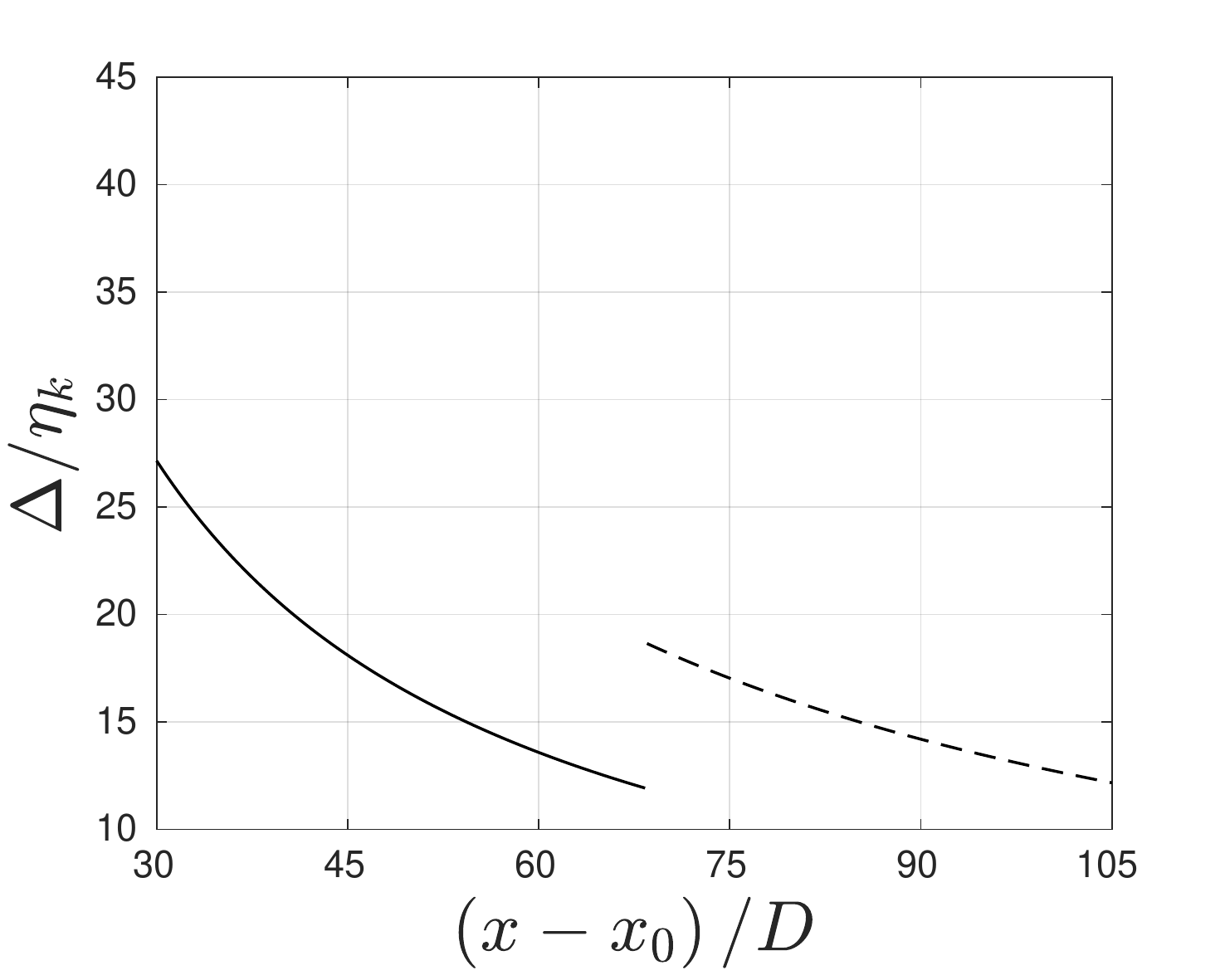}\label{fig:resolution_azur}}
\subfloat[]{\includegraphics[width=0.5\linewidth]{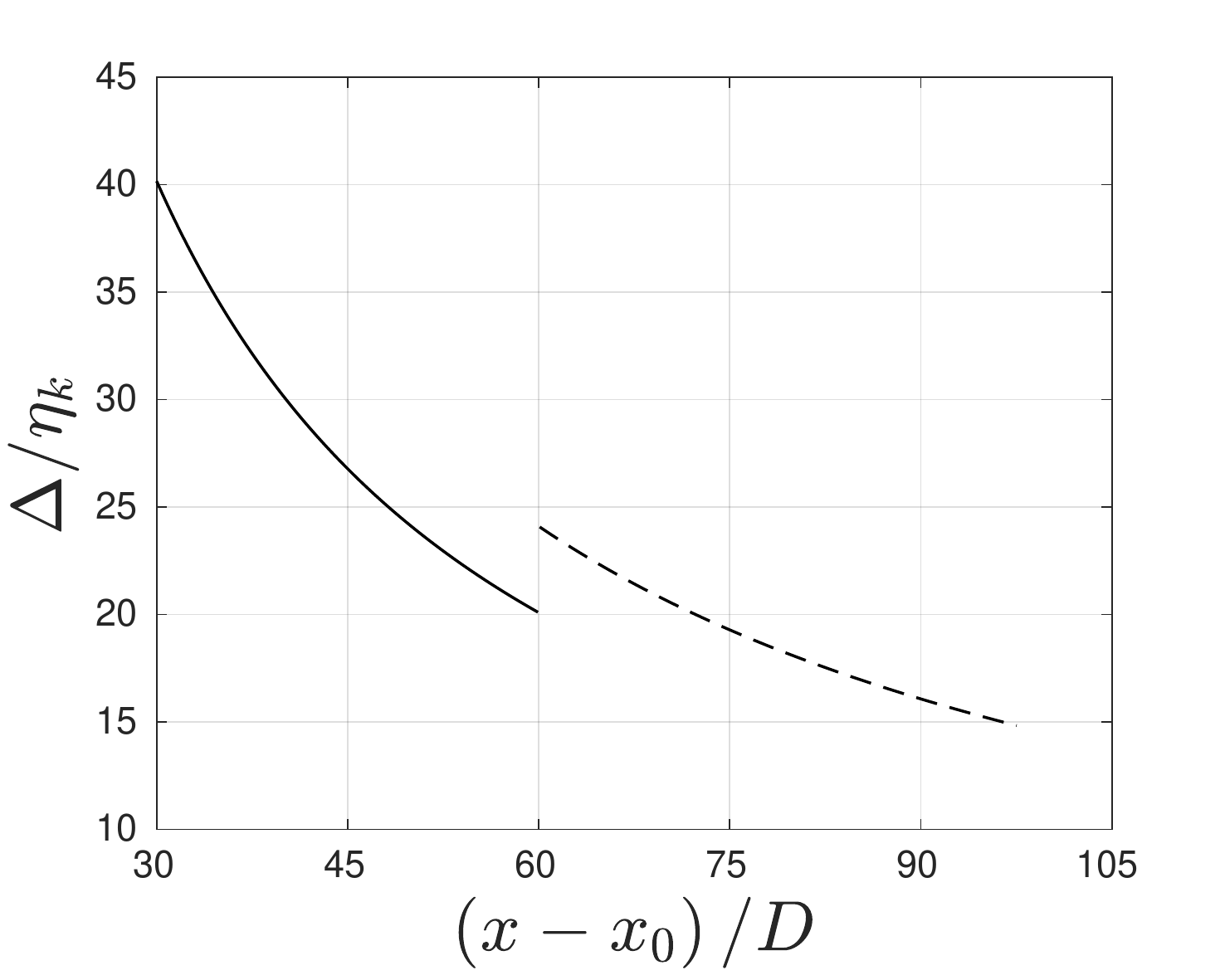}\label{fig:resolution_maja}}
\caption{The effective spatial resolution ratio of the $16,\mathrm{Mpix}$ ($E_1$) and $4\,\mathrm{Mpix}$ ($E_2$) cameras seen in (a) and (b), normalized by the Kolmogorov microscale estimate. The discontinuity is caused by the relative distances between the laser sheet and the cameras.\label{fig:resolution}}
\end{figure}

\noindent\\
Applying equilibrium similarity analysis, \cite{George1989}, \cite{Hussein1994}, to the one-dimensional energy spectrum the following expression for the Kolmogorov microscale can be deduced in the fully developed region of the jet
\begin{equation}
\eta_k\left(x\right) = {(\tilde{\varepsilon}{(B_u\Rey_D)}^3)}^{-\frac{1}{4}}(x-x_0),\label{eq:kolmogorov}
\end{equation}
where the non-dimensional dissipation, ${\tilde{\varepsilon} = 0.36}$, was estimated from \cite{Hussein1994}. For $E_1$ and $E_2$ the centerline velocity decay rates were ${B_u=5.76}$, and ${B_u=5.72}$, respectively and the virtual origins were evaluated to be ${x_0=3.1D}$ and ${x_0=2.4D}$ for the two experiments. These values were obtained from a non-linear optimization scheme of a Gaussian fit to the mean stream-wise velocities. The resolutions of the two experiments normalized by the Kolmogorov microscale, $\Delta/\eta_k$, are seen in figure \ref{fig:resolution} and show the effective resolution variation due to the increase of scales with downstream position. The discontinuity, due to the transition between fields-of-view of the two cameras only affects small scale estimates, as these are below the cut-off wavenumber caused by spatial filtering. This threshold value for $E_1$ can be estimated from figure \ref{fig:resolution_azur} and \eqref{eq:kolmogorov} to ${\tilde{\kappa}_{c,s} = \pi/\Delta_{c,s}}$, where the dimensionless cut-off wavenumber can be estimated to be ${\kappa_{c,s} \approx 560}$. 
\FloatBarrier
\noindent

Since the discontinuous ``jump" must represent a $\kappa$-value greater than $\kappa_{c,s}$, it means that the corresponding effects on the spectrum are located after the roll-off wavenumber, caused by spatial filtering. The effect of the discontinuity on the results can therefore be neglected in the present analysis. 
\section{Experimental results}
In the following, the single-point statistics are presented herein the components from the Reynolds stress tensor as well as the turbulence kinetic energy production across the jet. The statistics of $E_1$ are compared to those of $E_2$ after which the streamwise energy spectra are constructed based on the SADFM derived in section \ref{sec:LD_in_SSC}. Finally, the numerical eigenfunctions from the LD are discussed.
\subsection{Single-point statistics}
\begin{figure}[h]
 \centering
\subfloat[]{\includegraphics[width=0.6\linewidth]{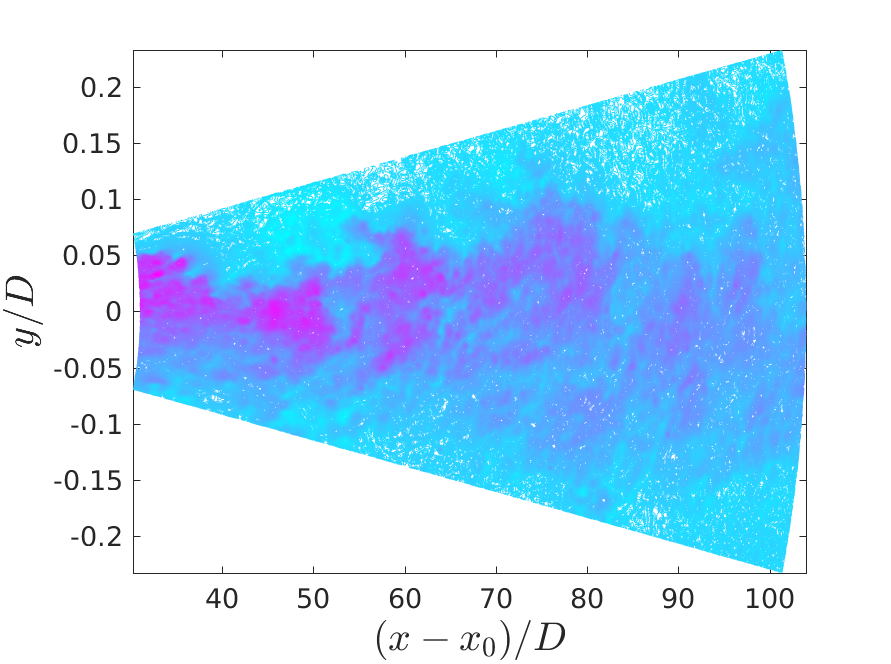}\label{fig:instantaneous_SSC}}\\
\subfloat[]{\includegraphics[width=0.6\linewidth]{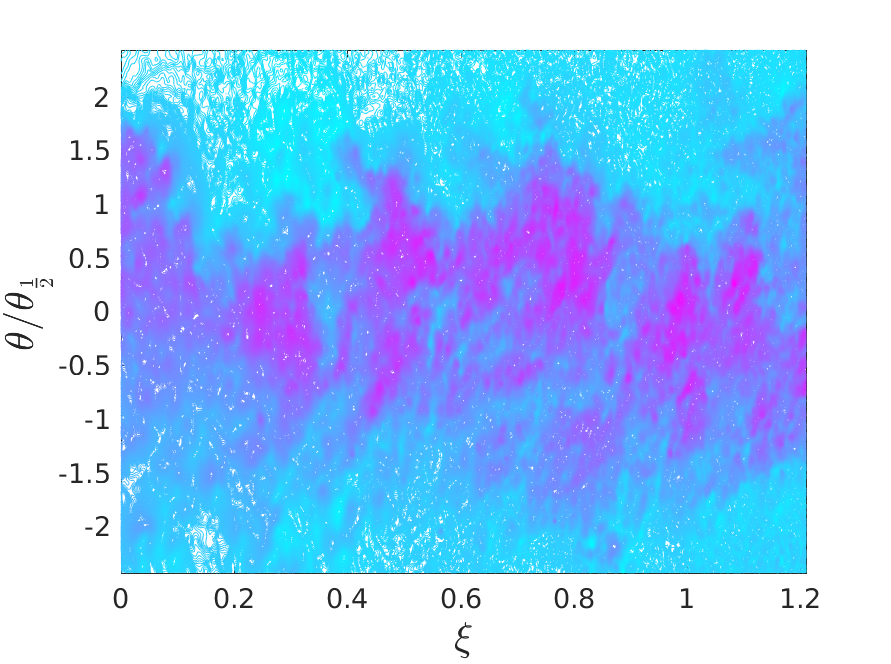}\label{fig:instantaneous_scaled_SSC}}
\caption{(a): Velocity magnitude in SSC and (b): velocity magnitude in SSC scaled with the centerline velocity, $U_c$, depicting a homogeneous turbulent field. \label{fig:instantaneous_velocities_SSC}}
\end{figure}

\begin{figure}[htp]
\centering   
\subfloat[]{\includegraphics[width=0.40\linewidth]{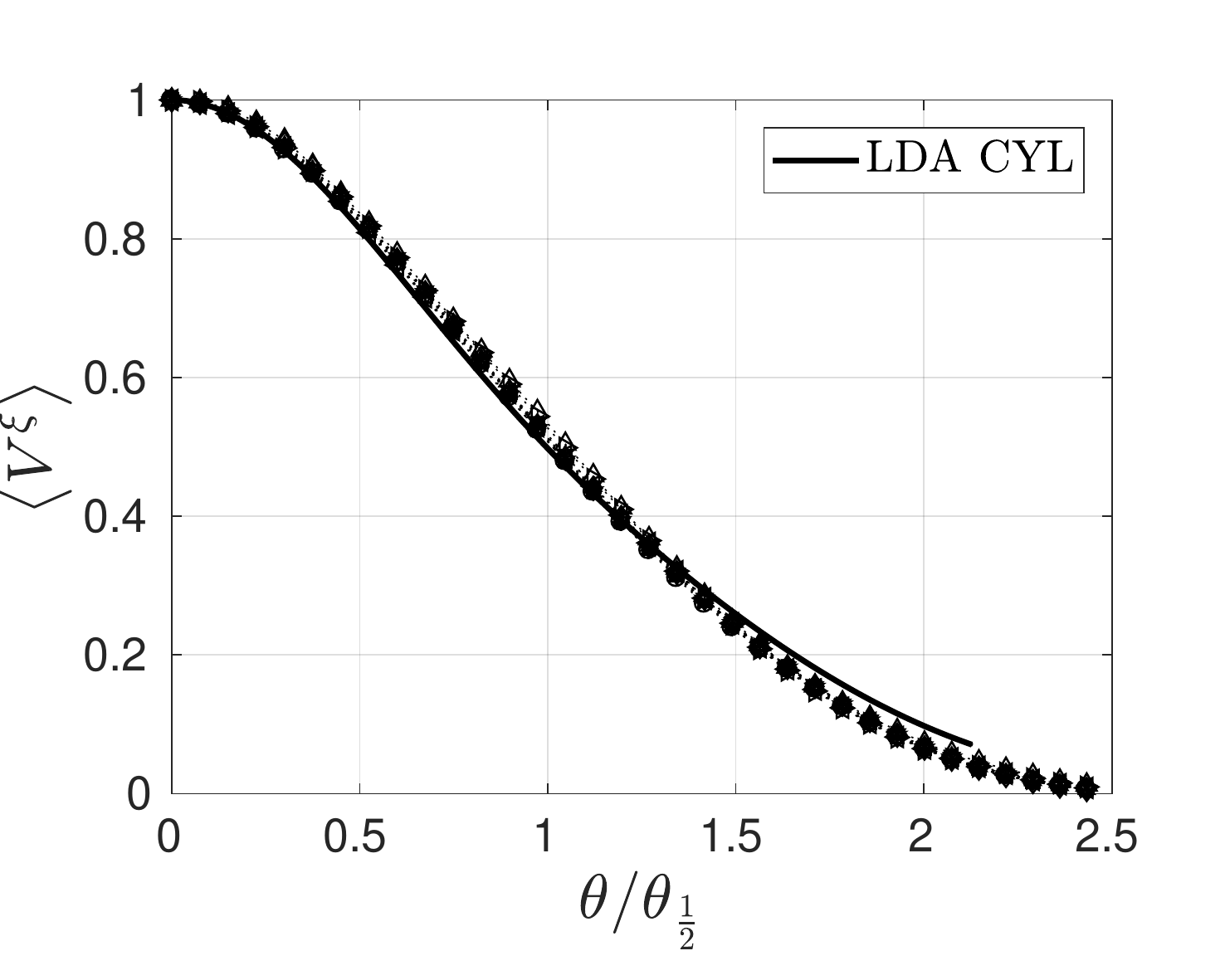}\label{fig:Um_multiple_SSC}}
\subfloat[]{\includegraphics[width=0.40\linewidth]{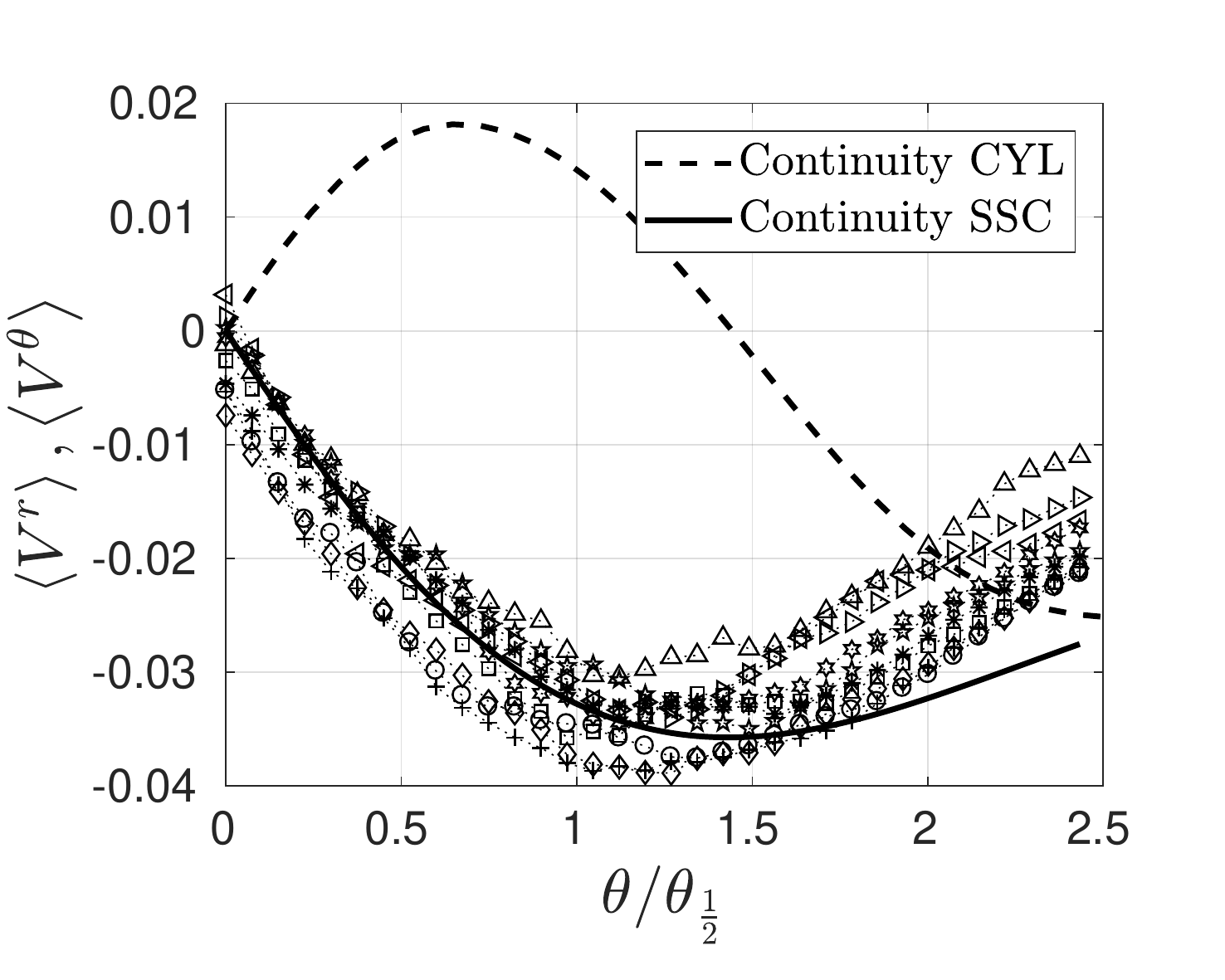}\label{fig:Vm_multiple_SSC}}\\
\subfloat[]{\includegraphics[width=0.40\linewidth]{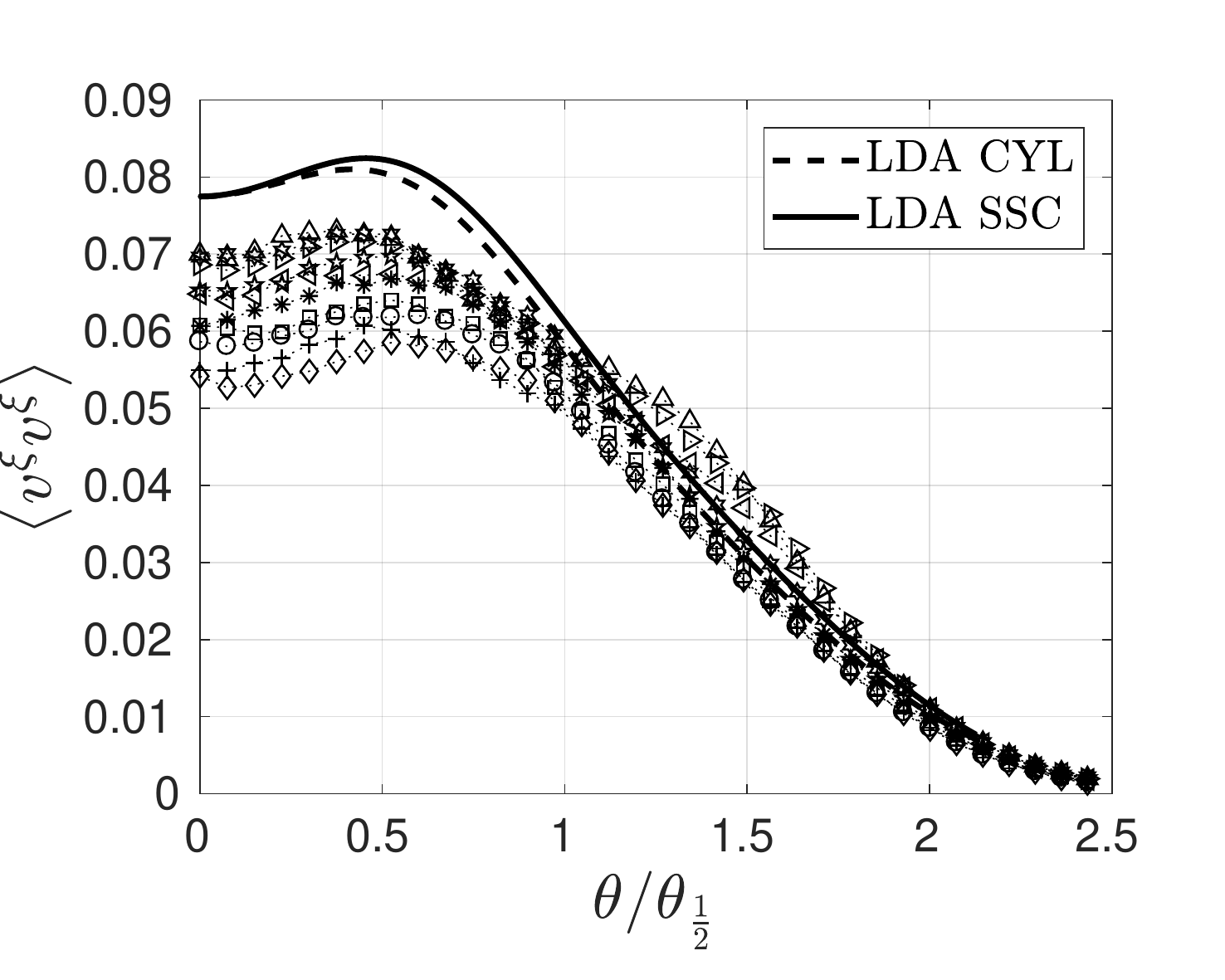}\label{fig:vxi_vxi_multiple_SSC}}
\subfloat[]{\includegraphics[width=0.40\linewidth]{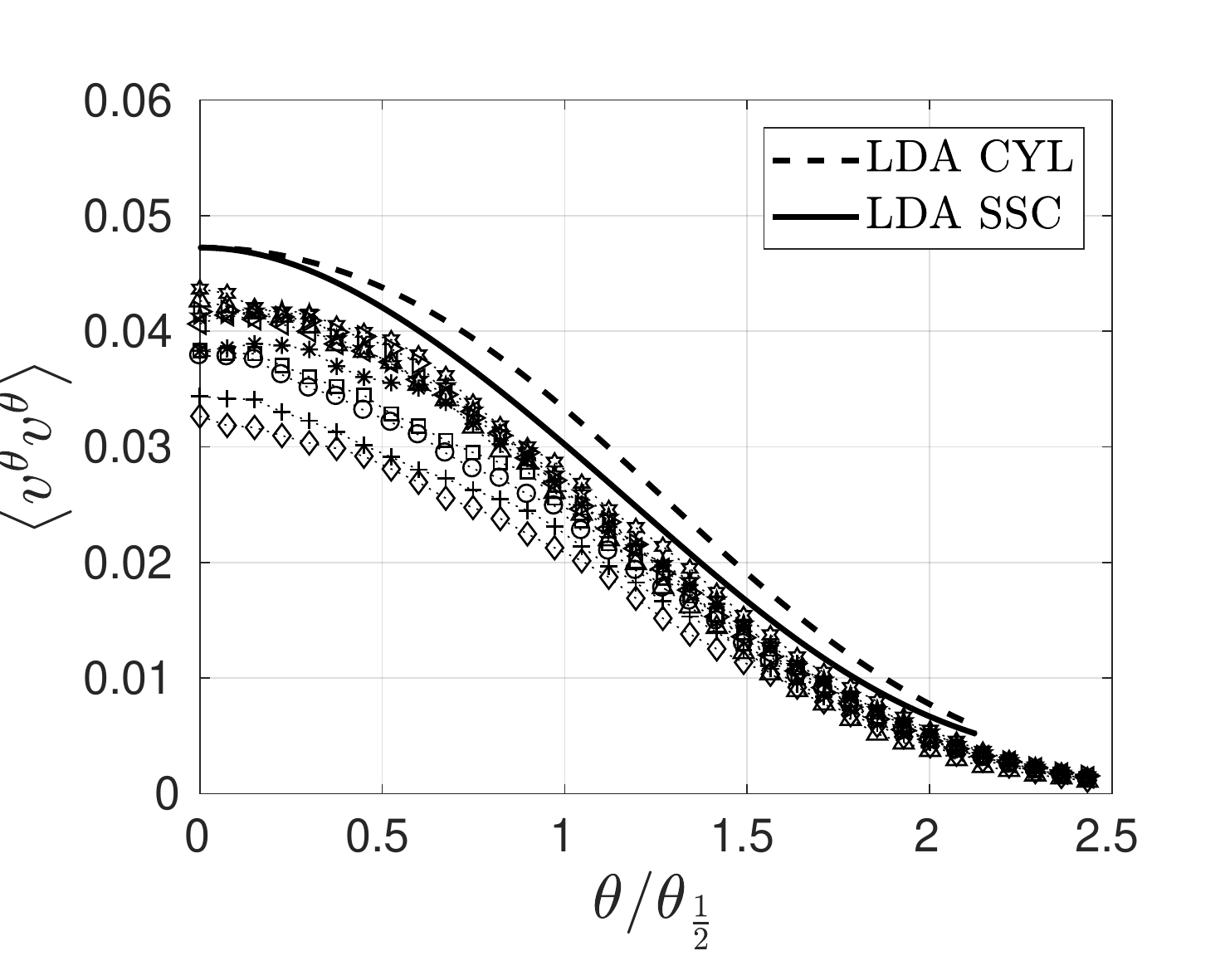}\label{fig:vtheta_vtheta_multiple_SSC}}\\
\subfloat[]{\includegraphics[width=0.40\linewidth]{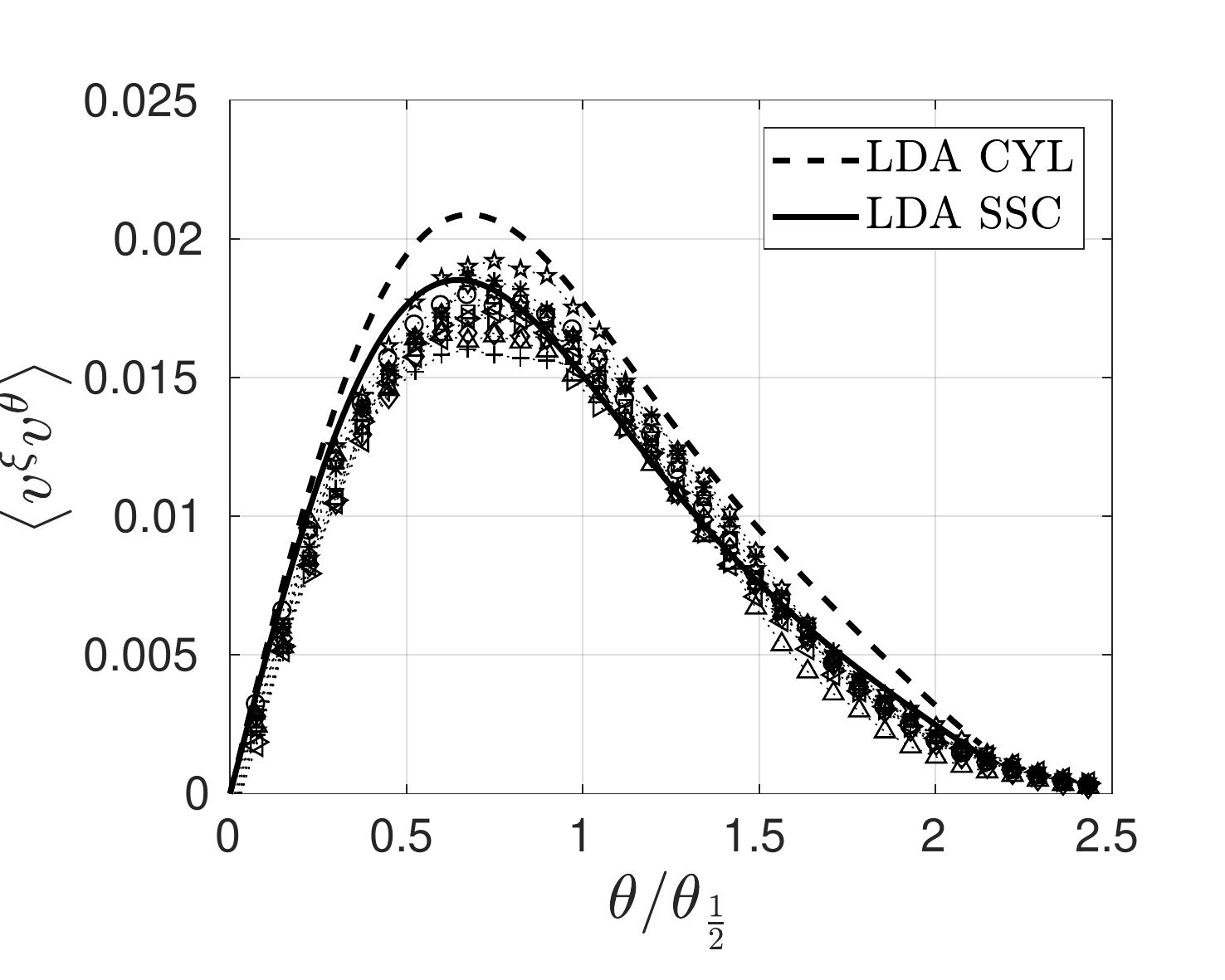}\label{fig:vxi_vtheta_multiple_SSC}}
\caption{Single-point statistics from $E_1$ sampled at the following streamwise coordinates $(x-x_0)/D=[31.0,35.2,40.0,45.5,51.7,58.7,66.7,75.7,86.1,97.8]$. (a): Mean streamwise velocity, (b): mean radial velocity, (c): normal stresses in the streamwise direction, (d): normal stresses in the radial direction, (e): shear-stresses\label{fig:single-point_statistics_multiple_SSC}}
\end{figure}

The instantaneous SSC velocities are seen in figure \ref{fig:instantaneous_velocities_SSC}, where the arc of constant $\xi$ is seen at the beginning and end of the domains in figure \ref{fig:instantaneous_SSC}. Figure \ref{fig:instantaneous_scaled_SSC} shows the instantaneous velocity field scaled with the local centerline velocity. For a constant $\xi$-coordinate in a self-similar flow, the deviation of the single-point statistics between the representation in SSC and cylindrical coordinates is strictly due to the rotation of the coordinate system. Therefore, identical profiles to those seen in figure \ref{fig:single-point_statistics_multiple_SSC} could theoretically be captured from data sampled from a plane perpendicular to the centerline of the jet combined with a rotation of the coordinate system.
\FloatBarrier
\noindent
This, however, is not true in the case of two-point statistics, since the two-point correlation is dependent on the instantaneous field at multiple points in space. Even for a self-similar flow, the correlation between two points, $P_A=(\xi_1,\theta_1,0)$ and $P_B=(\xi_2,\theta_2,0)$, would not be the same as between point $P_A$ and $P_C=(\xi_1+\Delta\xi,\theta_2,0)$ - where $\Delta\xi$ is chosen such that point $P_C$ is located perpendicular to the centerline directly above point $(\xi_1,0,0)$. This is due to the inhomogeneity of the turbulence represented by the scaled flow field. 

The $U_c$-scaled physical mean velocity profiles and the Reynolds stresses from $E_1$ are shown in figures \ref{fig:single-point_statistics_multiple_SSC}. Note that the scale for the mean $\theta$-component differs by a factor of $1.0/0.06$ from the streamwise velocity. This component is rarely (if ever) shown in most reports of experimental data, so its concurrence with the continuity estimate using the mean streamwise velocity is quite gratifying. The negative $\left\langle V^\theta\right\rangle$-component is obtained from the continuity equation in SSC using the following expression for a constant density flow
\begin{equation}
\left\langle V^\theta\right\rangle = -\frac{1}{\tan\theta}\int_0^{\theta}\left\langle V^\xi\right\rangle\tan\tilde{\theta} d\tilde{\theta}\label{eq:Vtheta_SSC},
\end{equation}
which is shown in figure \ref{fig:Vm_multiple_SSC} with the measured $\left\langle V^\theta\right\rangle$-components. The simplicity of \eqref{eq:Vtheta_SSC} should not be taken lightly. An optimal coordinate system simplifies the expression of the dynamics that it depicts. For the far-field region of the jet the SSC seems to depict the velocity field in a simpler way than any other classical coordinate system. The simpler the statistical profiles of the contravariant velocity components are the closer the coordinate system is to the velocity field itself. In addition to this, when the coordinate system itself has a simple geometrical formulation such as \eqref{eq:x_SSC}-\eqref{eq:z_SSC}, much intuition can be gained by viewing the turbulent jet velocity field in terms of a spherical coordinate system. 

The prominent feature of this depiction is the mean $\theta$-component which indicates the entrainment of the surrounding fluid across the jet. It is seen from \eqref{eq:Vtheta_SSC} that $\left\langle V^\theta\right\rangle$ is negative for $\theta>0$ telling us that the jet on average is continuously entraining fluid across its entire width. We note that the profiles collapse well when taking into consideration that $\left\langle V^\theta\right\rangle$ is an order of magnitude smaller than $\left\langle V^\xi\right\rangle$. The shapes of the profiles seen in \ref{fig:Vm_multiple_SSC} stand in stark contrast to the shape of the radial velocity component, $\left\langle V^r\right\rangle$, in cylindrical coordinates also shown in the figure \ref{fig:Vm_multiple_SSC}. In figure \ref{fig:Vm_multiple_SSC} the profile from \eqref{eq:Vtheta_SSC} is also shown to agree well with the measured averaged $\left\langle V^\theta\right\rangle$-profiles. After approximately $\theta/\theta_{\frac{1}{2}}>1$ the profiles deviate from the profile obtained from \eqref{eq:Vtheta_SSC}, most likely due to back flow caused by the finite confinement, \cite{Hussein1994}. 
\begin{figure}[htp]
\centering   
\subfloat[]{\includegraphics[width=0.40\linewidth]{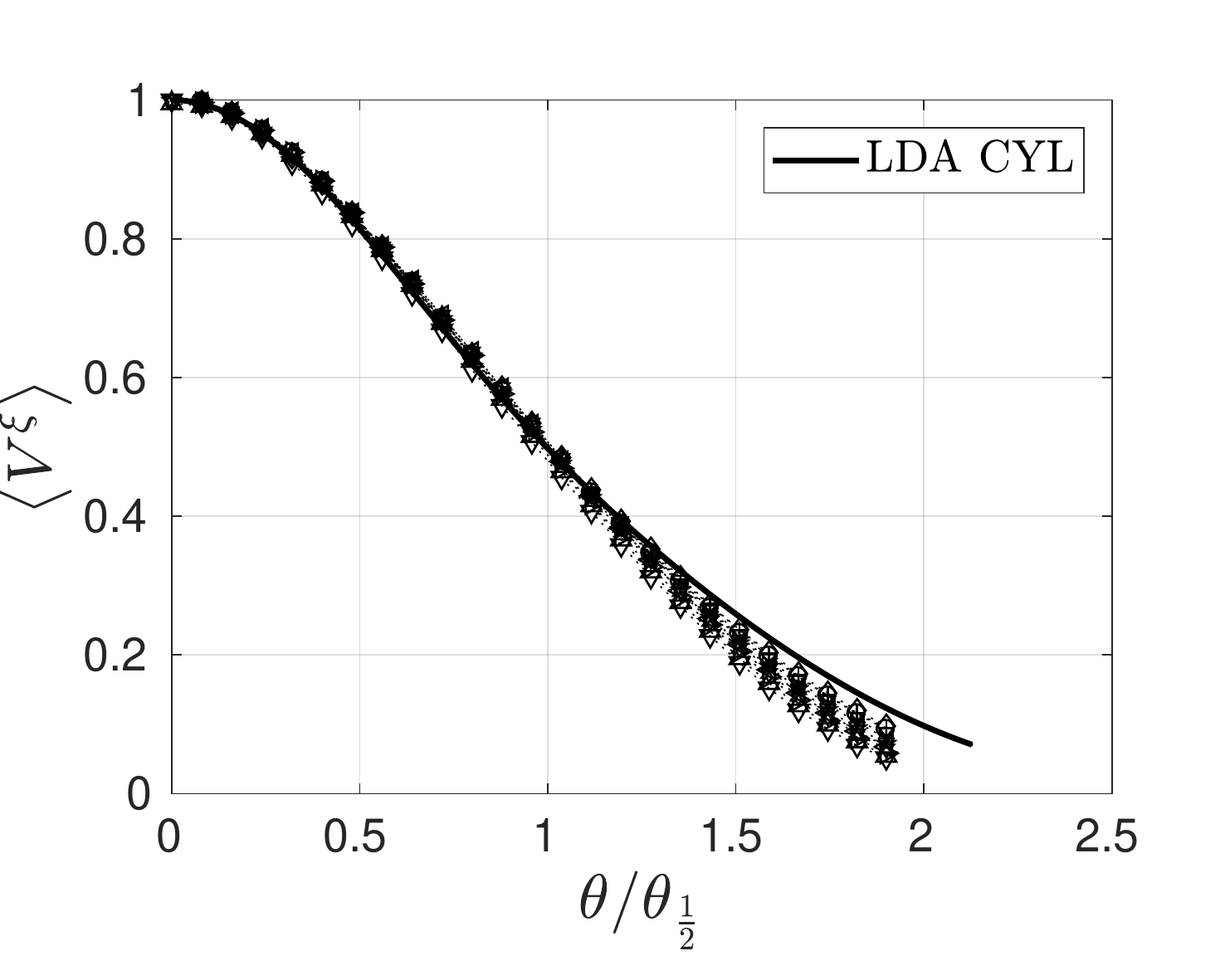}\label{fig:Um_multiple_SSC_maja}}
\subfloat[]{\includegraphics[width=0.40\linewidth]{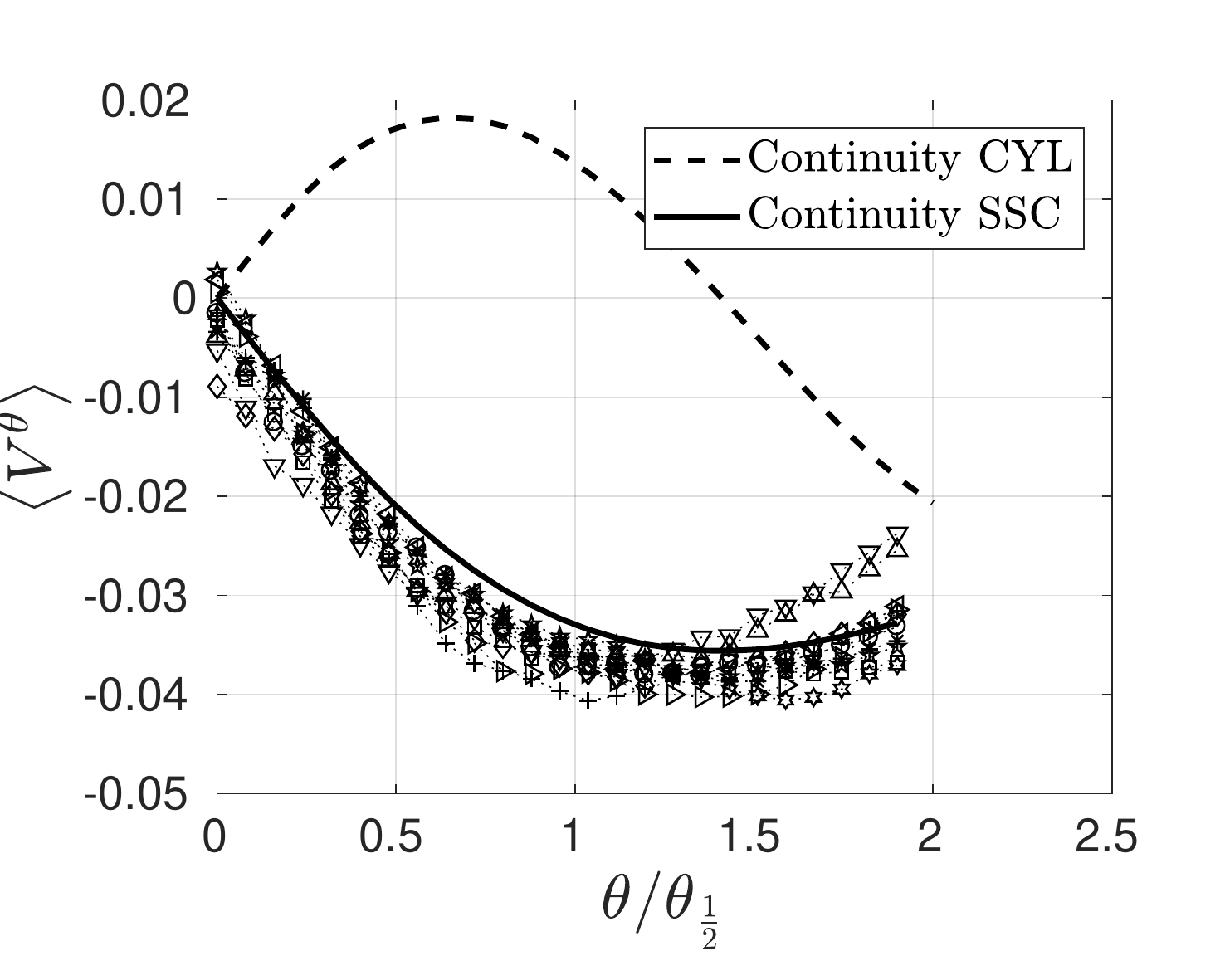}\label{fig:Vm_multiple_SSC_maja}}\\
\subfloat[]{\includegraphics[width=0.40\linewidth]{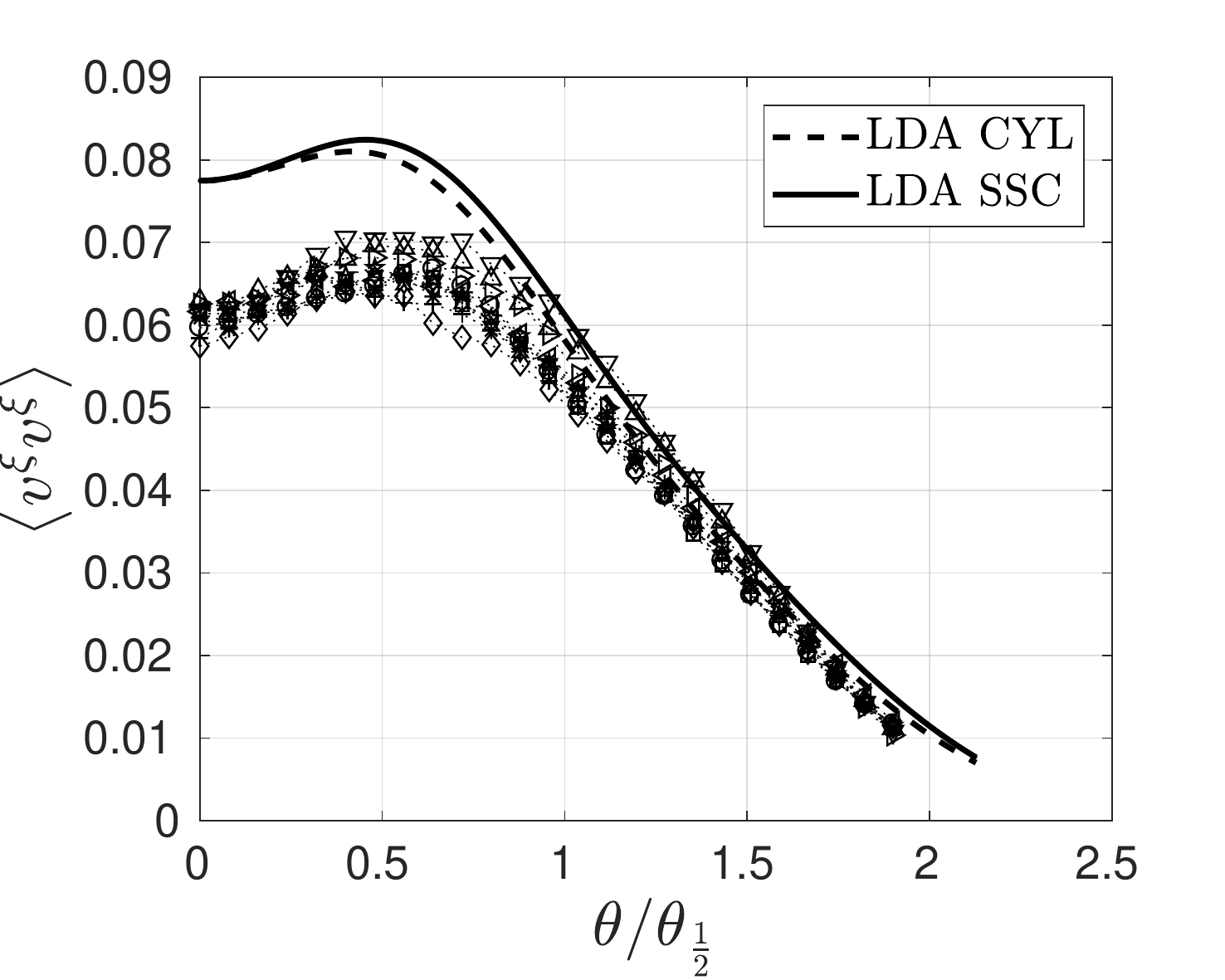}\label{fig:vxi_vxi_multiple_SSC_maja}}
\subfloat[]{\includegraphics[width=0.40\linewidth]{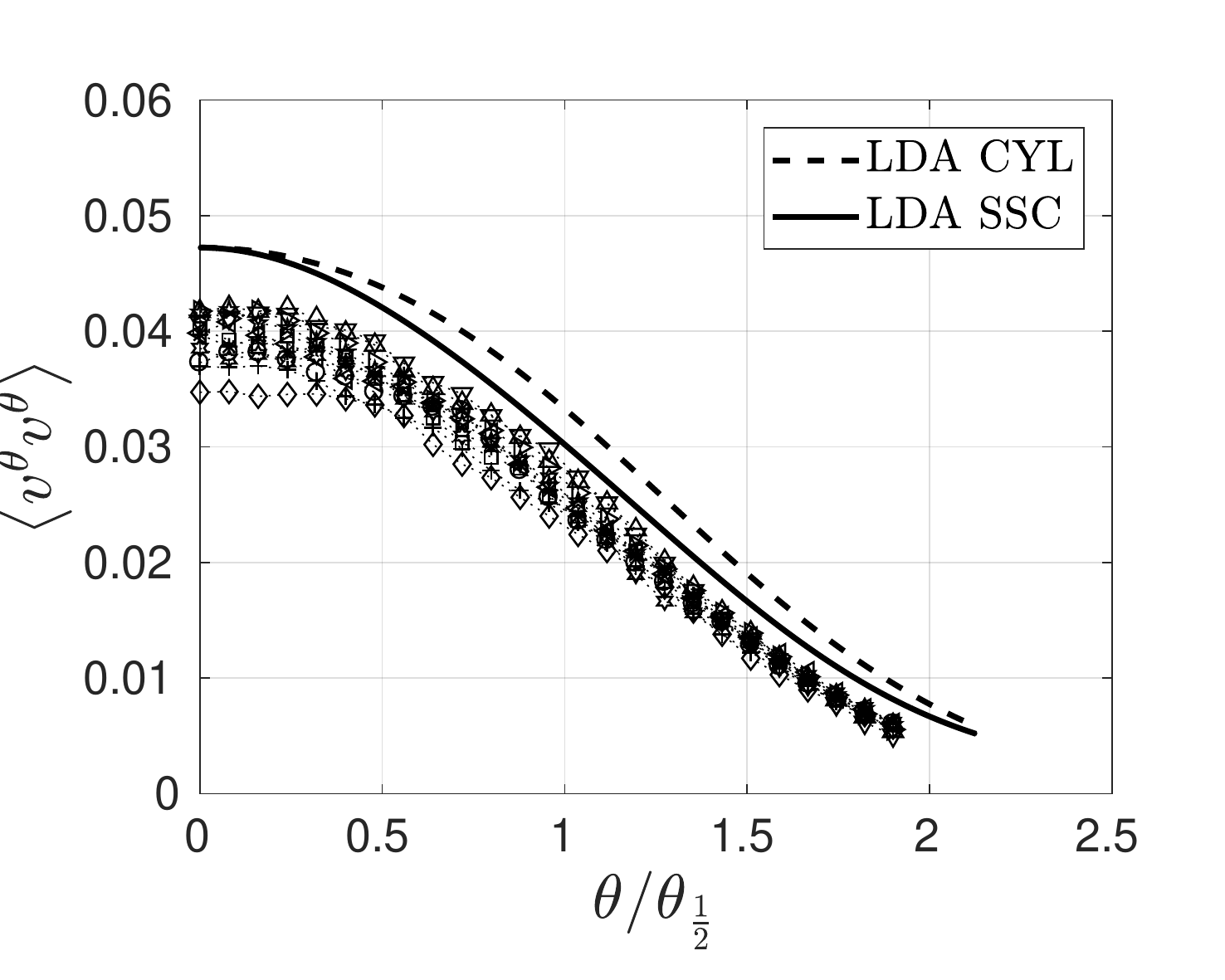}\label{fig:vtheta_vtheta_multiple_SSC_maja}}\\
\subfloat[]{\includegraphics[width=0.40\linewidth]{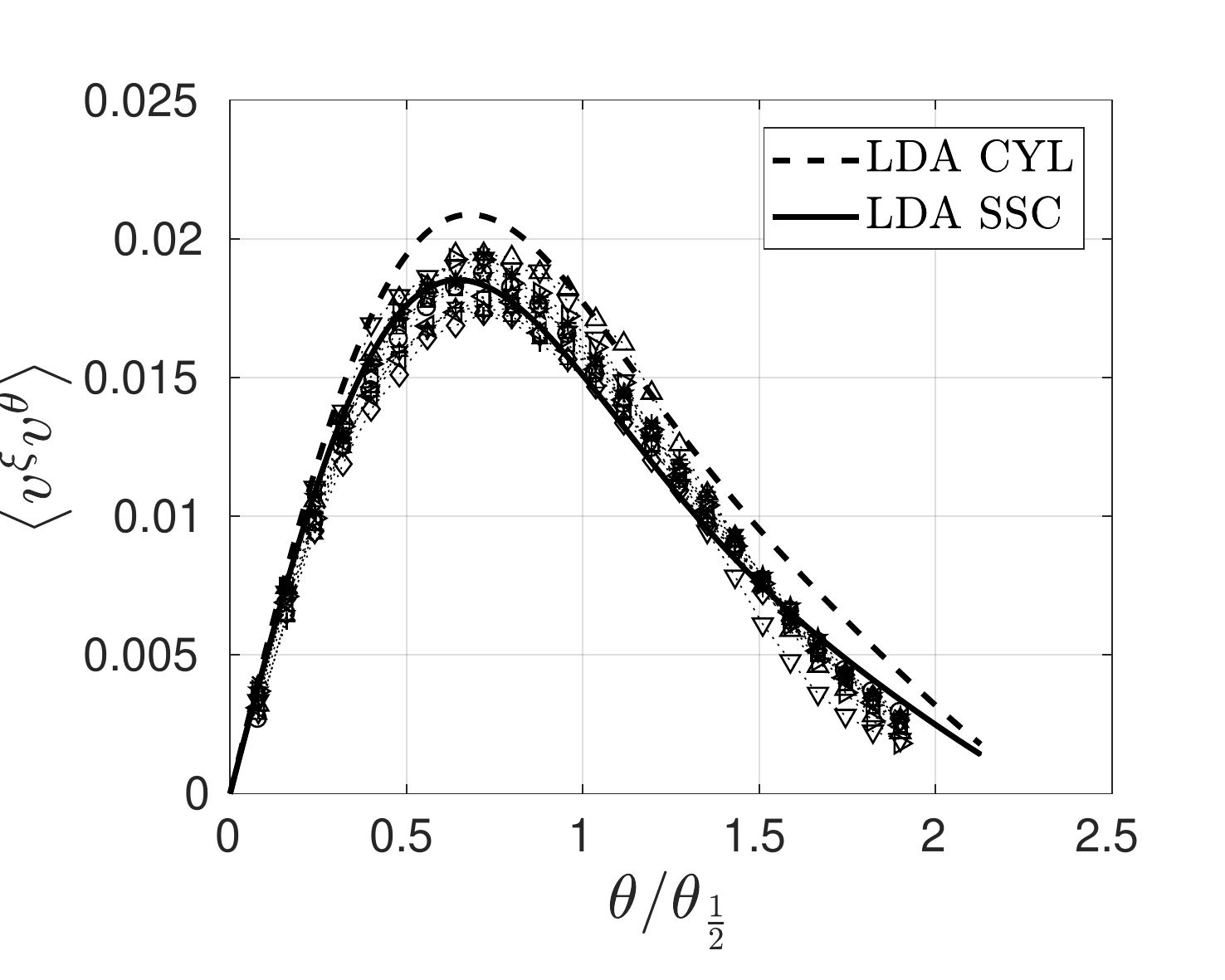}\label{fig:vxi_vtheta_multiple_SSC_maja}}
\caption{Single-point statistics from $E_2$ sampled at the following streamwise coordinates $(x-x_0)/D=[30.5, 34.2, 38.4,43.1,48.4,54.3,60.9,68.4,76.7,86.1,96.6]$. (a): Mean streamwise velocity, (b): mean radial velocity, (c): normal stresses in the streamwise direction, (d): normal stresses in the radial direction, (e): shear-stresses\label{fig:single-point_statistics_multiple_SSC_maja}}
\end{figure}

The Reynolds stresses at different streamwise locations are shown in figures \ref{fig:vxi_vxi_multiple_SSC}-\ref{fig:vxi_vtheta_multiple_SSC}. The profiles in figure \ref{fig:Um_multiple_SSC}, \ref{fig:vxi_vxi_multiple_SSC}, \ref{fig:vtheta_vtheta_multiple_SSC} and \ref{fig:vxi_vtheta_multiple_SSC} are shown together with corresponding LDA profiles from \cite{Hussein1994}. These profiles are depicted both in cylindrical but also in SSC such that they can be used with the current data. The conversion relations from cylindrical to SSC along the centerline plane are given by
\begin{eqnarray}
\left\langle v^\xi v^\xi\right\rangle &=& \left\langle v^x v^x\right\rangle\cos^2\theta+\left\langle v^r v^r\right\rangle\sin^2\theta+2\left\langle v^x v^r\right\rangle\cos\theta\sin\theta,\\
\left\langle v^\theta v^\theta\right\rangle &=& \left\langle v^x v^x\right\rangle\sin^2\theta+\left\langle v^r v^r\right\rangle\cos^2\theta-2\left\langle v^x v^r\right\rangle\cos\theta\sin\theta,\\
\left\langle v^\xi v^\theta\right\rangle &=& \left\langle v^x v^r\right\rangle\left(1-2\sin^2\theta\right)+\left(\left\langle v^r v^r\right\rangle-\left\langle v^x v^x\right\rangle\right)\cos\theta\sin\theta.
\end{eqnarray}
The $\left\langle v^\xi v^\xi\right\rangle$ and $\left\langle v^\theta v^\theta\right\rangle$ profiles are narrower and wider, respectively, than the profiles depicted in cylindrical coordinates. The biggest impact of the coordinate transformation is reflected in the shear-stresses as seen in figure \ref{fig:vtheta_vtheta_multiple_SSC}. The deviation of the Reynolds stresses in SSC from those in cylindrical coordinates shows that the turbulence in the jet is locally anisotropic, since the transformation from cylindrical coordinates to SSC is locally a rotation of the cylindrical coordinate system. Nevertheless the profiles show a good collapse and compare well with the transformed LDA profiles from \cite{Hussein1994}. It is also noticeable that the Reynolds stress profiles increase with downstream position. This is because of the increase in the effective resolution as the scales grow downstream. Figure \ref{fig:single-point_statistics_multiple_SSC_maja} shows the corresponding single-point statistics of $E_2$, which is in good agreement with the profiles seen in figure \ref{fig:single-point_statistics_multiple_SSC}, especially since the two experiments were conducted completely independently and using different experimental equipment (c.f. section \ref{sec:experimental_procedure}). 

\subsection{Energy production}
\begin{figure}[h]
\centering
\subfloat[]{\includegraphics[width=0.40\linewidth]{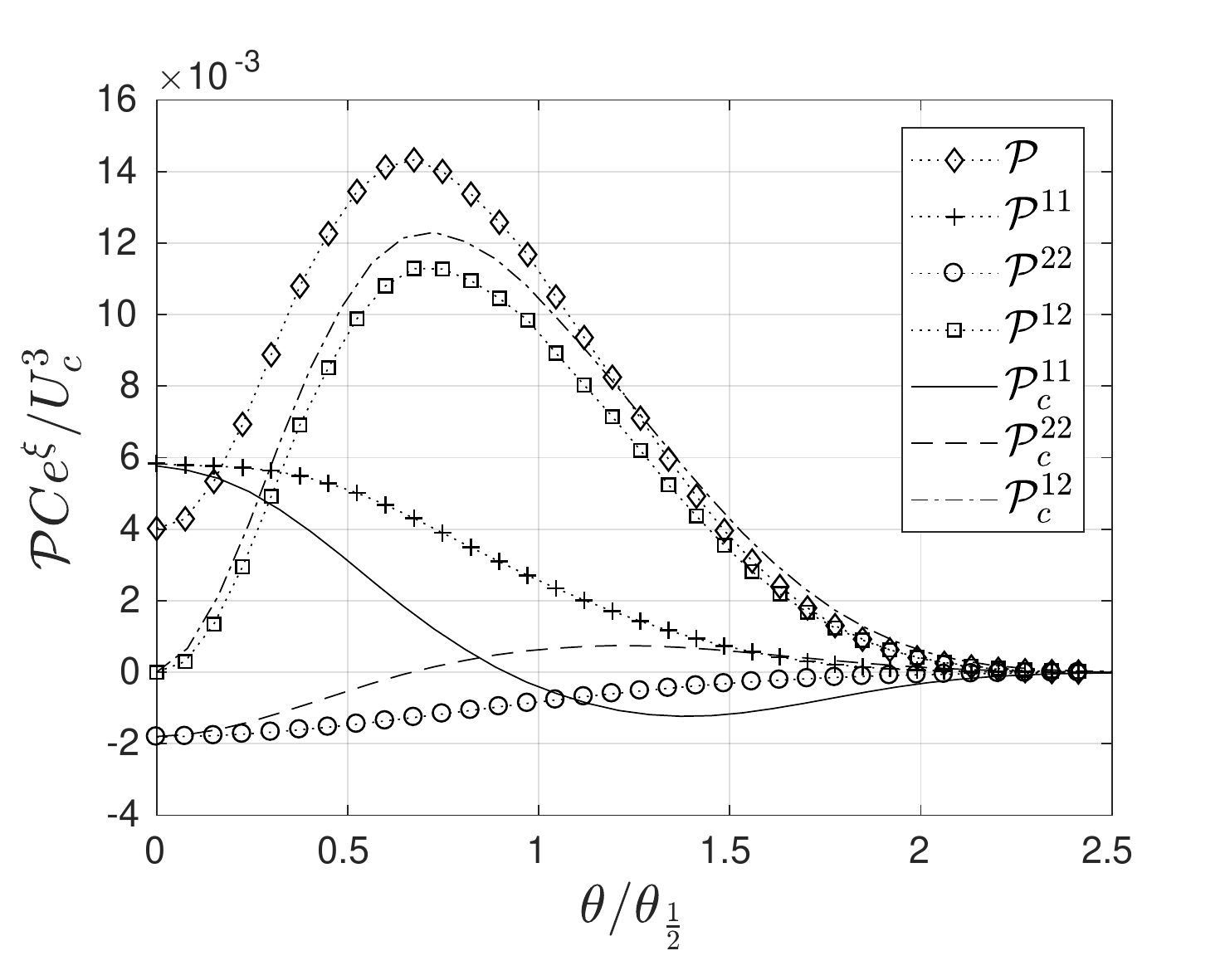}\label{fig:production_budget_SSC}}
\subfloat[]{\includegraphics[width=0.40\linewidth]{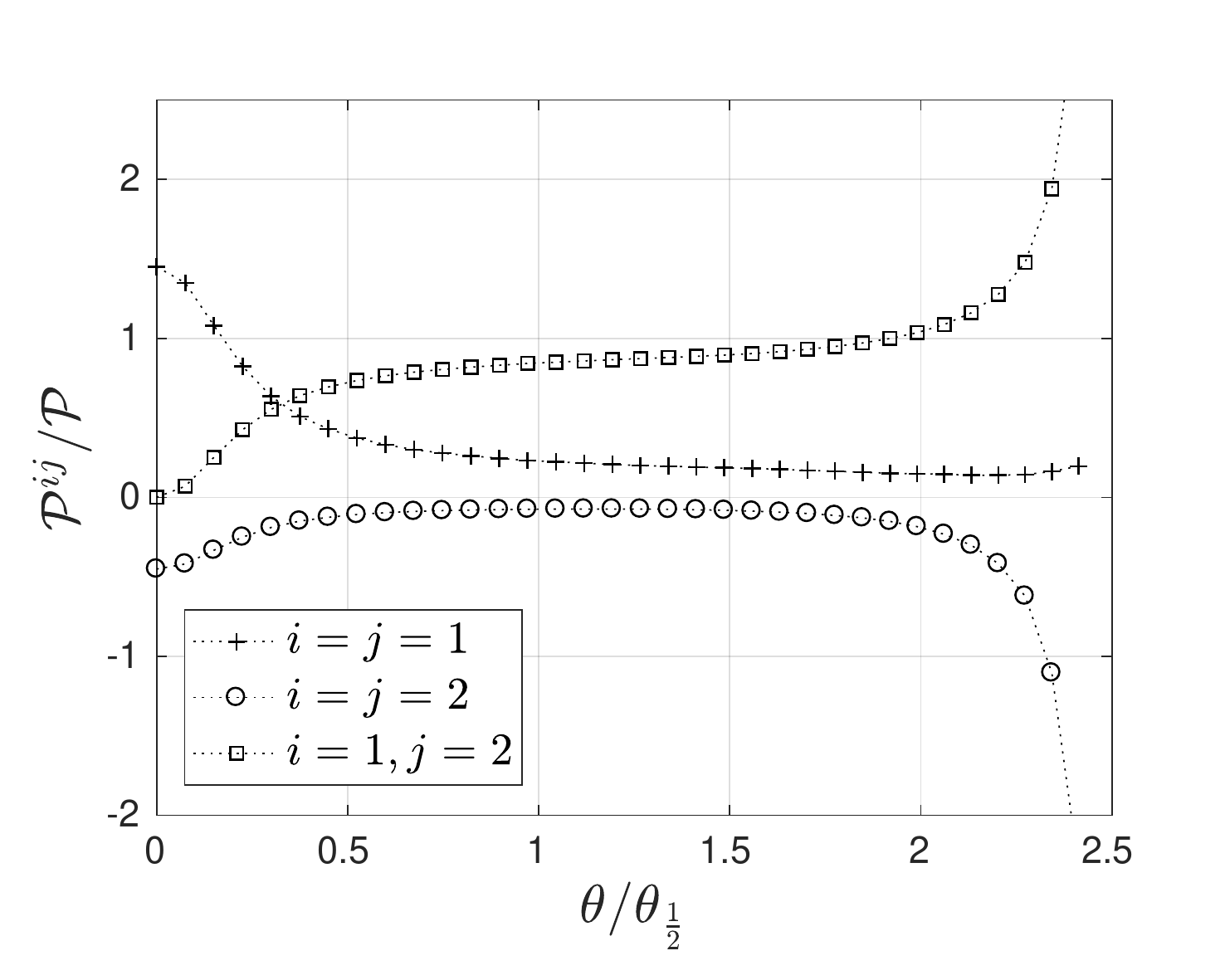}\label{fig:production_budget_local_contribution_SSC}}\\
\subfloat[]{\includegraphics[width=0.40\linewidth]{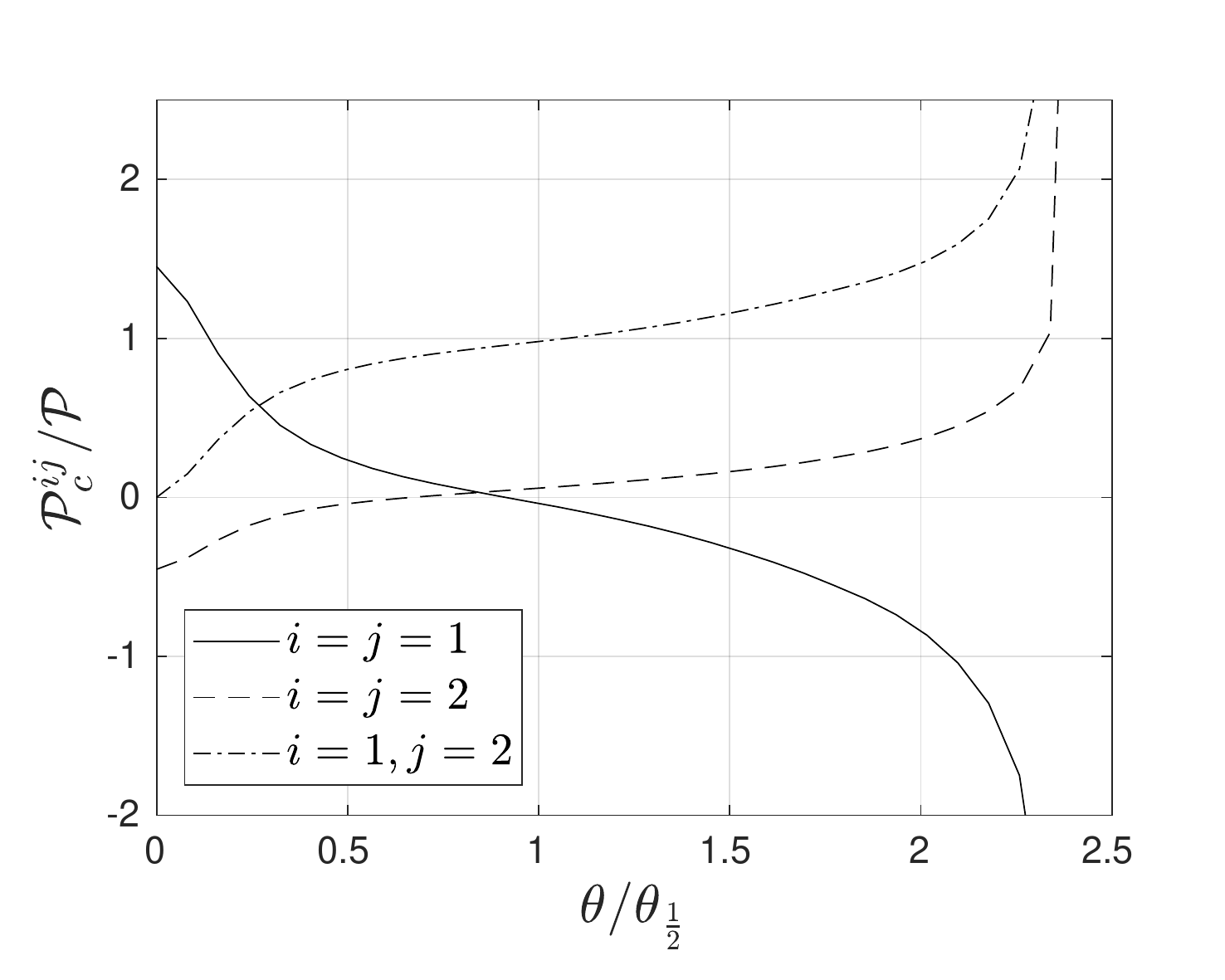}\label{fig:production_budget_local_contribution_CYL}}
\subfloat[]{\includegraphics[width=0.40\linewidth]{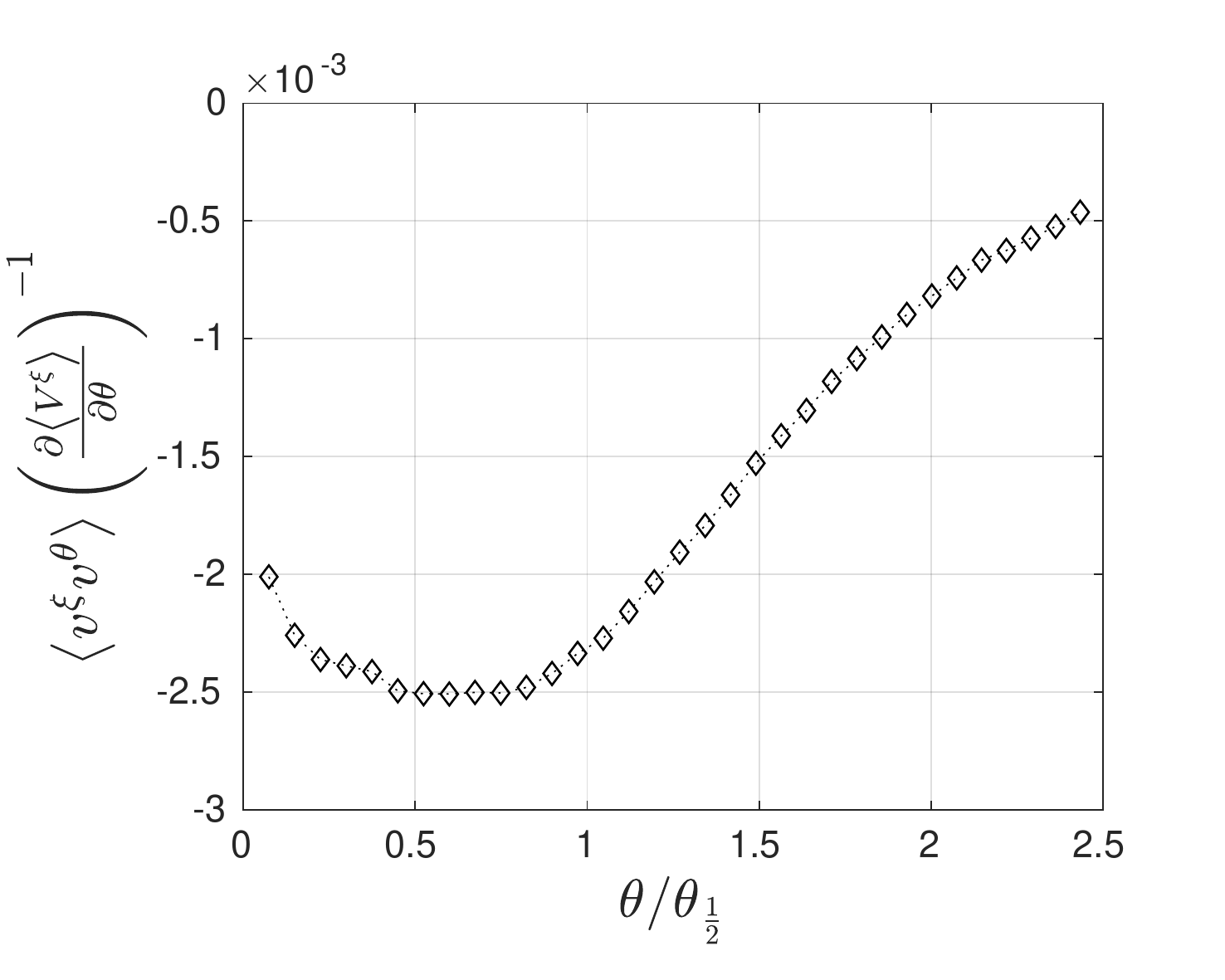}\label{fig:stress_meanvel_grad_SSC}}%
\caption{(a): TKE production components for SSC, $\mathcal{P}^{ij}Ce^\xi/U^3_c$, obtained from \eqref{eq:production_budget_SSC} and correspondingly in cylindrical coordinates, $\mathcal{P}^{ij}_cCe^\xi/U^3_c$, (b): Relative contributions to the turbulence kinetic energy production of normal and shear-stresses in SSC, (c): Relative contributions to the turbulence kinetic energy production of normal and shear-stresses in cylindrical coordinates, (d): Proportionality of shear-stresses and mean velocity gradient.\label{fig:production_budget}}
\end{figure}

Figure \ref{fig:production_budget_SSC} depicts the energy production term, $II$, in \eqref{eq:turb_energy_terms}, which can be expanded such that the terms obtain the following form in SSC
\begin{eqnarray}
\mathcal{P} &=& \underbrace{z_{11}\left\langle v^1v^1\right\rangle \nabla_1\left\langle V^1\right\rangle}_{\mathcal{P}^{11}}+\underbrace{z_{22}\left\langle v^2v^2\right\rangle \nabla_2\left\langle V^2\right\rangle}_{\mathcal{P}^{22}}\label{eq:production_budget_SSC}\\
&+&\underbrace{\left\langle v^1v^2\right\rangle (z_{11}\nabla_2\left\langle V^1\right\rangle +z_{22}\nabla_1\left\langle V^2\right\rangle)}_{\mathcal{P}^{12}}.\nonumber
\end{eqnarray}
These production contributions obtained from $E_1$ are seen in figure \ref{fig:production_budget_SSC}, where they have been normalized by $Ce^\xi/U^3_c$. It is seen that the production contributions are shifted in SSC compared to the cylindrical contributions. Figures \ref{fig:production_budget_local_contribution_SSC} and \ref{fig:production_budget_local_contribution_CYL} are the production contributions normalized by the total production, $\mathcal{P}$, in \eqref{eq:production_budget_SSC}. Figure \ref{fig:production_budget_local_contribution_SSC} depicts a continuously negative contribution to the energy production of $\mathcal{P}^{22}$, unlike the representations in cylindrical coordinates in figure \ref{fig:production_budget_local_contribution_CYL}. 
\FloatBarrier
\noindent
The energy production from mean shear is also smaller than in cylindrical coordinates, naturally due to the decreased Reynolds shear stresses as seen in figures \ref{fig:vxi_vtheta_multiple_SSC} and \ref{fig:vxi_vtheta_multiple_SSC_maja}. The energy contributions of the various components to the total energy production are $27.9\%$, $-10.2\%$ and $82.3\%$ for $\mathcal{P}^{11}$, $\mathcal{P}^{22}$, and $\mathcal{P}^{12}$ respectively. Therefore, a majority of the energy production is related to the shear-stresses, representing a $16.7\%$ decrease in energy production compared to the shear-stresses in cylindrical coordinates. Figure \ref{fig:production_budget_local_contribution_SSC} shows the relative contributions of the terms in \eqref{eq:production_budget_SSC} normalized by the total energy. The variations of the contributions between $\theta/\theta_{\frac{1}{2}}=0.5$ and $2$ are striking, since the normalized energy production components are nearly constant for $\mathcal{P}^{ij}$ over this range. This shows that the components of energy production in \eqref{eq:production_budget_SSC} have approximately the same percentwise contribution to the total energy production across the entire flow except at the centerline and at the periphery of the flow in SSC. Note that this simple relation is not observed in the relative contributions to the production in cylindrical coordinates, as seen in figure \ref{fig:production_budget_local_contribution_CYL}, indicating that the relative production in the jet also exhibits a spherical symmetry.  

The gradual increase of the shear-stress contribution from approximately $70\%$ to $105\%$ over the same jet-span is due to the corresponding decrease of $\mathcal{P}^{11}$. This means that moving away from the centerline of the jet, the contribution of the shear-stresses to the turbulence kinetic energy production increases, while the role of the normal stresses decreases. At the periphery of the jet the shear-stress contribution increases immensely due to the dominance of the second terms in \eqref{eq:physical_velocity_xi_SSC} and \eqref{eq:physical_velocity_theta_SSC} for large $\theta$. It is evident from this that the most dynamic region in the turbulent jet far-field is the region around one half-width. Many of the characteristic profiles have their extrema around this region and the depiction of the mean spanwise velocity component, \eqref{eq:Vtheta_SSC}, indicates that large turbulent structures are located in this region. These may very well have a significant influence on the location of the extremum of the $\left\langle V^\theta\right\rangle$, and thereby the entrainment rate. 
\FloatBarrier
\subsection{SADFM spatial spectra\label{sec:Energy_density_spectra_in_SSC}}
One advantage of the PIV technique is that it allows true experimental spatial spectra to be computed for high Reynolds number flows along tilted coordinate axes, as in the case of the SSC. Similar multi-component measurements would be difficult, if not impossible to perform using hot-wires. 

Away from the centerline high turbulence intensities would reduce the validity of Taylor's hypothesis and make effective separation of the velocity components more challenging. Having confirmed that the SADFM are orthogonal with respect to the $L^2(\Omega,\mathbb{C}^3)$-inner product they can be used to expand the flow field. Analogously to the classical trigonometric polynomials the SADFM coefficients are obtained by projecting the physical components of the fluctuating field, $v^\xi$, and $v^\theta$ onto the SADFM (see figure \ref{fig:example_of_mode}) 
\begin{eqnarray}
c^\xi\left(\kappa,\theta\right)&=&\frac{1}{\sqrt{L_\xi C^3\sin\theta}}\int_0^{L_\xi} v^\xi\left(\xi,\theta\right)e^{-i\kappa\xi-\frac{3}{2}\xi}\sqrt{Z}d\xi \label{eq:SADFM_coefficients_xi},\\
c^\theta\left(\kappa,\theta\right)&=&\frac{1}{\sqrt{L_\xi C^3\sin\theta}}\int_0^{L_\xi} v^\theta\left(\xi,\theta\right)e^{-i\kappa\xi-\frac{3}{2}\xi}\sqrt{Z}d\xi \label{eq:SADFM_coefficients_theta},
\end{eqnarray}
where the volume element for the entire three dimensional spatial domain, \eqref{eq:volume_element_SSC}, has been included in the integral over $\xi$. In the computation of the coefficients the FFT can be 
applied in $\xi$-coordinates by multiplying the streamwise decaying velocity components $v^\xi$ and $v^\theta$ with $e^{\frac{3}{2}\xi}$ prior to applying the FFT to the signal.

\begin{figure}[h]
    \centering   
    \subfloat[]{\includegraphics[width=0.40\linewidth]{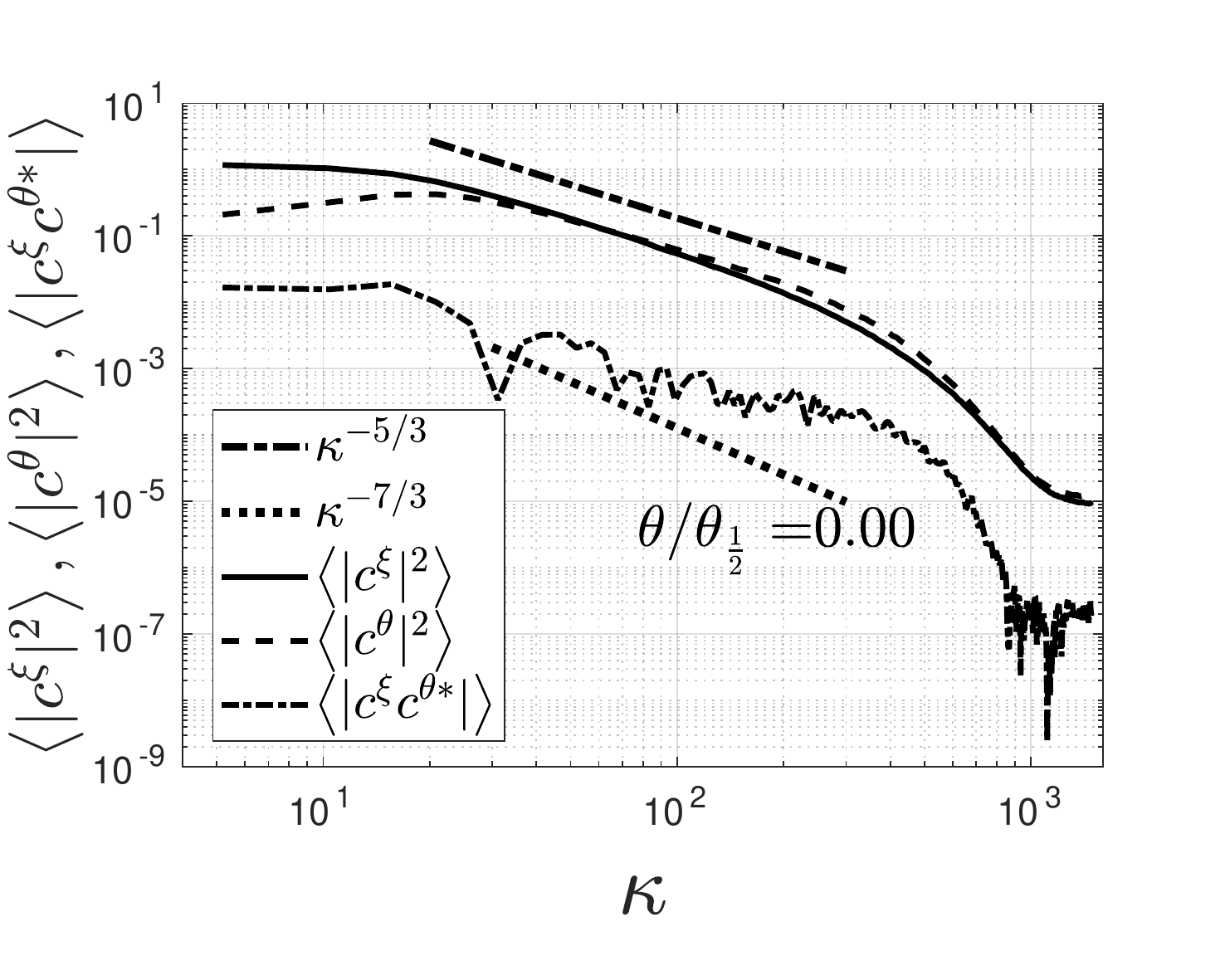}\label{fig:spectra_0}}
    \subfloat[]{\includegraphics[width=0.40\linewidth]{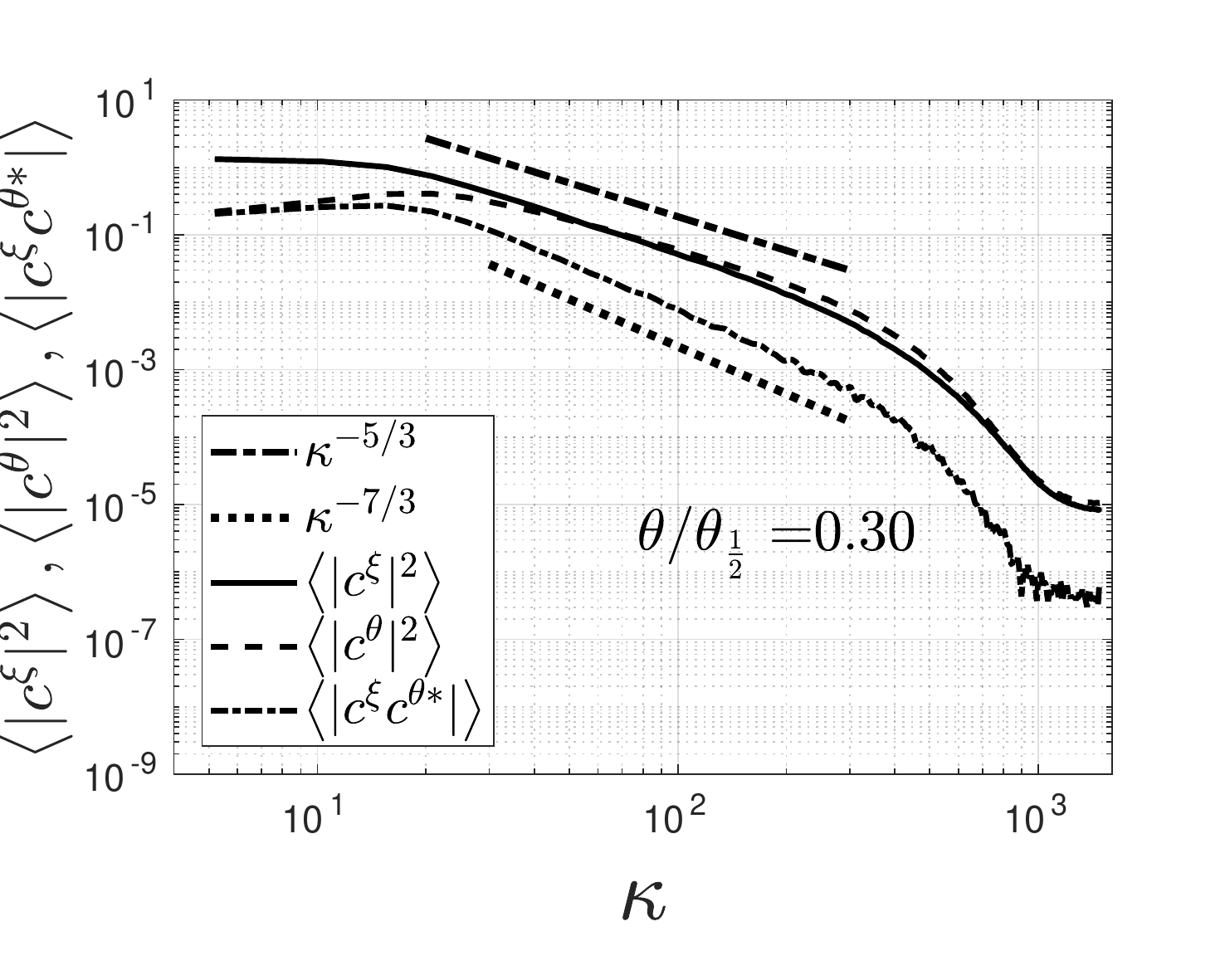}\label{fig:spectra_30_SSC}}\\
    \subfloat[]{\includegraphics[width=0.40\linewidth]{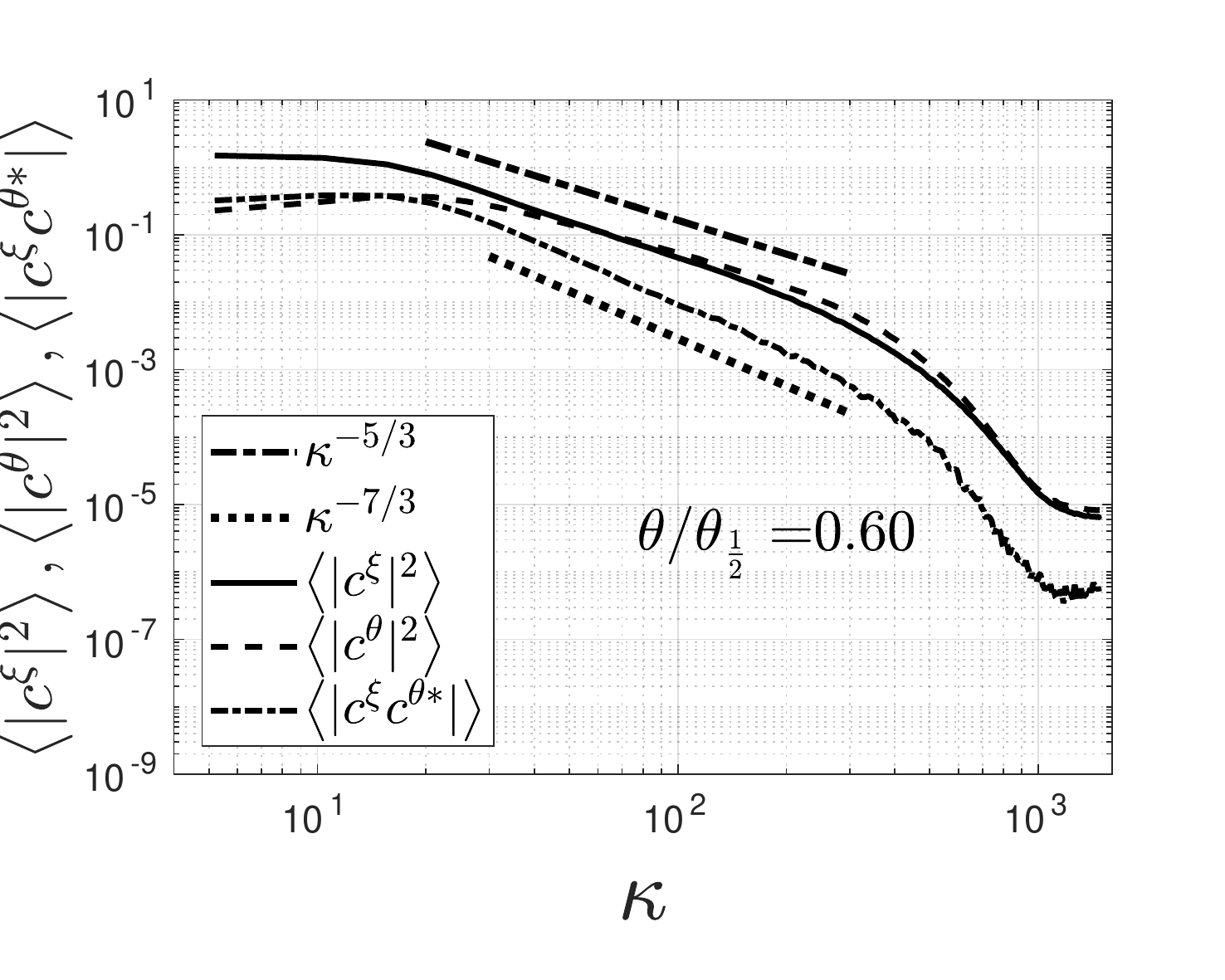}\label{fig:spectra_60_SSC}}
    \subfloat[]{\includegraphics[width=0.40\linewidth]{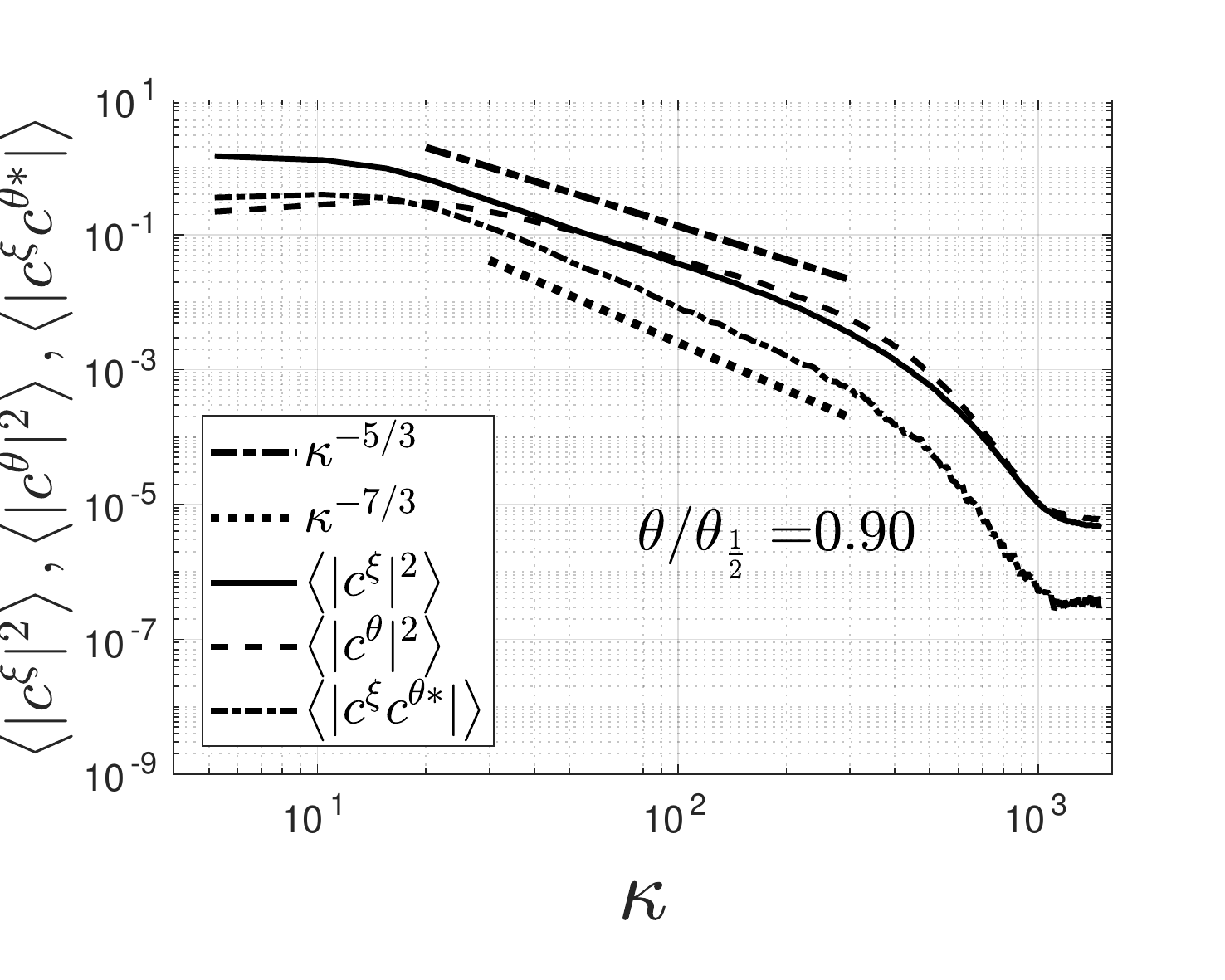}\label{fig:spectra_90_SSC}}\\
    \subfloat[]{\includegraphics[width=0.40\linewidth]{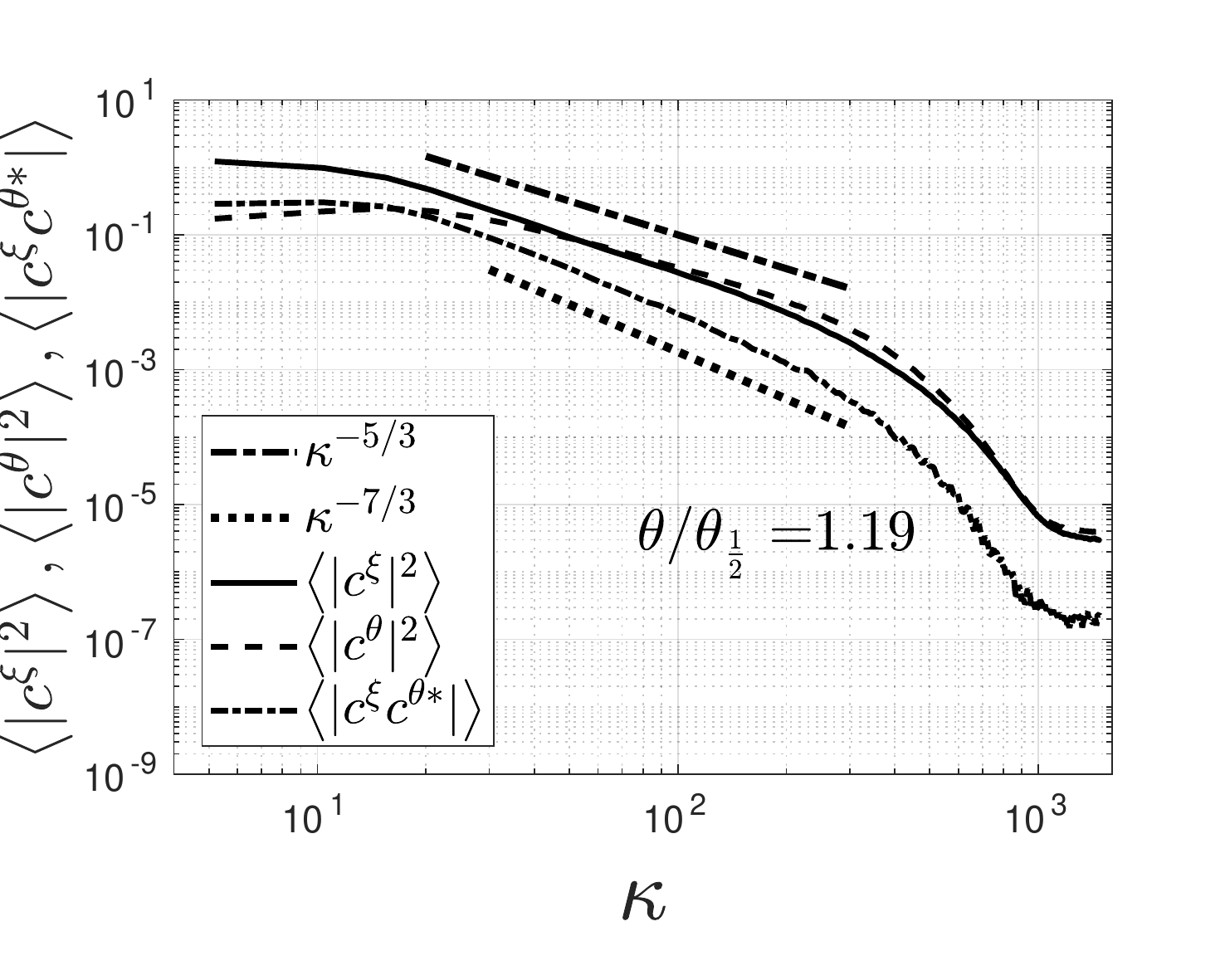}\label{fig:spectra_120_SSC}}        
    \subfloat[]{\includegraphics[width=0.40\linewidth]{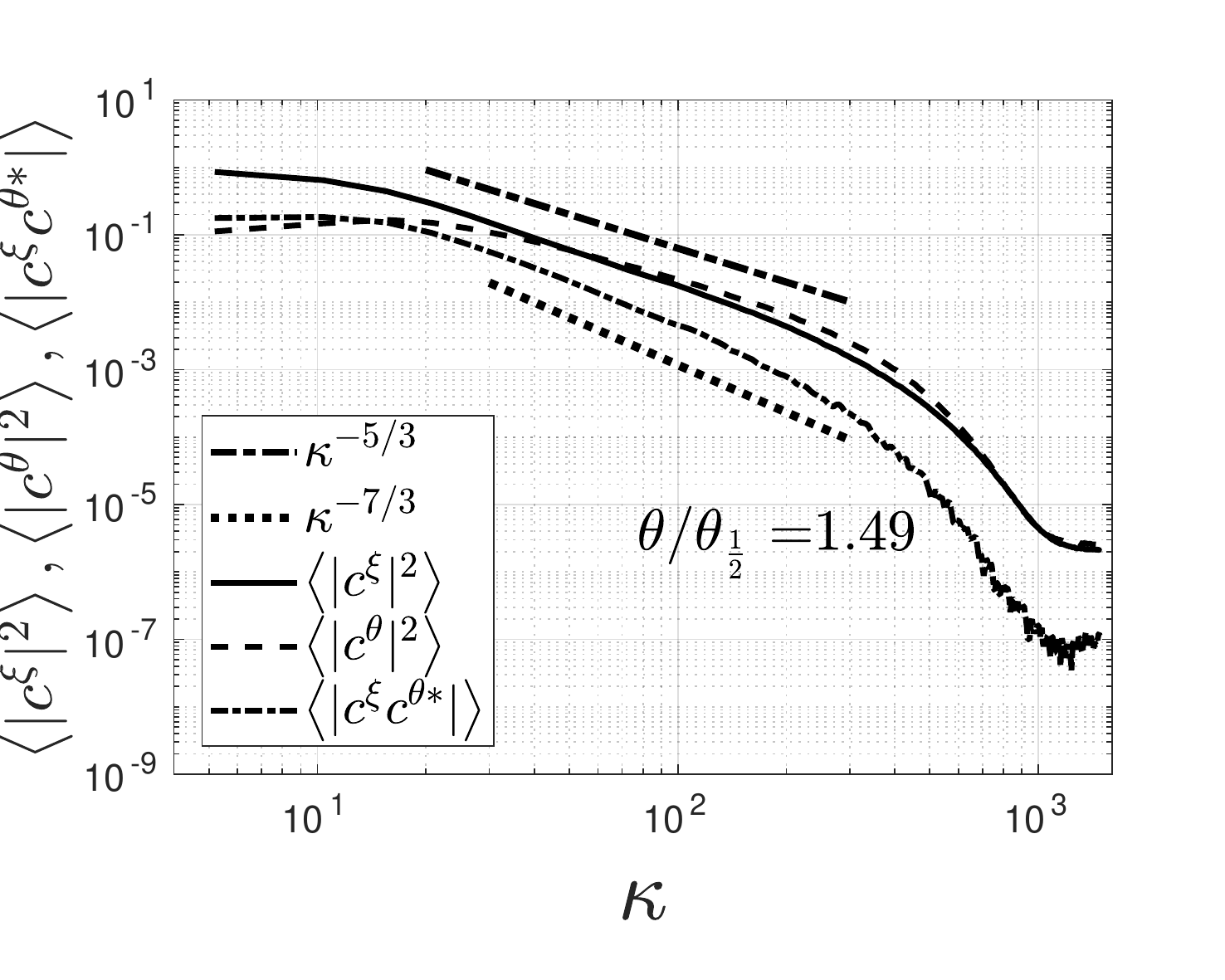}\label{fig:spectra_149_SSC}}\\
    \subfloat[]{\includegraphics[width=0.40\linewidth]{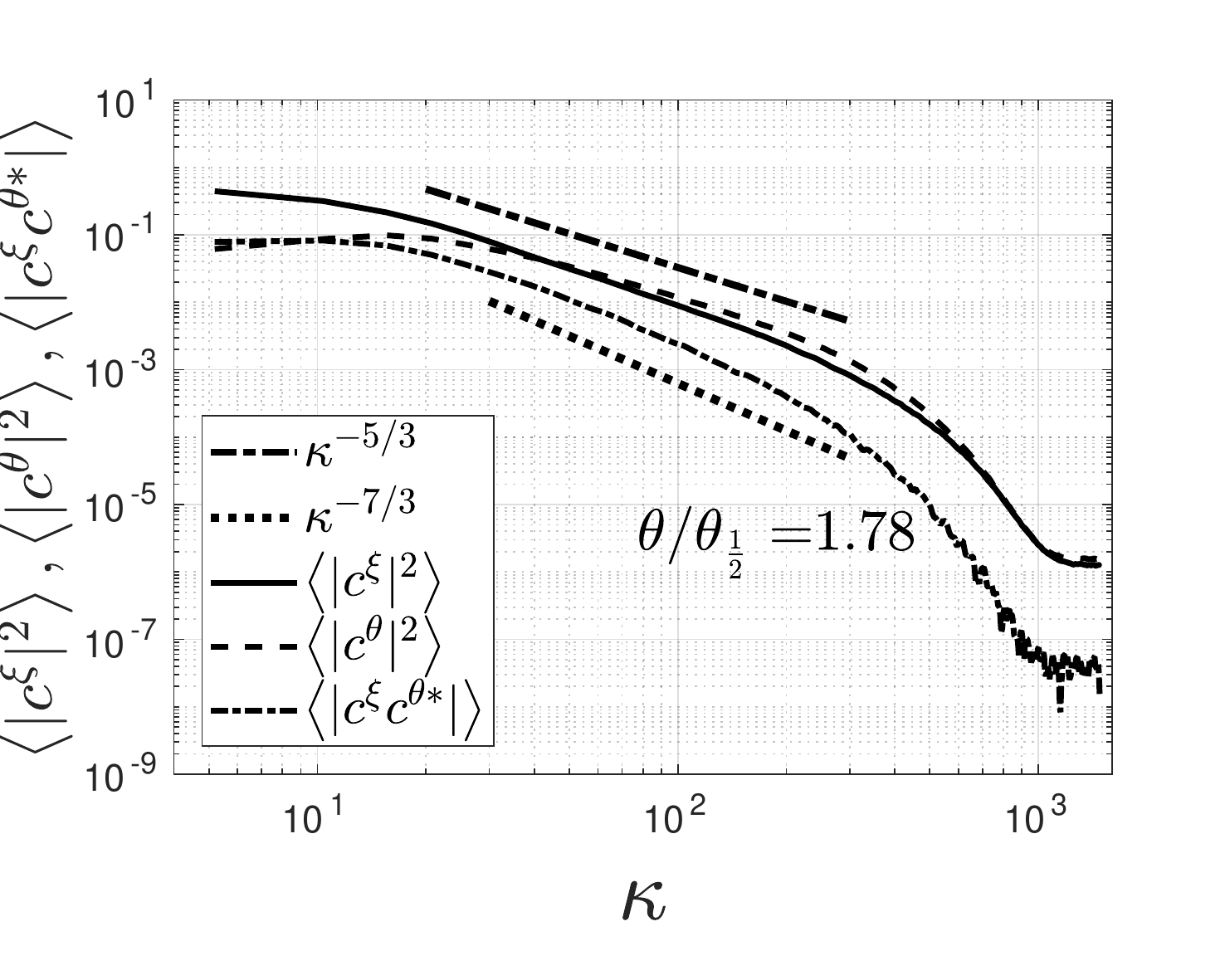}\label{fig:spectra_178_SSC}}
    \subfloat[]{\includegraphics[width=0.40\linewidth]{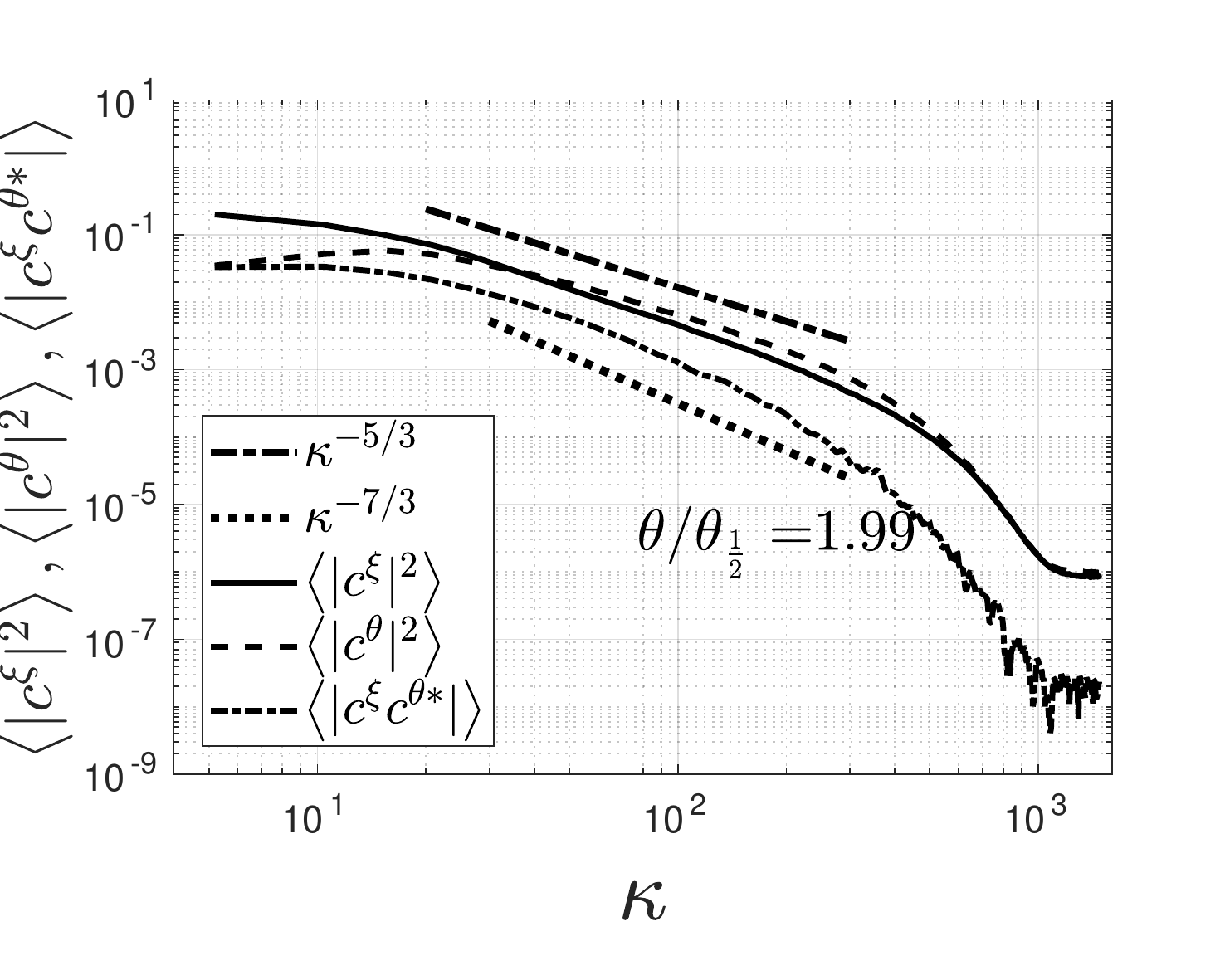}\label{fig:spectra_200_SSC}}    
\caption{Spatial spectra in SSC from various spanwise coordinates, $\theta/\theta_{\frac{1}{2}}$. \label{fig:spectra_SSC}}
\end{figure}

The one-dimensional energy density spectra obtained from the projection coefficients (i.e. $|c^\xi|^2$, $|c^\theta|^2$, $|c^\xi c^{\theta *}|$) in \eqref{eq:SADFM_coefficients_xi}, \eqref{eq:SADFM_coefficients_theta} are shown together with the cross-spectra for various $\theta$-coordinates in figure \ref{fig:spectra_SSC} from $E_1$. Similar spectra were reported by \cite{Wanstrom2009} by applying a regular Fourier transform of the velocity components normalized by the centerline velocity. In the current case, a Parzen-window has been applied in order to reduce the spectral leakage at higher wavenumbers related to the finite window length. It is worth noting that the energy spectra shown in Figure \ref{fig:spectra_SSC} are based on amplitude-decaying stretched Fourier basis functions, \eqref{eq:chi_xi}-\eqref{eq:chi_theta}, illustrated in figure \ref{fig:example_of_mode}, and not on regular trigonometric polynomials. It was noted by \cite{Wanstrom2009} from observing spectra obtained from a Fourier transform of scaled velocity components that the consequences of such spectra are quite profound, as similar trends can be seen to what has been observed in homogeneous turbulence. 
\FloatBarrier
\noindent
Figure \ref{fig:spectra_SSC} shows a $-5/3$-range as well as a $-7/3$-slope for the cross spectra, even though those are spectra of streamwise \textit{decaying} fluctuating fields with respect to the basis functions, $e^{i\kappa\xi-\frac{3}{2}\xi}$, as seen from the coefficients in \eqref{eq:SADFM_coefficients_xi} and \eqref{eq:SADFM_coefficients_theta}. The spectra exhibit the $-5/3$- and $-7/3$-slopes already from $\theta/\theta_{\frac{1}{2}}=0.07$ and the $-5/3$-region is estimated to be in $\kappa\in[20:300]$. 

The $-7/3$-slope in the cross-spectra was predicted to exist in flows where turbulence was sustained by a uniform mean shear gradient by \cite{Lumley1967}. Lumley's model was based on the assumption that shear-stresses were proportional to the mean shear, which in the current case means that
\begin{equation}
\left\langle v^\xi v^\theta\right\rangle\propto \frac{\partial\left\langle V^\xi\right\rangle}{\partial\theta},
\label{eq:proportionality_SSC}
\end{equation}
where the shear-stresses are obtained by integrating the cross-spectrum, and arguing
that the Reynolds shear-stress production is dependent exclusively on $\kappa$, the turbulent
energy spectral flux, $\epsilon$, and the mean shear. Dimensional analysis then implies that the
shear-stress production is manifested in the $-5/3$-range in terms of a $-7/3$-slope of the cross-spectrum, \cite{Lumley1967}
\begin{equation}
\left\langle\left| c^\xi c^{\theta *}\right|\right\rangle \propto\epsilon^{1/3}\kappa^{-7/3}\frac{\partial\left\langle V^\xi\right\rangle}{\partial\theta}, \label{eq:-7/3-equation}
\end{equation}
Note that the coefficient will be universal only if the turbulence at these wavenumbers is as well, an idea which is very much in dispute, \cite{Champagne1978}, \cite{George2014}. The cross-spectra in figure \ref{fig:spectra_SSC} indeed exhibit a $-7/3$-slope for wavenumbers between $20$ and $250$, with the exception of the spectra in the vicinity of the centerline, where the low spectral energies are in correspondence with the negligible shear-stresses in this region. The slope increases rapidly away from the centerline due to the increase of mean shear as seen from figure \ref{fig:stress_meanvel_grad_SSC}. 
\begin{figure}[h]
\centering
\includegraphics[width=0.7\linewidth]{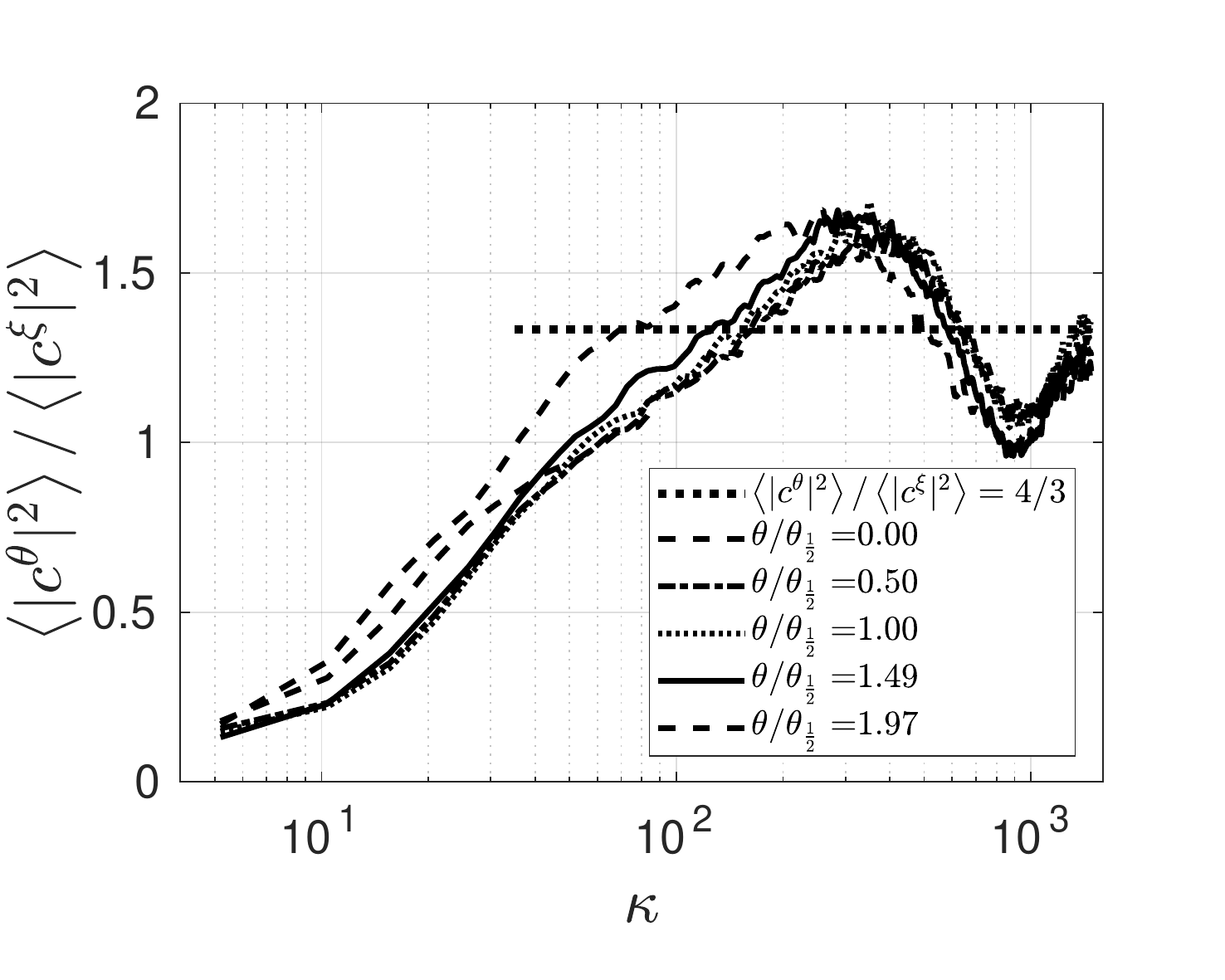}
\caption{Ratio of the component spectra. The assumption of isotropy requires ${\left\langle|c^\theta|^2\right\rangle/\left\langle|c^\xi|^2\right\rangle=4/3}$, \cite{Hinze1975}. \label{fig:4_3_ratio_spectra}}
\end{figure}
\FloatBarrier
\noindent

Beyond $\theta/\theta_{\frac{1}{2}}=0.72$, the $-7/3$ range is slowly eradicated for large $\kappa$. The reason for this can be understood by examining the assumptions underlying Lumley's model leading to the prediction of the $-7/3$-range. The manifestation of the $-7/3$-range traces back to regions in which the shear-stresses are proportional to the mean gradient, corresponding to the assumption \eqref{eq:proportionality_SSC}. Although the $-5/3$-slope of the energy spectra can be identified across the entire span of the jet, the $-7/3$-slope diminishes with increasing $\theta$, as the underlying assumption of proportionality in \eqref{eq:proportionality_SSC} weakens. Figure \ref{fig:stress_meanvel_grad_SSC} reveals that the proportionality described by \eqref{eq:proportionality_SSC} is valid in the range $0.25<\theta/\theta_{\frac{1}{2}}<1$. This correlates with the diminishing $-7/3$-slope for high $\kappa$-values as one moves beyond one jet half-width. 

The existence of the $-7/3$-range for small $\theta$-coordinates ($\theta/\theta_{\frac{1}{2}}\approx 0.5$) indicates, that in the region where shear-stresses are proportional to the mean velocity gradient, a wide range of scales contribute to the energy production through shear-stresses - see \eqref{eq:production_budget_SSC} and figure \ref{fig:production_budget_SSC}. Note that most of the contribution to the energy production from shear-stresses is obtained from low wavenumbers. In the absence of three-dimensional spectra, however, it is difficult to draw conclusions from this since the low wavenumbers suffer from aliasing from the missing dimensions.

It is worth mentioning that the notion of isotropy is not clearly manifested in the current spectra. Although both energy component spectra exhibit the $-5/3$-slope these are manifested in different wavenumber regions, which means that the ratio between the component energy spectra is varying with wavenumber. A prerequisite for isotropy in the $-5/3$-range is, \cite{Hinze1975}
\begin{equation}
\left\langle|c^\theta|^2\right\rangle/\left\langle|c^\xi|^2\right\rangle= 4/3.
\end{equation}
Figure \ref{fig:4_3_ratio_spectra} shows the noticeable variability of ${\left\langle|c^\theta|^2\right\rangle/\left\langle|c^\xi|^2\right\rangle}$ around the value $4/3$ for various positions across the jet indicating that the turbulence is not isotropic. The reason for this may be that the Reynolds number of $20\,000$ is too low in order for this flow to exhibit isotropy as assumed by the Kolmogorov theory. Note, however, that the decomposition applied in the current work does not assume isotropy for any turbulent scale.
\FloatBarrier
\subsection{Decomposition of the Velocity field with respect to LD modes\label{sec:Decomposition_of_the_velocity_field}}
\begin{figure}[h]
\centering   
\subfloat[]{\includegraphics[width=0.40\linewidth]{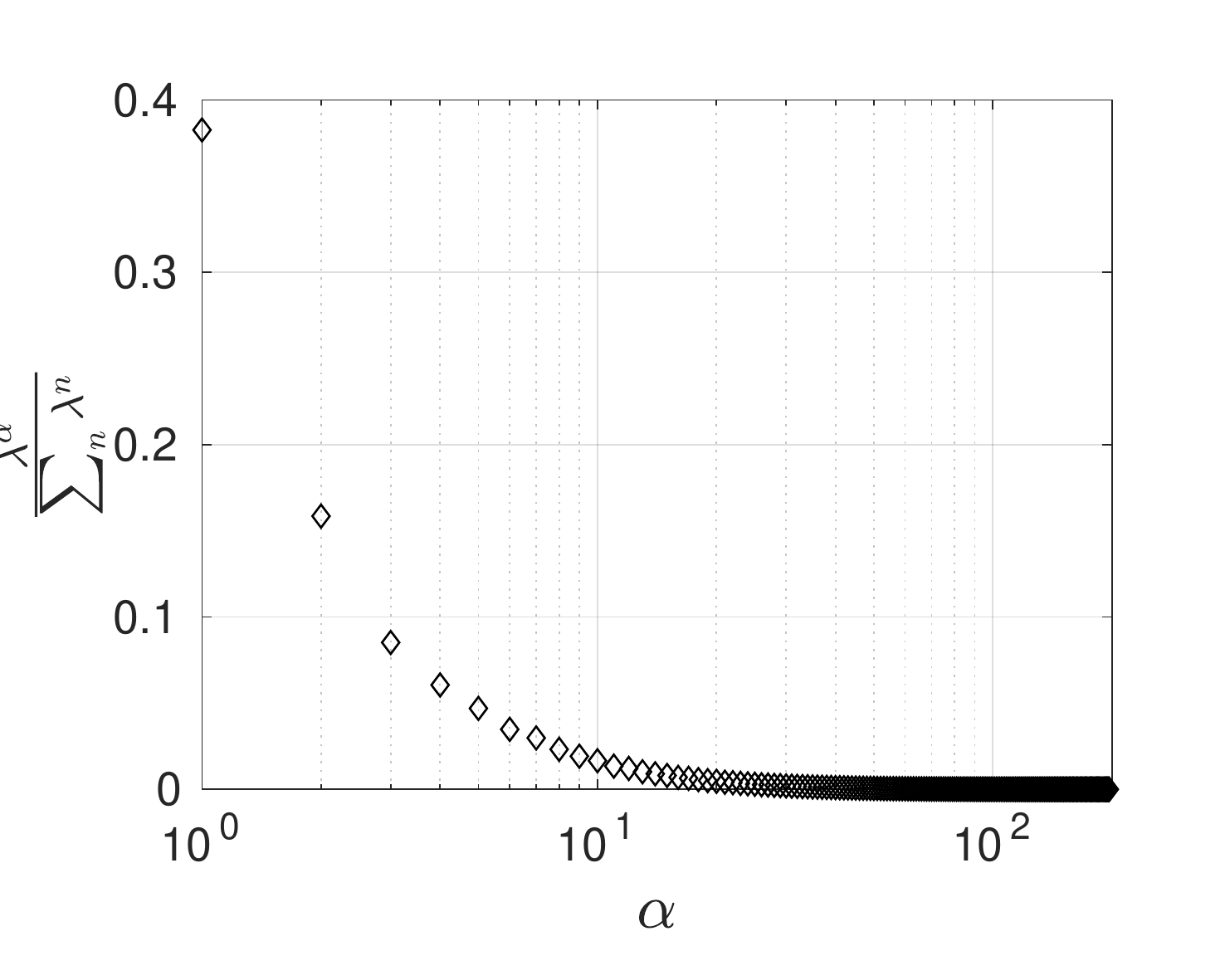}\label{fig:e_values_rel}}
\subfloat[]{\includegraphics[width=0.40\linewidth]{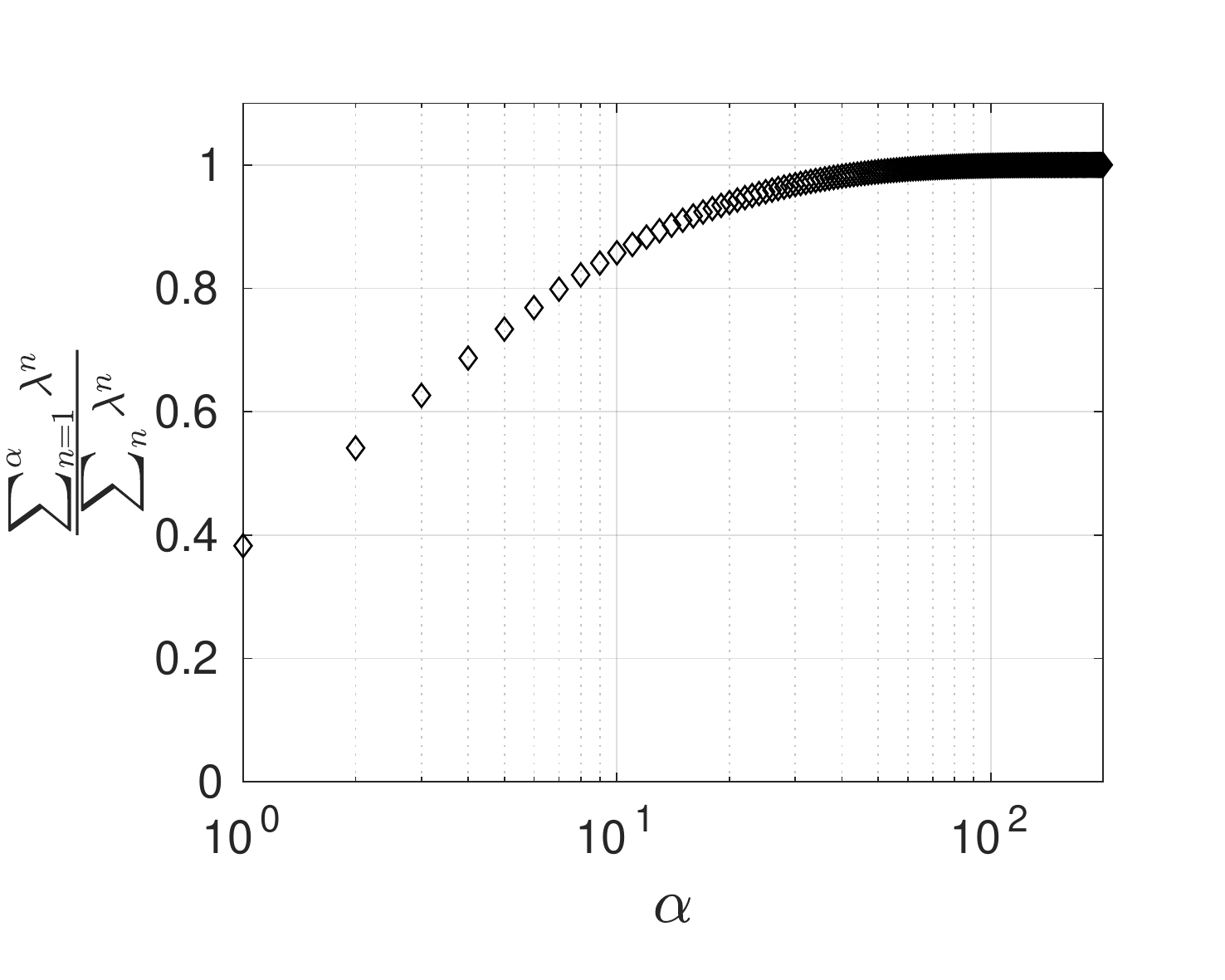}\label{fig:e_values_rel_cumsum}}\\
\subfloat[]{\includegraphics[width=0.40\linewidth]{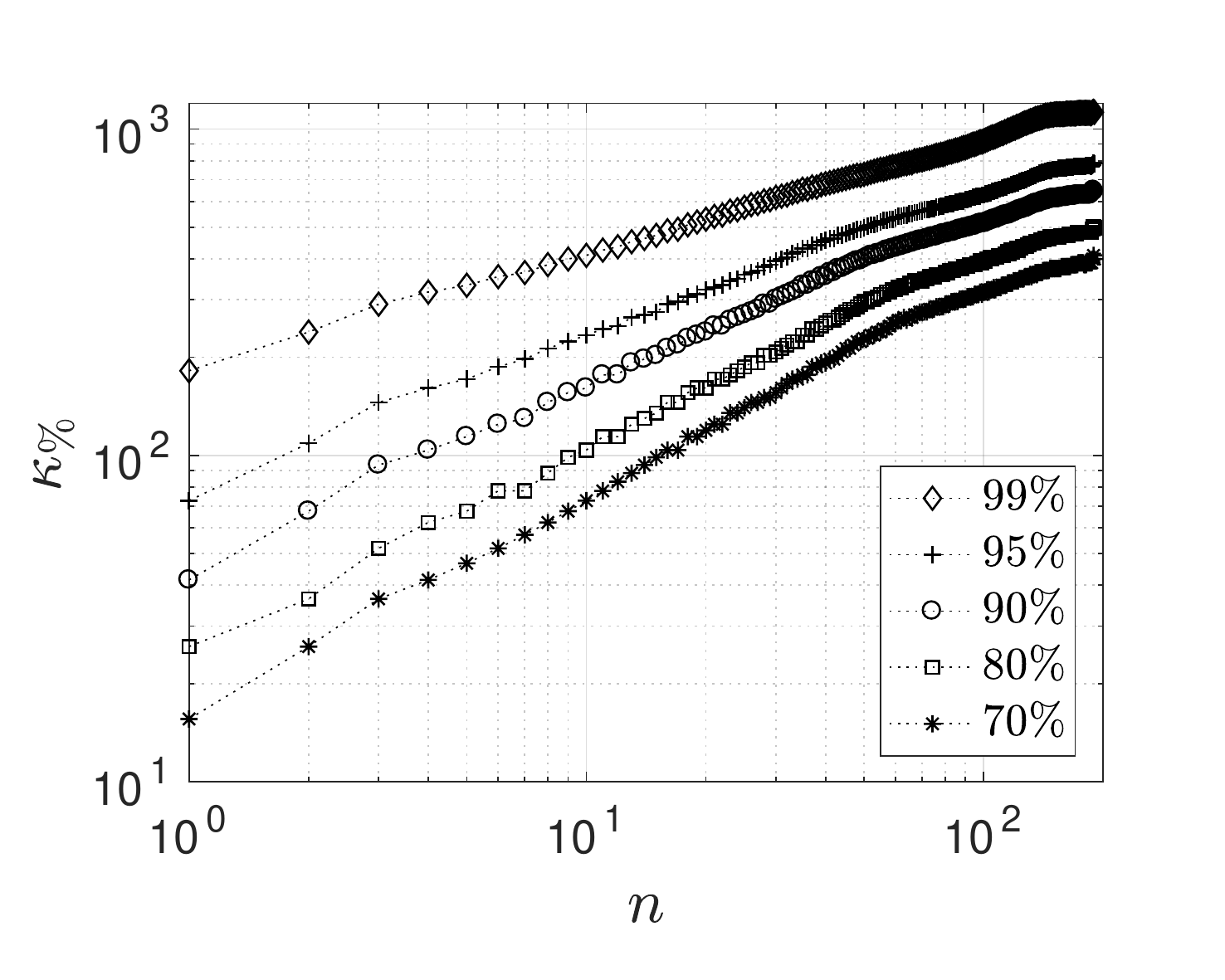}\label{fig:kappa_percentiles}}
\caption{(a): Normalized eigenvalues integrated over wavenumbers, (b): cumulative sum over normalized eigenvalues, (c): wavenumbers defining percentiles of energy across mode numbers, $n$.\label{fig:eigenvalues}}
\end{figure}

The LD was performed by discretizing the $\theta$-coordinate into $n=100$ points between $\theta/\theta_\frac{1}{2}=0$ and $2.475$. This resulted in a total of $200$ LD modes for each Fourier coefficient. The normalized energy distribution of the LD modes are shown in figure \ref{fig:e_values_rel}, and the cumulative sum is shown in figure \ref{fig:e_values_rel_cumsum}. The first mode contains $38.3\%$ of the total energy of the field while the first seven modes contain $80\%$ of the total energy in the flow. Figure \ref{fig:kappa_percentiles} shows further that $80\%$ of the energy of the first seven modes is contained in the region below $\kappa=80$. The relative energies and the cumulative sums, are summarized in Table \ref{tab:modal_energies} for the first eight modes. Each eigenfunction was rotated in the complex plane in order to minimize the imaginary part. This leads to an objective way of comparing the real- and imaginary parts of eigenfunctions across mode numbers. 

\begin{figure}[h]
\centering   
\subfloat[$|\Re\left\lbrace\psi^\xi_1\left(\kappa,\theta\right)\right\rbrace|$]{\includegraphics[width=0.40\linewidth]{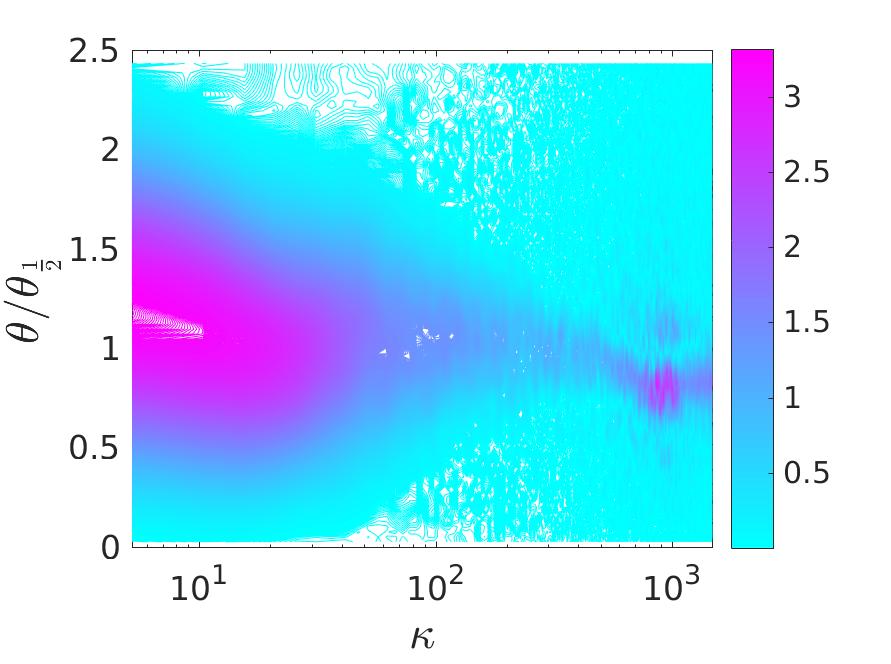}\label{fig:mode_u_1real}}
\subfloat[$|\Re\left\lbrace\psi^\xi_2\left(\kappa,\theta\right)\right\rbrace|$]{\includegraphics[width=0.40\linewidth]{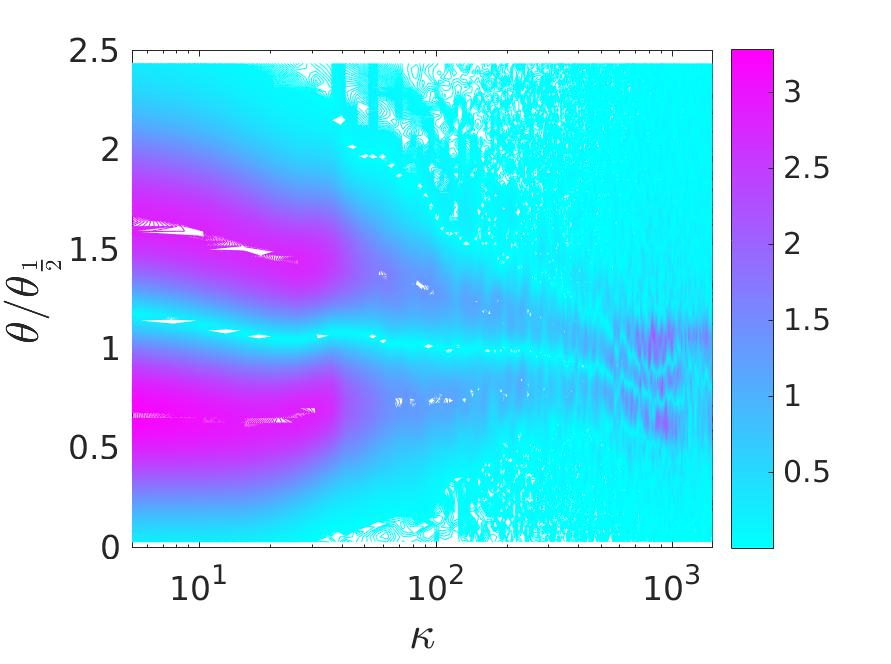}\label{fig:mode_u_2real}}\\
\subfloat[$|\Re\left\lbrace\psi^\theta_1\left(\kappa,\theta\right)\right\rbrace|$]{\includegraphics[width=0.40\linewidth]{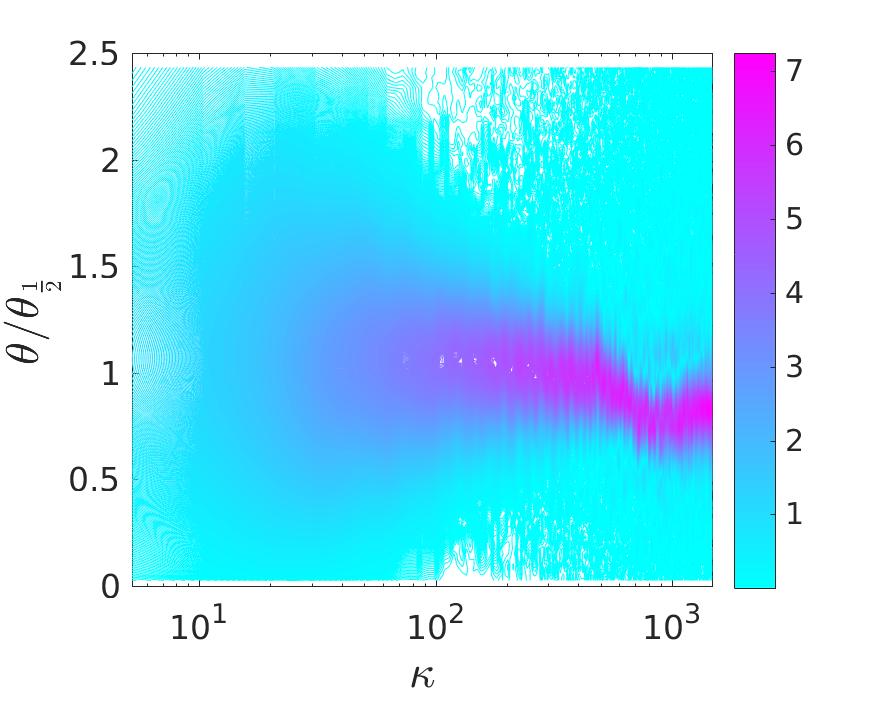}\label{fig:mode_v_1real}}
\subfloat[$|\Re\left\lbrace\psi^\theta_2\left(\kappa,\theta\right)\right\rbrace|$]{\includegraphics[width=0.40\linewidth]{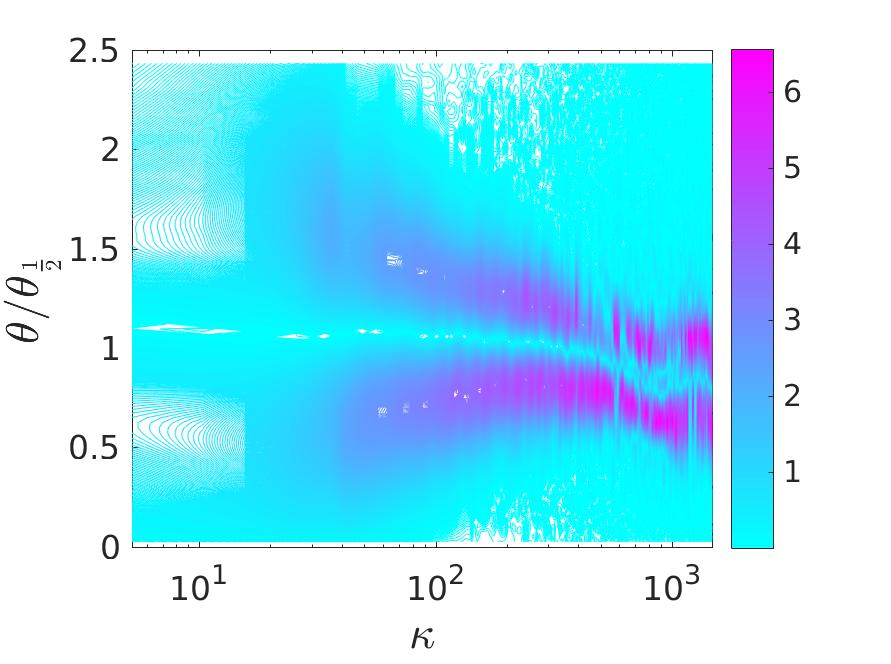}\label{fig:mode_v_2real}}\\
\subfloat[$|\Im\left\lbrace\psi^\xi_1\left(\kappa,\theta\right)\right\rbrace|$]{\includegraphics[width=0.40\linewidth]{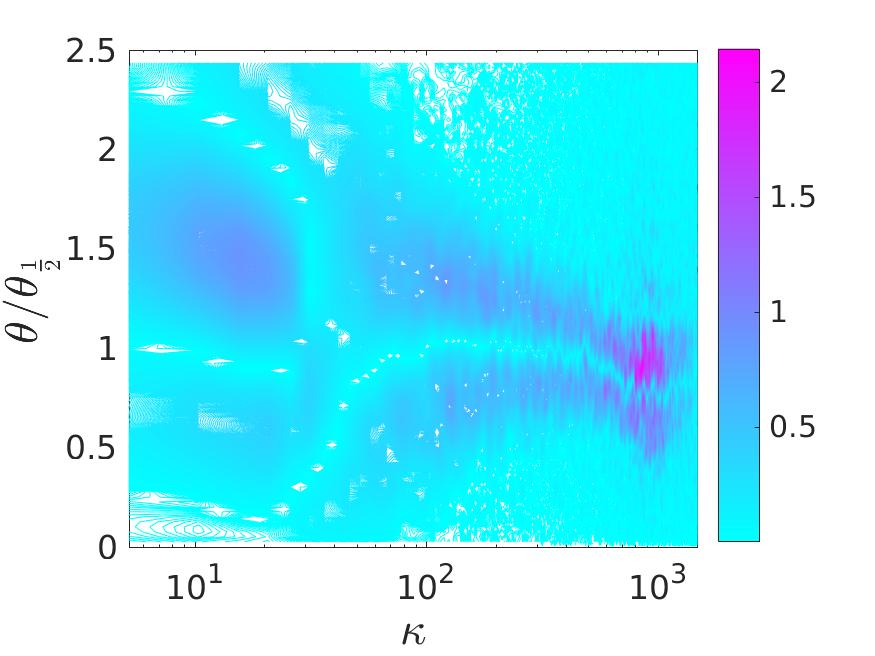}\label{fig:mode_u_1imag}}
\subfloat[$|\Im\left\lbrace\psi^\xi_2\left(\kappa,\theta\right)\right\rbrace|$]{\includegraphics[width=0.40\linewidth]{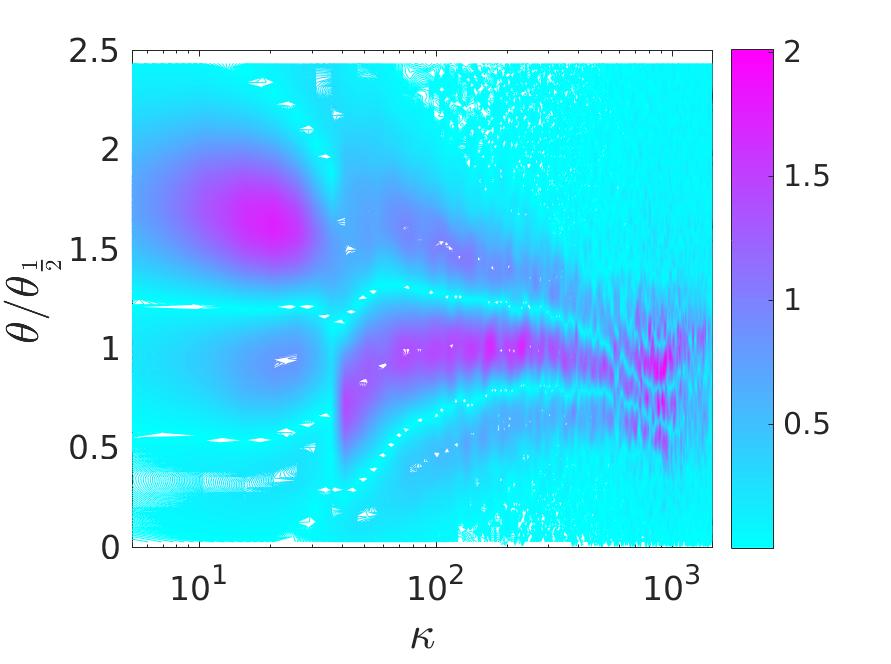}\label{fig:mode_u_2imag}}\\
\subfloat[$|\Im\left\lbrace\psi^\theta_1\left(\kappa,\theta\right)\right\rbrace|$]{\includegraphics[width=0.40\linewidth]{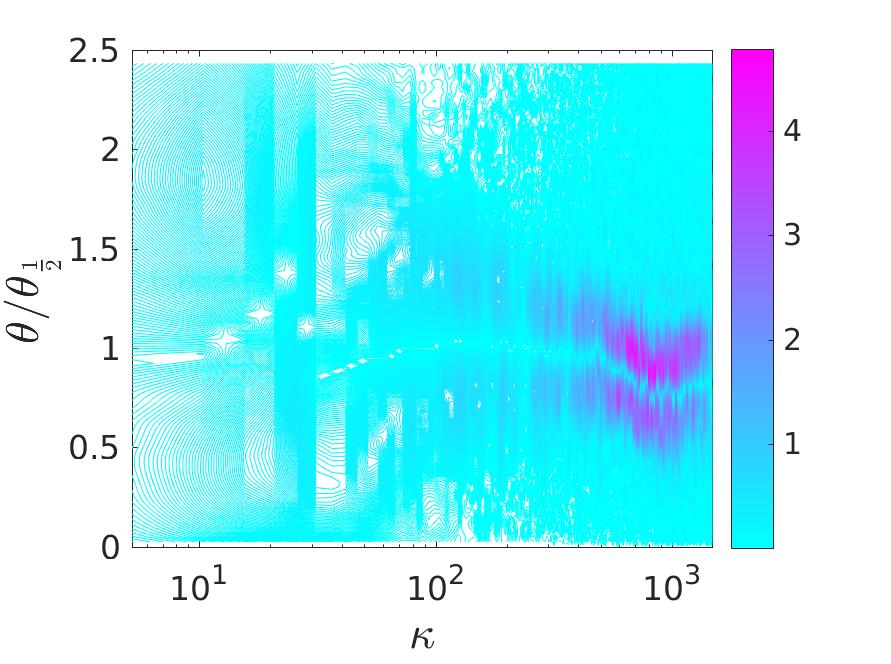}\label{fig:mode_v_1imag}}
\subfloat[$|\Im\left\lbrace\psi^\theta_2\left(\kappa,\theta\right)\right\rbrace|$]{\includegraphics[width=0.40\linewidth]{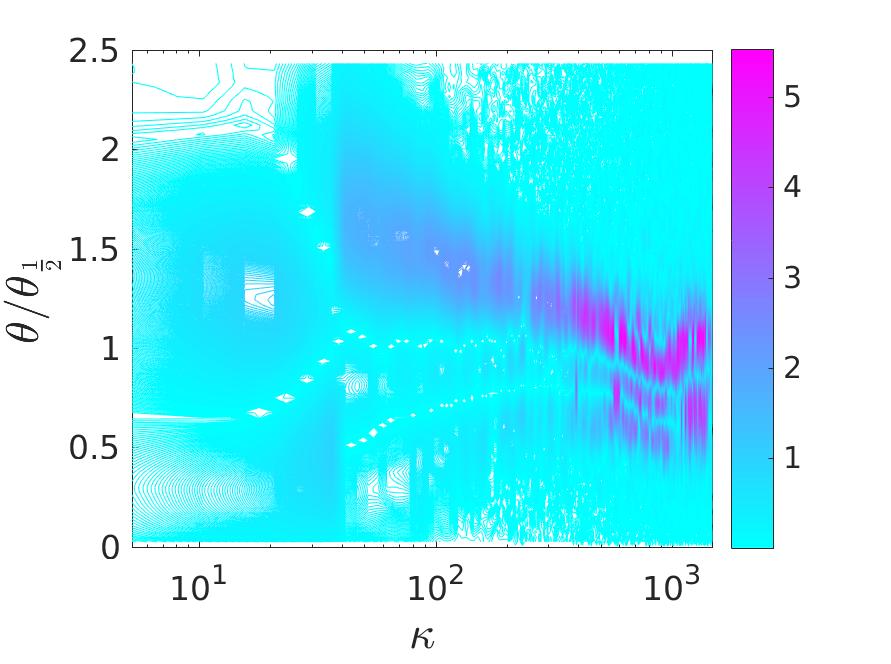}\label{fig:mode_v_2imag}}
\caption{The absolute real- and imaginary parts of $\xi$- and $\theta$ components of the LD modes $\alpha=1-2$, as a function of dimensionless wavenumber, $\kappa$. The superscript indicates the directional component $\xi$ or $\theta$, and the subscript denotes the LD mode number. \label{fig:LD_modes_contours}}
\end{figure}

The real and imaginary parts of the first two modes are seen in figure \ref{fig:LD_modes_contours}. The absolute real parts of the first eight modes for the components $\xi$- and $\theta$ are seen in figures \ref{fig:app_LD_modes_real_u_1} and \ref{fig:app_LD_modes_real_v_1} in Appendix \ref{app:LD_modes_as_function_of_wavenumber} where the subscript indicates the LD mode number. As expected, most of the energy is contained in the low wavenumber region as it is seen that the contours subside for increasing $\kappa$-values. Part of this is due to the aliasing from the missing spatial direction, \cite{george201750}. 

\FloatBarrier
\noindent

The Parzen filter effectively removes the spectrally leaked energy to high wavenumbers due to the windowing effect - this effect was seen in \cite{Wanstrom2009}. The contours between the current experiment and that of \cite{Wanstrom2009} show a great resemblence although they were performed independently using different experimental setups and that the scaled cylindrical velocity components in similarity coordinates were applied in \cite{Wanstrom2009}. 

\begin{figure}[h]
\centering   
\subfloat[]{\includegraphics[width=0.40\linewidth]{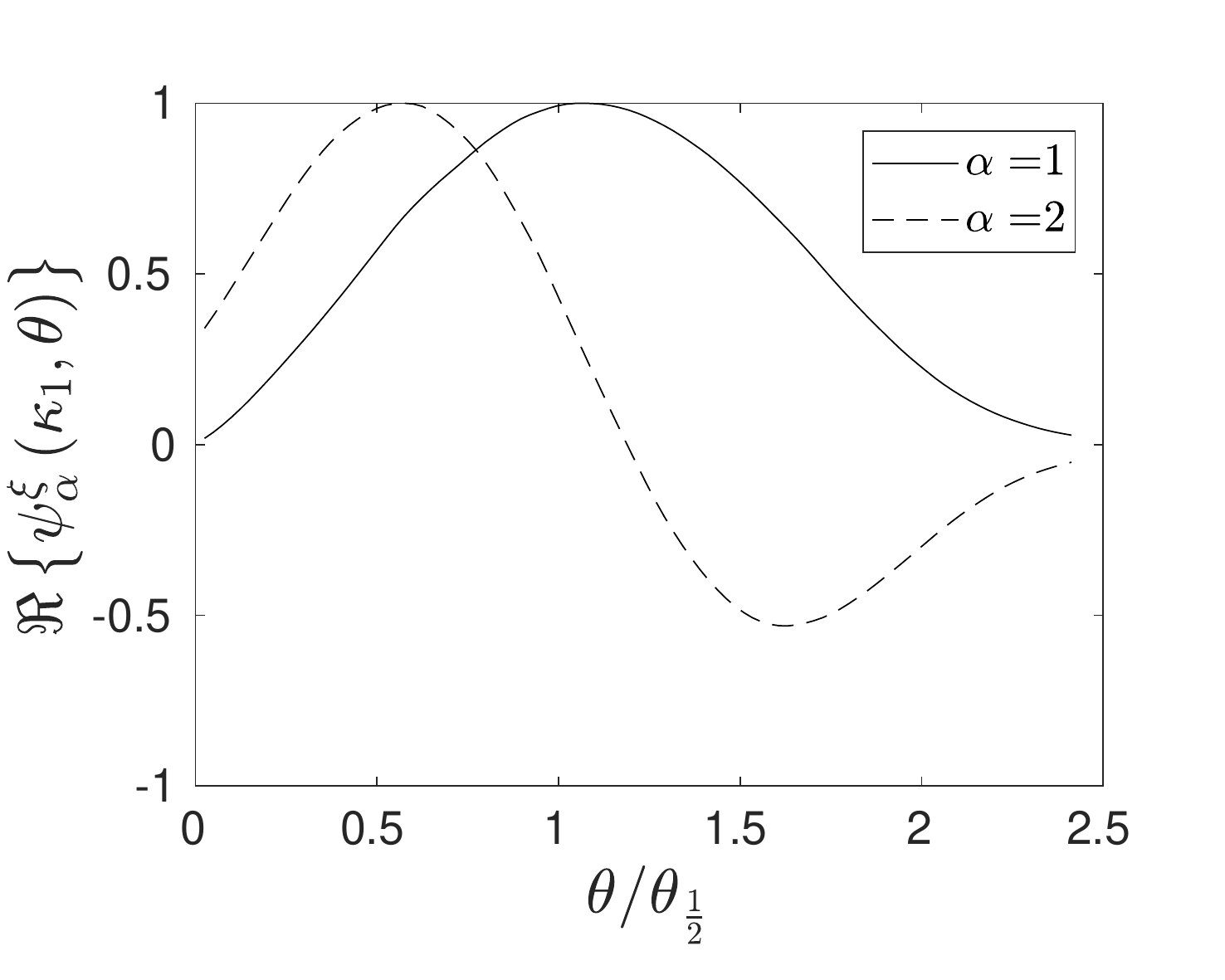}\label{fig:mode_u_real_n_1_2}}
\subfloat[]{\includegraphics[width=0.40\linewidth]{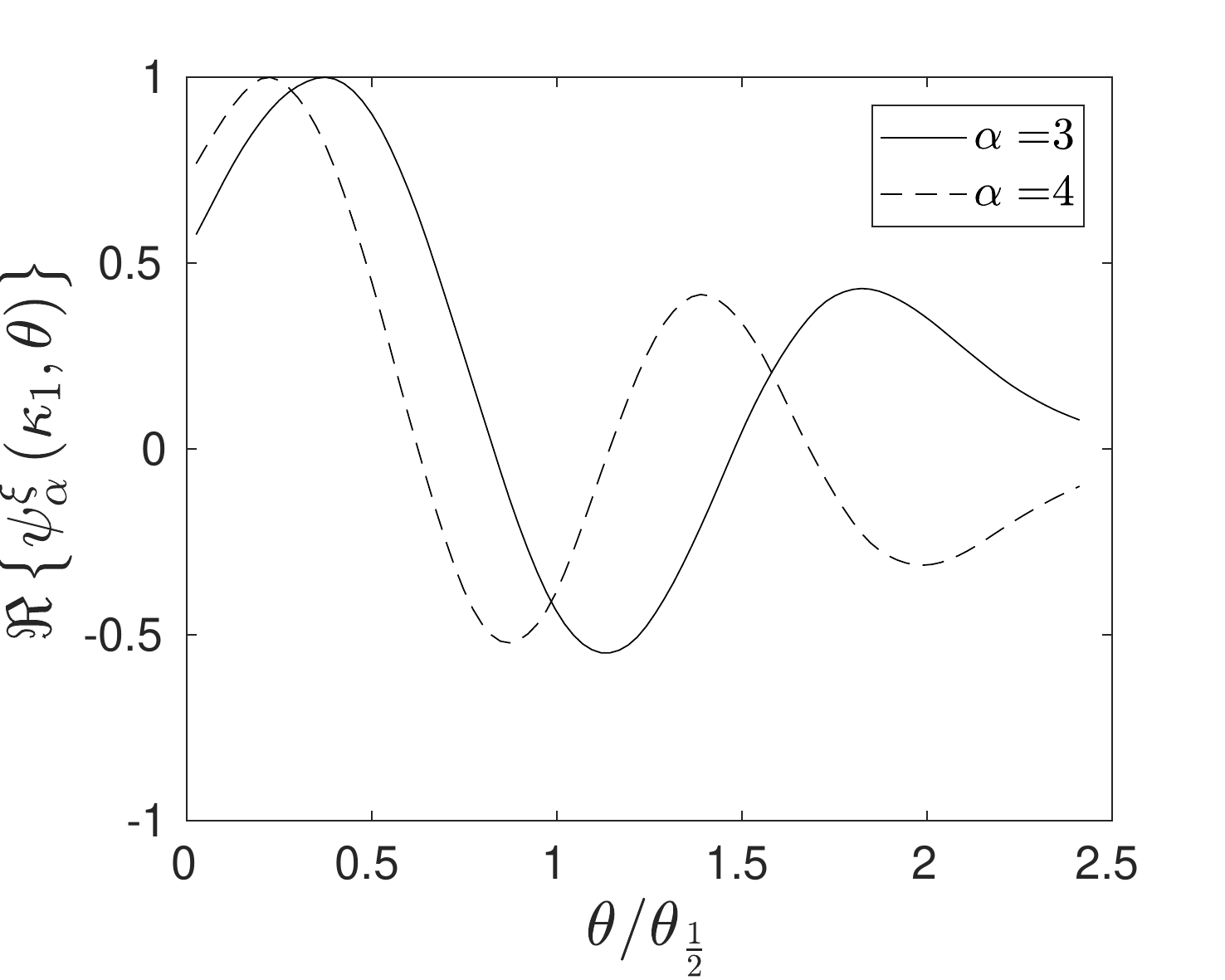}\label{fig:mode_u_real_n_3_4}}\\
\subfloat[]{\includegraphics[width=0.40\linewidth]{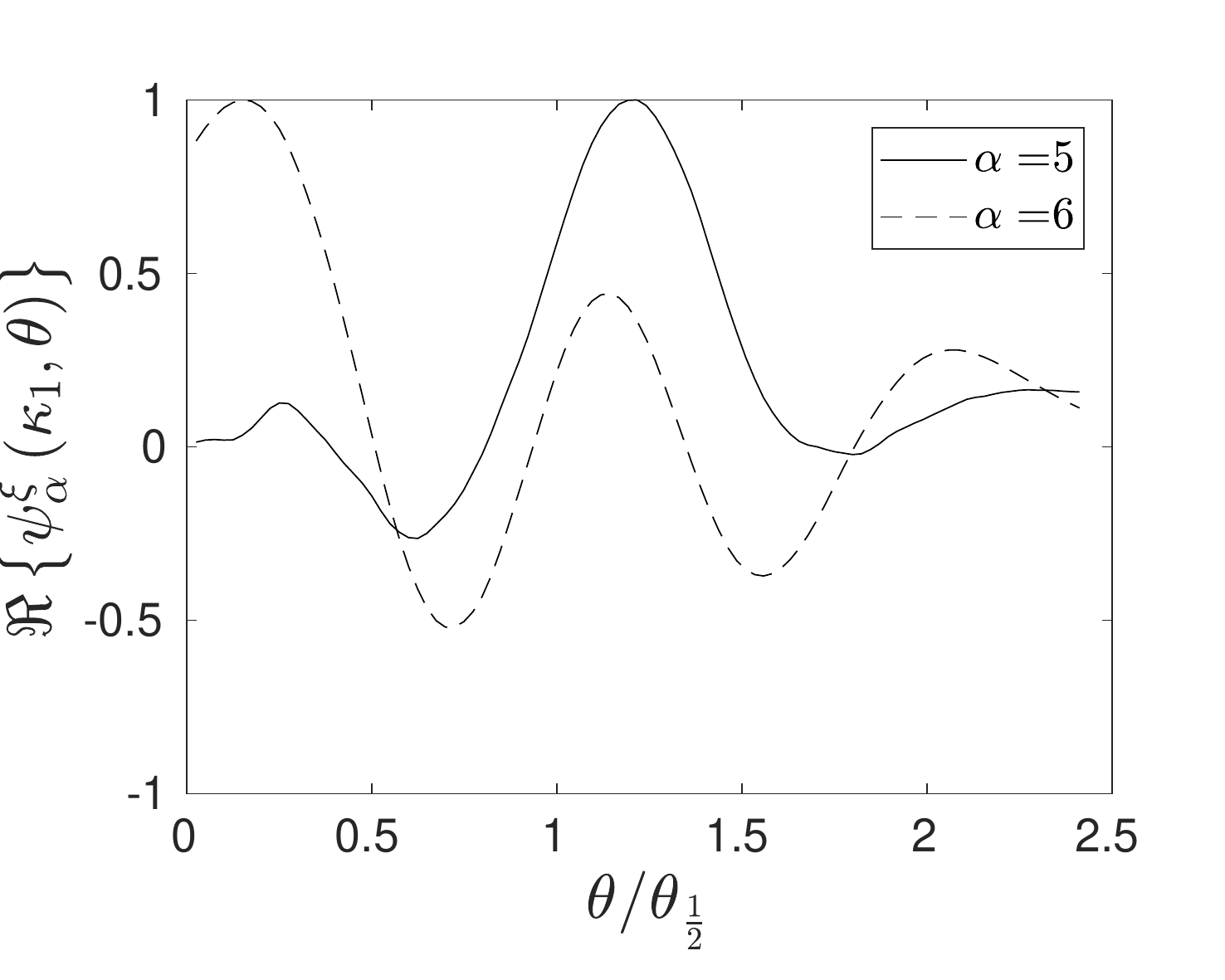}\label{fig:mode_u_real_n_5_6}}
\subfloat[]{\includegraphics[width=0.40\linewidth]{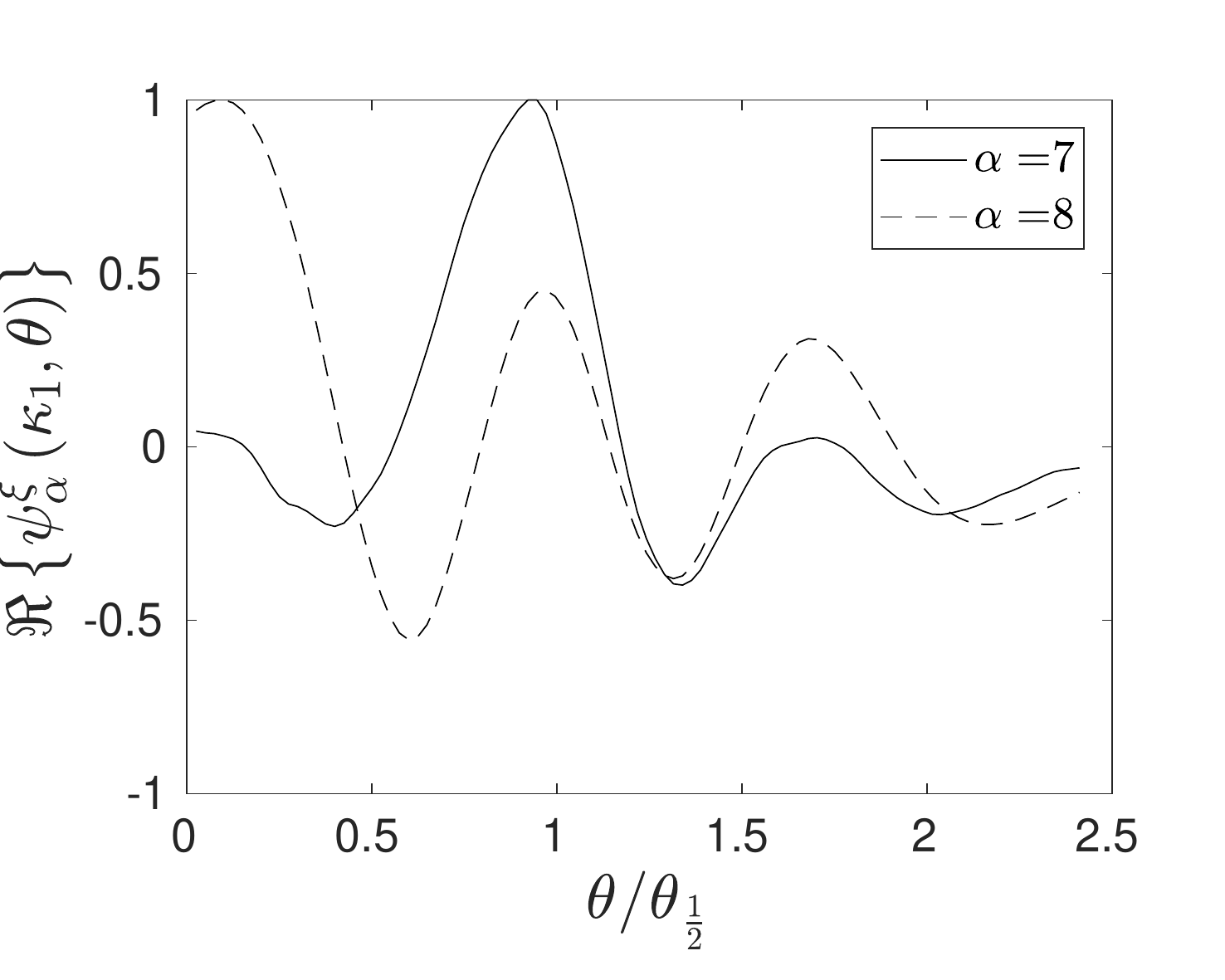}\label{fig:mode_u_real_n_7_8}}
\caption{The real parts of the $\xi$-components of LD modes $\alpha=1-8$ related to the first wavenumber, $\Re\left\lbrace\psi^\xi_\alpha\left(\kappa_1,\theta\right)\right\rbrace$. The superscript indicates the directional component $\xi$ or $\theta$ and the subscript denotes the LD mode number.\label{fig:LD_modes_standard_real_u_1}}
\end{figure}

\begin{figure}[h]
\centering   
\subfloat[]{\includegraphics[width=0.40\linewidth]{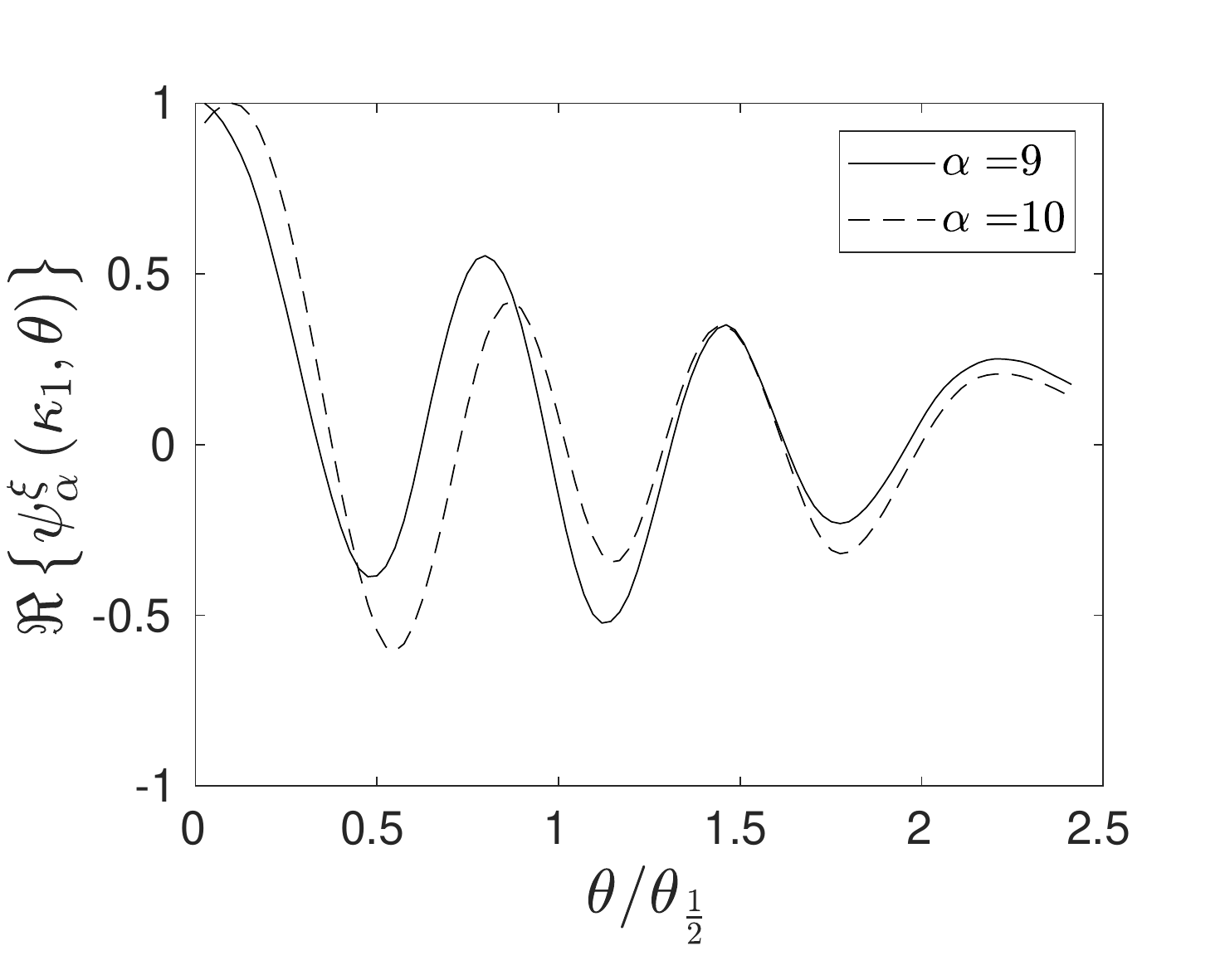}\label{fig:mode_u_real_n_9_10}}
\subfloat[]{\includegraphics[width=0.40\linewidth]{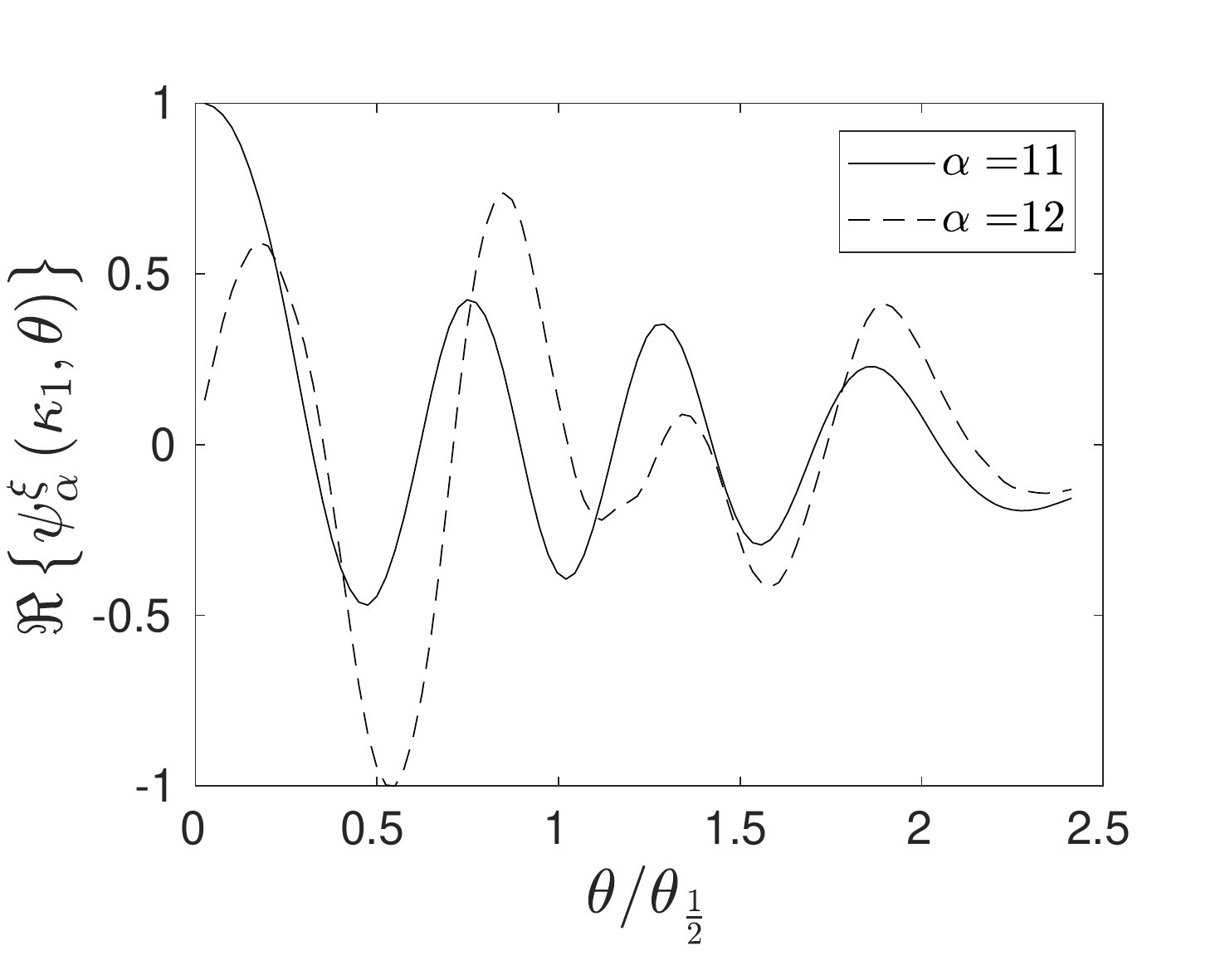}\label{fig:mode_u_real_n_11_12}}\\
\subfloat[]{\includegraphics[width=0.40\linewidth]{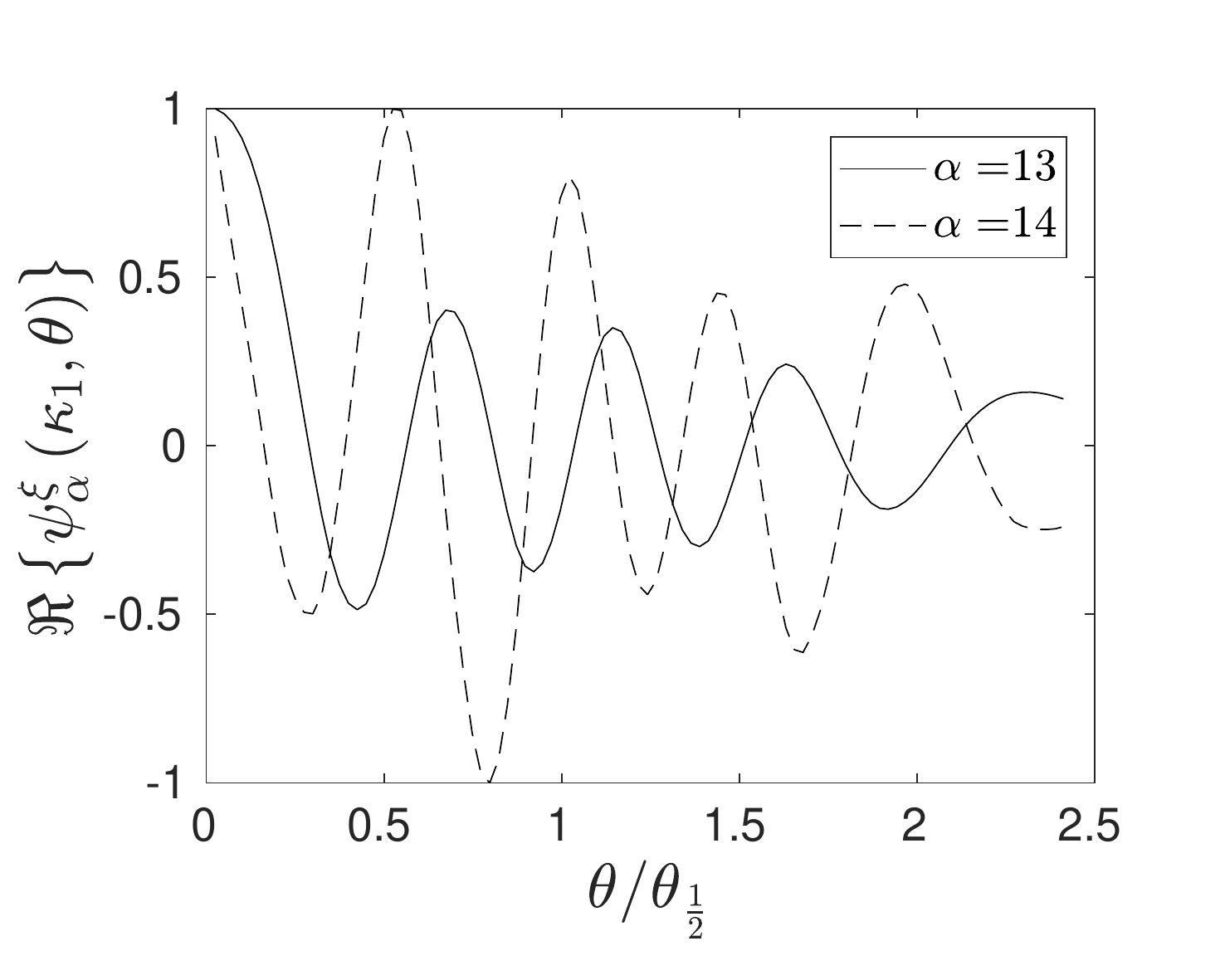}\label{fig:mode_u_real_n_13_14}}
\subfloat[]{\includegraphics[width=0.40\linewidth]{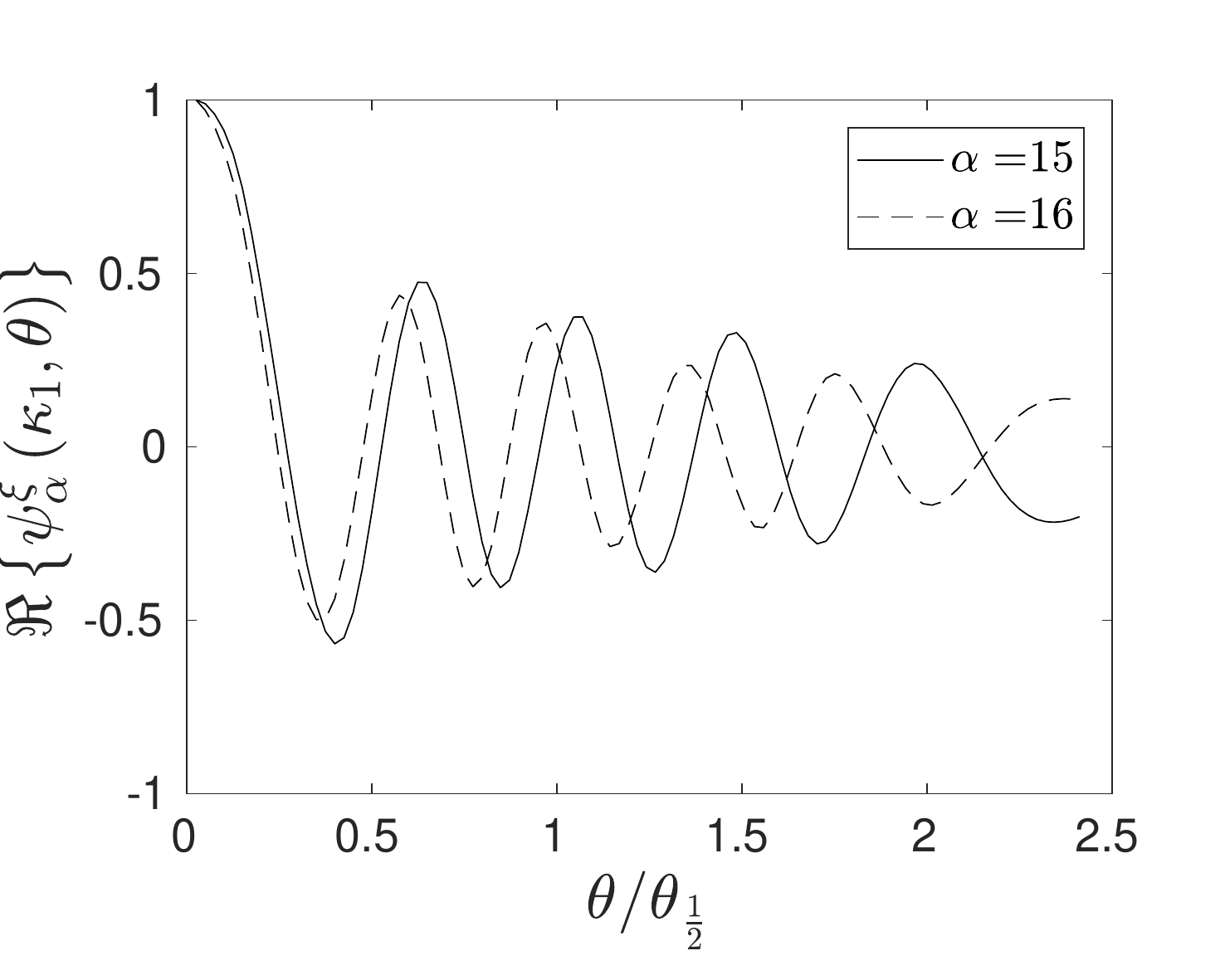}\label{fig:mode_u_real_n_15_16}}
\caption{The real parts of the $\xi$-components of LD modes $\alpha=9-16$ related to the first wavenumber, $\Re\left\lbrace\psi^\xi_\alpha\left(\kappa_1,\theta\right)\right\rbrace$. The superscript indicates the directional component $\xi$ or $\theta$ and the subscript denotes the LD mode number.\label{fig:LD_modes_standard_real_u_2}}
\end{figure}

\begin{figure}[h]
\centering   
\subfloat[]{\includegraphics[width=0.40\linewidth]{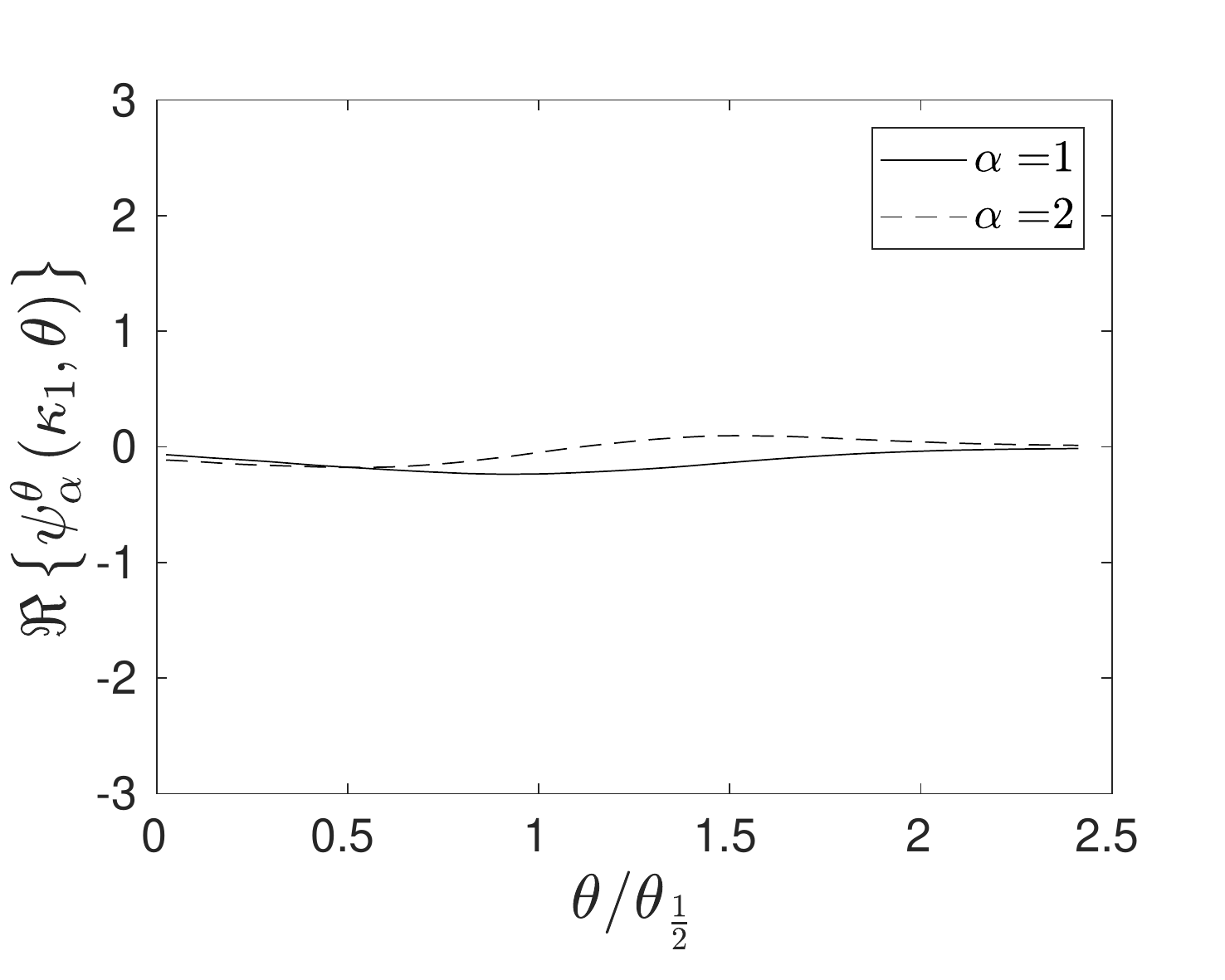}\label{fig:mode_v_real_n_1_2}}
\subfloat[]{\includegraphics[width=0.40\linewidth]{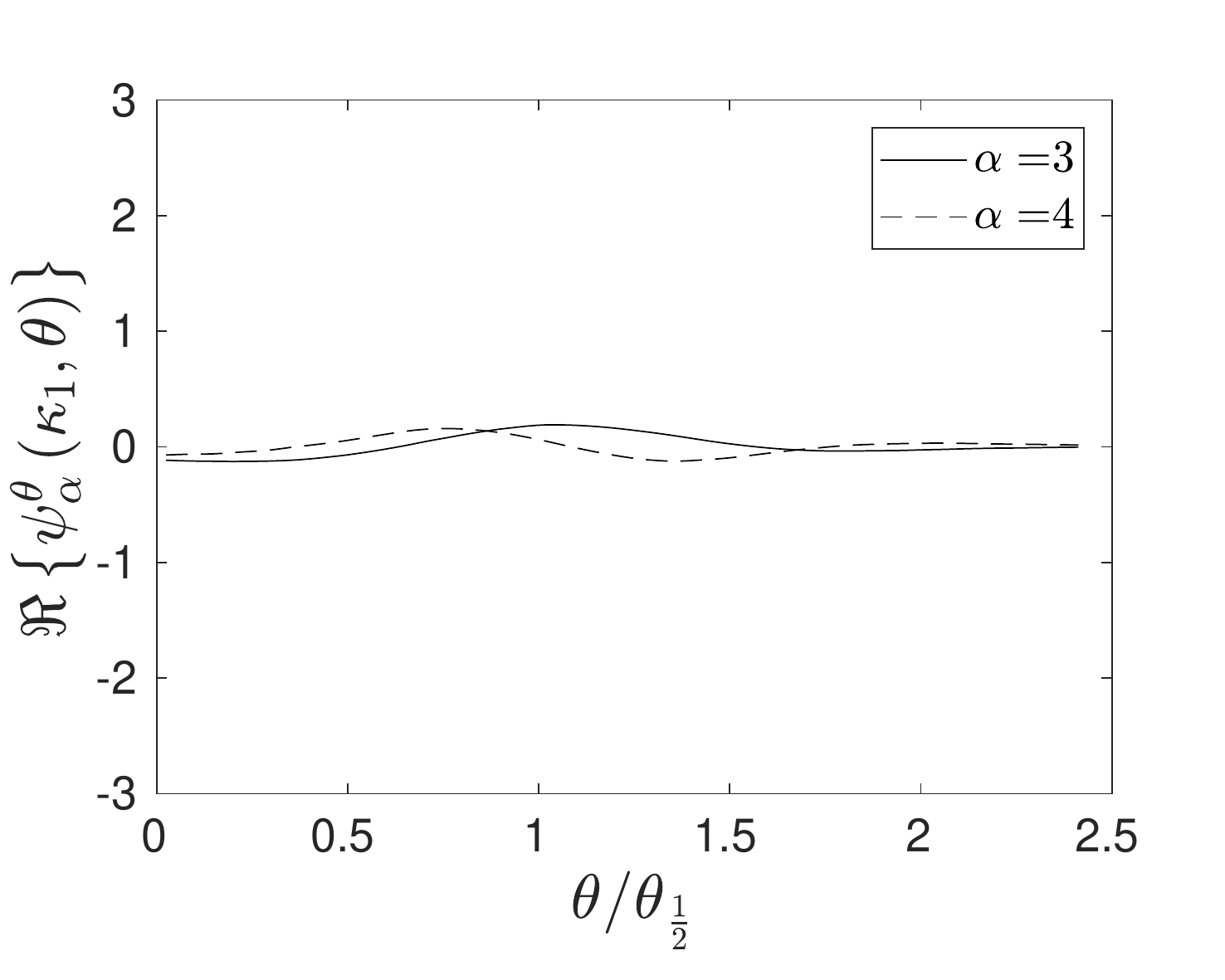}\label{fig:mode_v_real_n_3_4}}\\
\subfloat[]{\includegraphics[width=0.40\linewidth]{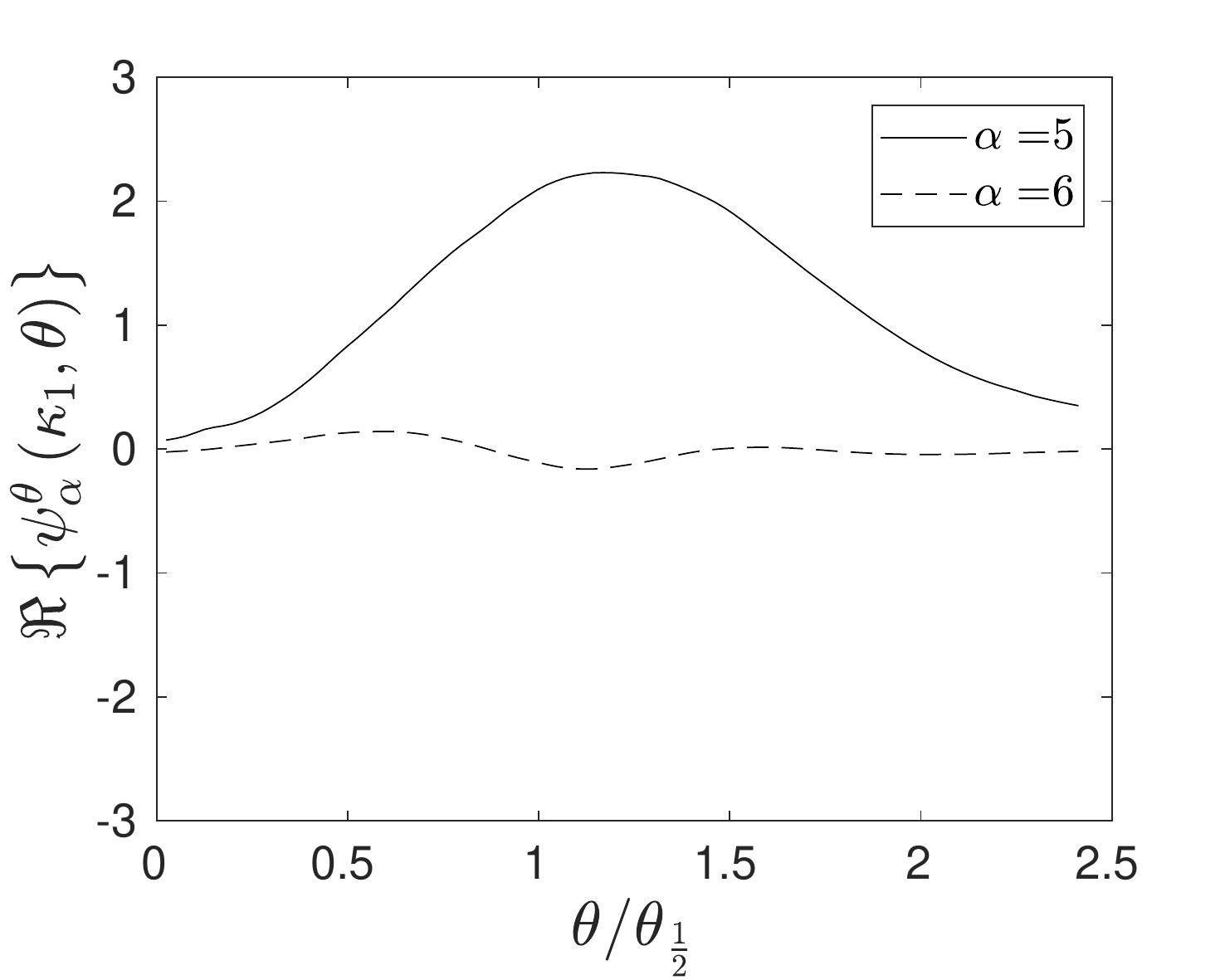}\label{fig:mode_v_real_n_5_6}}
\subfloat[]{\includegraphics[width=0.40\linewidth]{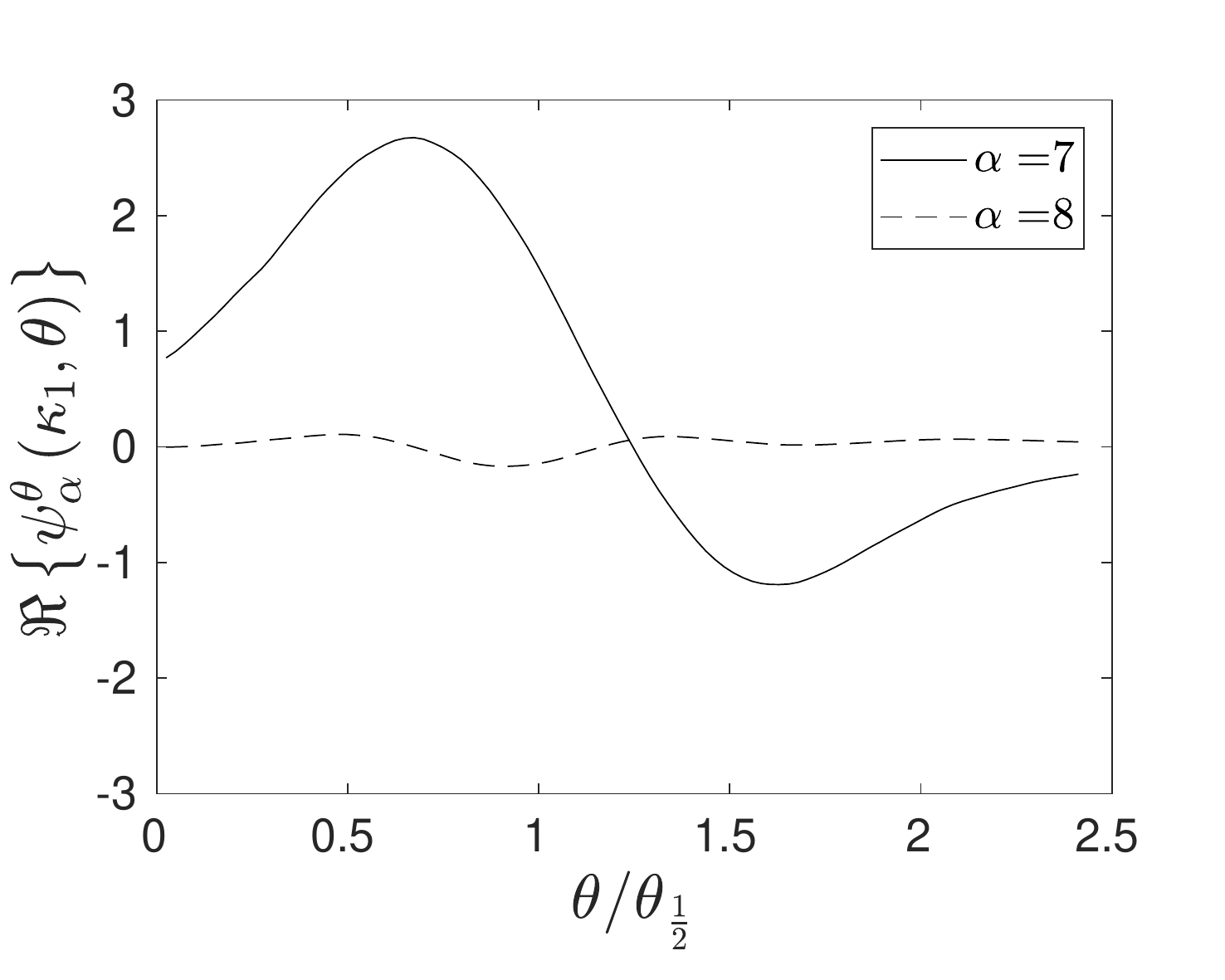}\label{fig:mode_v_real_n_7_8}}\\
\subfloat[]{\includegraphics[width=0.40\linewidth]{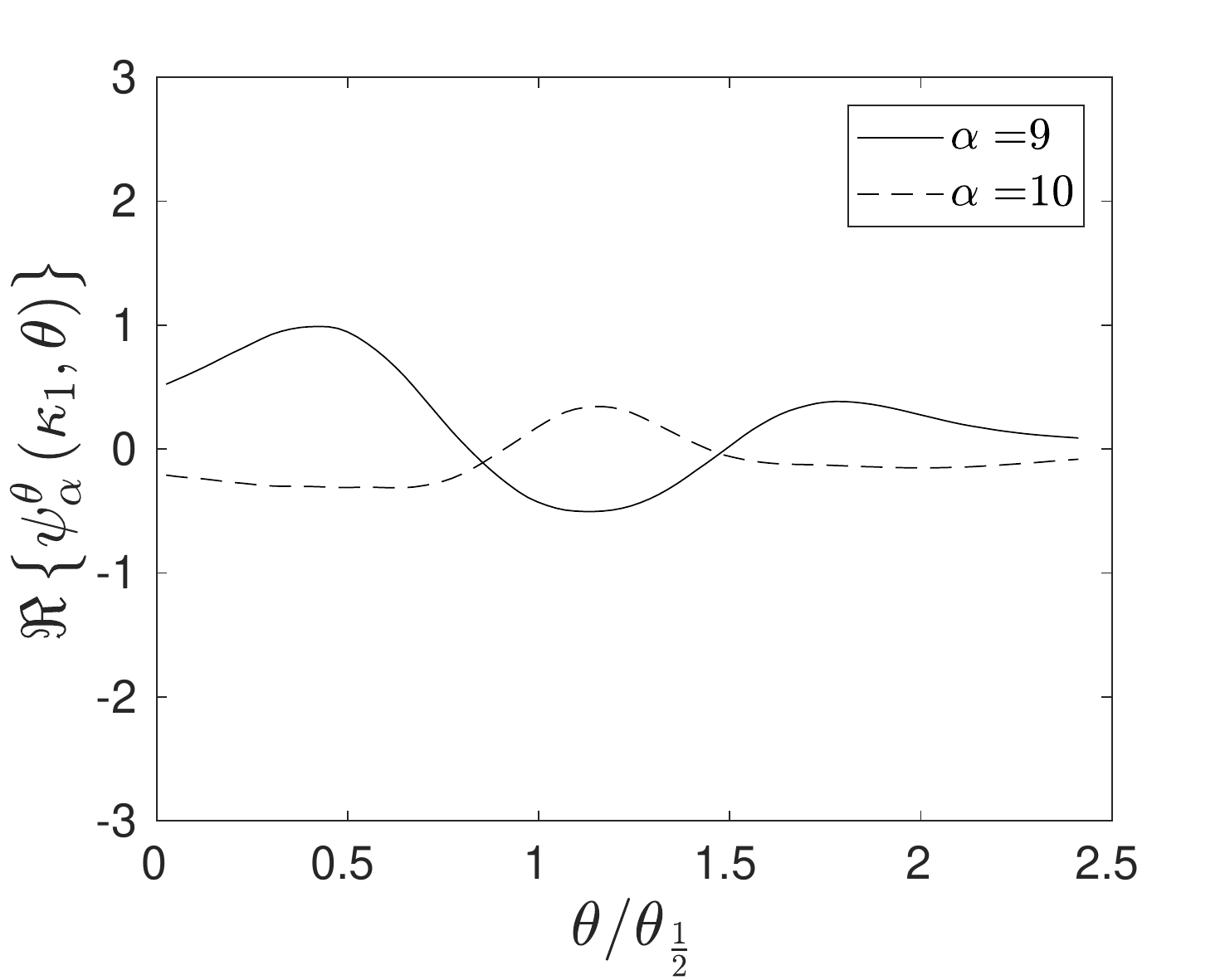}\label{fig:mode_v_real_n_9_10}}
\subfloat[]{\includegraphics[width=0.40\linewidth]{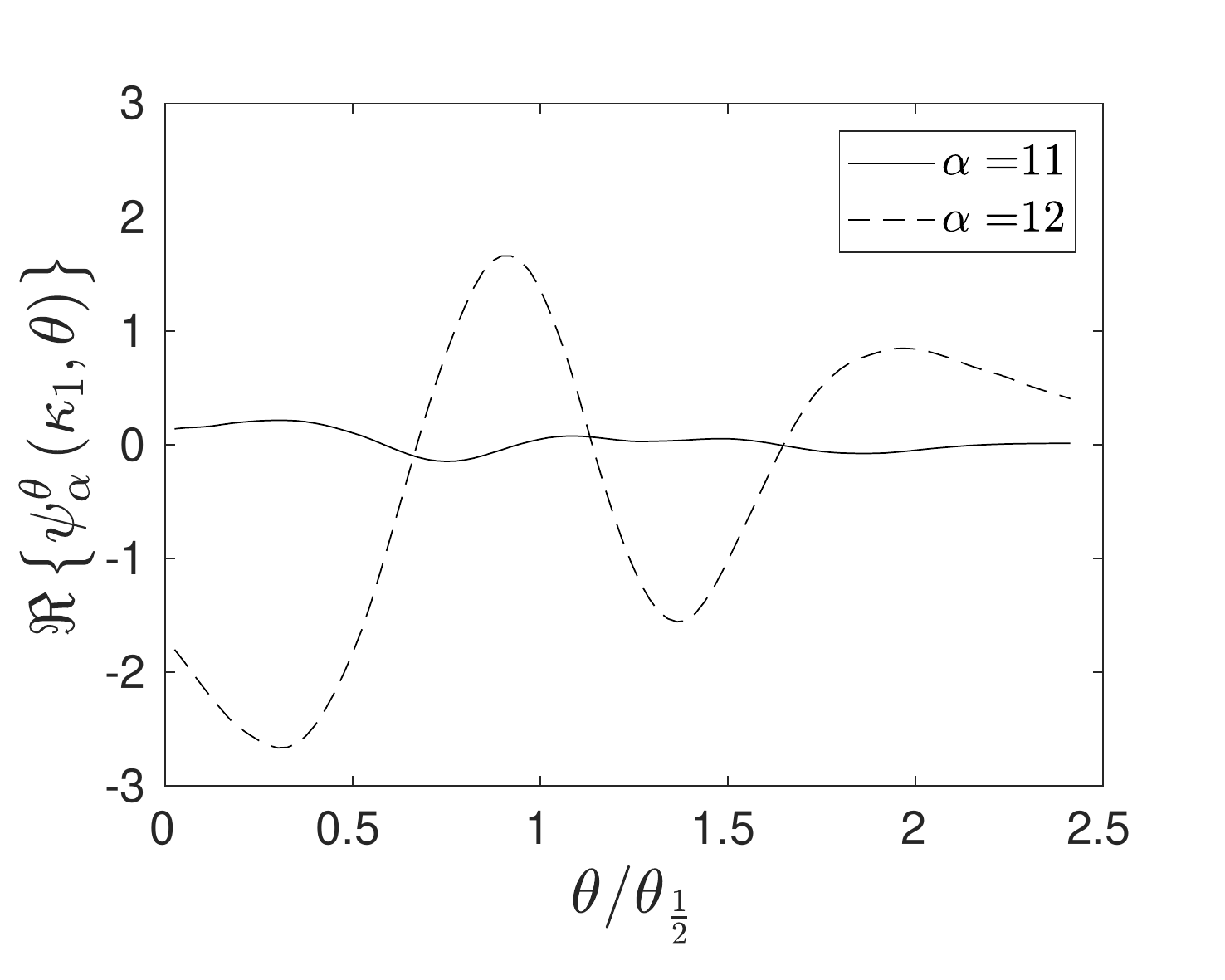}\label{fig:mode_v_real_n_11_12}}\\
\subfloat[]{\includegraphics[width=0.40\linewidth]{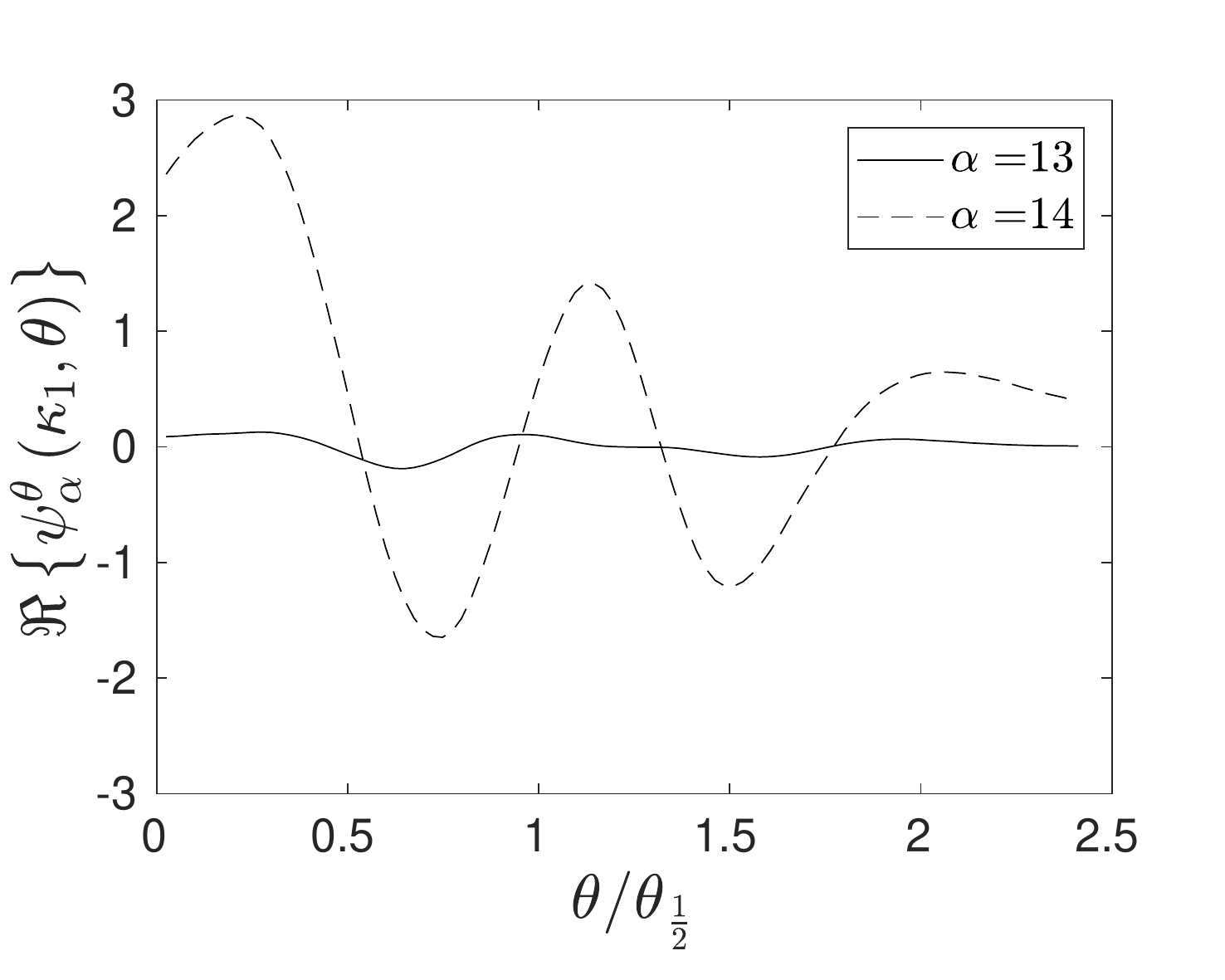}\label{fig:mode_v_real_n_13_14}}
\subfloat[]{\includegraphics[width=0.40\linewidth]{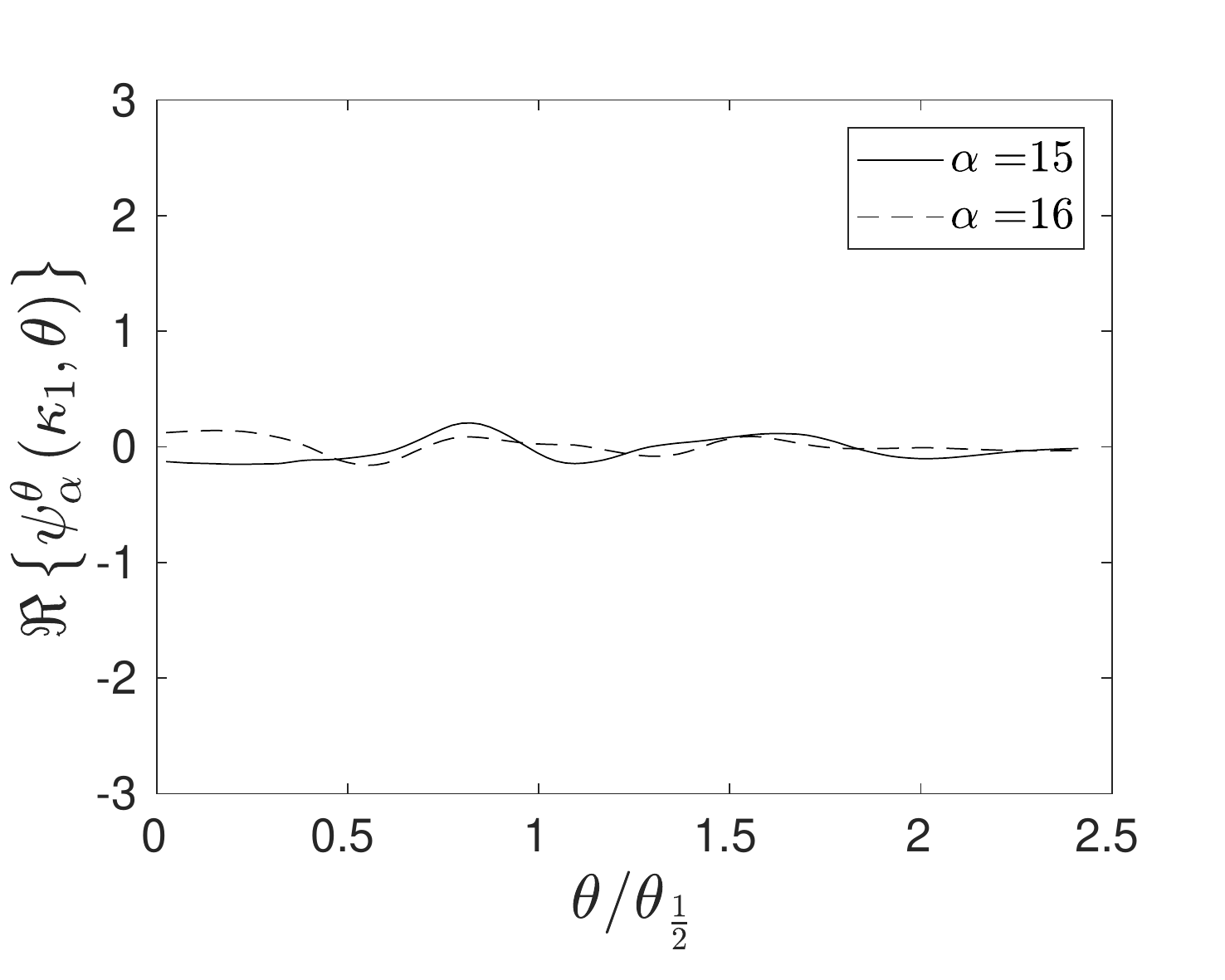}\label{fig:mode_v_real_n_15_16}}
\caption{The real parts of the $\theta$-components of LD modes $\alpha=1-16$ related to the first wavenumber, $\Re\left\lbrace\psi^\theta_\alpha\left(\kappa_1,\theta\right)\right\rbrace$. The superscript indicates the directional component $\xi$ or $\theta$ and the subscript denotes the LD mode number.\label{fig:LD_modes_standard_real_v_1}}
\end{figure}

From figures \ref{fig:mode_v_1real} and \ref{fig:mode_v_2real} showing the $\theta$-components of the LD modes 1 and 2, and the scaled eigenfunction components figures \ref{fig:LD_modes_standard_real_u_1}, \ref{fig:LD_modes_standard_real_u_2} and \ref{fig:LD_modes_standard_real_v_1} showing the real parts of the LD modes $1-16$, it is seen that the most energetic modes have negligible $\theta$-components at low wavenumbers. Note that the scaling of the LD modes does not affect their orthogonality, as seen from \eqref{eq:orthogonality_eigenfunctions}, since for any constant scaling of eigenvectors this multiplier can be taken out of the inner product. 
\FloatBarrier
\noindent
\begin{table}[h]
\caption{Relative modal energies and cumulative sums for the first eight modes.
}{\label{tab:modal_energies}}
 \begin{center}
 \def~{\hphantom{0}}
  \begin{tabular}{ccccccccc}
       $\alpha$ & 1  & 2 & 3 & 4 & 5 & 6 & 7 & 8\\
	   \midrule
       $\frac{\lambda^\alpha}{\sum_{n}\lambda^n}$ & 0.383 & 0.159 & 0.085 & 0.061 & 0.047 & 0.035 & 0.030 & 0.023\\
       \smallskip       
       $\frac{\sum_{n=1}^\alpha\lambda^n}{\sum_{n}\lambda^n}$ & 0.383 & 0.542 & 0.627 & 0.688 & 0.735 & 0.770 & 0.800 & 0.822\\
  \end{tabular}
 \end{center}
\end{table}
\FloatBarrier
This is expected since figures \ref{fig:production_budget_SSC} and \ref{fig:production_budget_local_contribution_SSC} illustrate that the $\theta$-component of velocity in general has negligible energy production and therefore can only play a key part in the energy transport across wavenumbers. This may be understood by perceiving the shear-stresses as the transport of $v^\xi$ by $v^\theta$ distributing the energy most effectively across the jet in regions where mean shear peaks, in a classical Boussinesq-type fashion. Since the low-wavenumber region of the energy spectra represents the region related to TKE production the $\theta$-components of the LD modes must necessarily be very limited in this same wavenumber region, as they would otherwise contribute significantly to the TKE production through the modal reconstruction of the cross-spectrum. This is confirmed by figures \ref{fig:app_LD_modes_real_v_1} in Appendix \ref{app:LD_modes_as_function_of_wavenumber} where significant amplitudes only appear in the $-5/3$-range for the $\theta$-component.

\FloatBarrier
It finally noted that the amplitude decay of $\psi^\xi_\alpha$ and $\psi^\theta_\alpha$ with increasing $\theta$ resembles the streamwise modes exemplified in figure \ref{fig:example_of_mode}. This is quite intriguing as the $\psi^i_\alpha$-components of $\overline{\chi}_\alpha$ are not analytical, but rather numerical. This provides some hope of developing methods for obtaining $\psi^i_\alpha$ in \textit{analytical} form, since the approach of finding analytical forms of the eigenfunctions need not be restricted to classical Fourier modes. 

The current presentation of the energy-optimized decomposition of the turbulent jet far-field resulted in the derivation of the SADFM as the streamwise components of the modal building blocks of the jet far-field. The current work lays the basis for further investigations of the role of these basis functions in the reconstruction of various elements of the flow field, which is addressed in \cite{Hodzic2019_part2}. In \cite{Hodzic2019_part2} the Galerkin Projection of the turbulence kinetic energy transport equation is presented in curvilinear coordinates allowing the investigation of any turbulent flow in any well-defined coordinate system to be projected onto a eigenfunction basis. This formulation is then used to investigate the energy transport properties of the jet in SSC for which the eigenfunctions presented in the current work are used - in particularly the capabilities of the eigenfunctions to extract energy directly from the mean flow, and thereby circumvent a Richardson-like energy cascade.

\FloatBarrier
\section{Conclusions}
In the current work a tensor calculus formulation of the weighted Lumley Decomposition (LD) was introduced and applied to the self-similar region of the turbulent axi-symmetric jet. This form of the LD was applied in order to express the jet in radially stretched spherical coordinates (SSC) in order to deduce the physical form of the eigenfunctions in the streamwise direction, $\xi$, denoted as stretched amplitude decaying Fourier modes (SADFM). With the introduction of the inner product weight function $w=e^{-\xi}$ it was possible to deduce the analytical form of the streamwise evolution of the eigenfunctions for the turbulent jet far-field, and it was demonstrated that the eigenfunctions decay with a $-3/2$-power with the distance from the virtual origin in three dimensional space and time. The spatial evolution of these eigenfunctions describe the wave scattering effect resembling reversed wave shoaling. Energy spectra were obtained from the SADFM, and it is demonstrated that these exhibit the $-5/3$-range as well as the $-7/3$-range for the cross-spectrum, normally affiliated to homogeneous turbulence. It is interesting to note that these slopes occur even for modes that are not regular trigonometric polynomials, the impact of which will be addressed in publications to follow. Lastly it was observed that the LD modes along the $\theta$-direction have a resembling characteristics to the SADFM as both eigenfunction components were decaying in amplitude, the former decaying with distance from the centerline. This symmetry is highly intriguing and should be explored further in order to characterize the coupling between the eigenfunction components.
\section*{Acknowledgments}
The authors would like to thank Dr. Maja W{\"a}nstr{\"o}m for sharing data. Her thesis, \cite{Wanstrom2009}, provided both a useful guidebook and reference, especially since her work was performed in the same facility.

The authors gratefully acknowledge the long-term support of the Department of Mechanical Engineering at the Technical University of Denmark. All the experimental work of this long research program was carried out in these laboratories. This generous support has made these investigations and their continuation possible. A considerable part of the current work is obtained from the PhD-dissertation of Azur Hod\v zi\' c, \cite{Hodzic2018b}.

Last, but not least, the final stages of the project would not have been possible without the support of the European Research Council: This project has received funding from the European Research Council (ERC) under the European Union’s Horizon 2020 research and innovation program (grant agreement No 803419).
\bibliographystyle{jfm}

\input{master.bbl}
%
\appendix
\section{Transport equations in curvilinear coordinates\label{app:Transport_equations}}
For completeness the energy equations of laminar, mean, and turbulent fields are deduced in curvilinear coordinates for a solenoidal, Newtonian fluid with constant material properties. Initially the Reynolds decompositions, $V^i=\left\langle V^i\right\rangle + v^i$ and, $P=\left\langle P\right\rangle + p$, are applied to the Navier-Stokes equations in \eqref{eq:NS-curvilinear}. After ensemble averaging the Reynolds averaged Navier-Stokes (RANS) equations are obtained
\begin{equation}
\frac{\partial \left\langle V^i\right\rangle}{\partial t}+\left\langle V^j\right\rangle\nabla_j\left\langle V^i\right\rangle+\nabla_j\left\langle v^iv^j\right\rangle=-\frac{1}{\rho}\nabla^i\left\langle P\right\rangle+\nu\nabla^j\nabla_j\left\langle V^i\right\rangle\label{eq:RANS},
\end{equation}
where, $\left\langle V^i\right\rangle$, and, $\left\langle P\right\rangle$, denote the ensemble averaged velocity components and pressure field and, $v^i$, and, $p$, denote the corresponding fluctuating fields. In order to obtain the transport equation for laminar flow we multiply the instantaneous momentum equation for the $i$-component with, $V^j$, and add to the the momentum equation for the $j$-component multiplied by, $V^i$. After collecting terms the following form of the transport equations for laminar flow is obtained 
\begin{eqnarray}
\frac{\partial\left(V^iV^j\right)}{\partial\tau}+V^k\nabla_kV^iV^j &=&-\frac{1}{\rho}\left(V^j\nabla^iP+V^i\nabla^jP\right)+\label{eq:laminar_transport}\\
&+&\nu\left(V^j\nabla^k\nabla_kV^i+V^i\nabla^k\nabla_kV^j\right).\nonumber
\end{eqnarray}
Applying the same procedure to the RANS equations and the ensemble averaged velocity components, we obtain the transport equation of the ensemble averaged field
\begin{eqnarray}
&&\frac{\partial\left(\left\langle V^i\right\rangle\left\langle V^j\right\rangle\right)}{\partial t}+\left\langle V^k\right\rangle\nabla_k \left\langle V^i\right\rangle\left\langle V^j\right\rangle +\nabla_k\left\langle V^j\right\rangle\left\langle v^iv^k\right\rangle\nonumber\\
&+&\nabla_k\left\langle V^i\right\rangle\left\langle v^jv^k\right\rangle 
-\left\langle v^iv^k\right\rangle\nabla_k\left\langle V^j\right\rangle -\left\langle v^jv^k\right\rangle\nabla_k\left\langle V^i\right\rangle=\label{eq:mean_transport}\\
&-&\frac{1}{\rho}\left(\left\langle V^j\right\rangle\nabla^iP+\left\langle V^i\right\rangle\nabla^jP\right)+\nu\left(\left\langle V^j\right\rangle\nabla^k\nabla_k\left\langle V^i\right\rangle+\left\langle V^i\right\rangle\nabla^k\nabla_k\left\langle V^j\right\rangle\right).\nonumber
\end{eqnarray}
The transport equation for the Reynolds-stresses is obtained by subtracting \eqref{eq:RANS} from \eqref{eq:NS-curvilinear} for the $i$-component and multiplying by, $V^j$, and adding this to the difference of \eqref{eq:RANS} and \eqref{eq:NS-curvilinear} for the $j$-component multiplied by, $V^i$. Ensemble averaging the resulting field and manipulating with terms yields the final form of the Reynolds-stress transport equation
\begin{eqnarray}
\frac{\partial \left\langle v^iv^j\right\rangle}{\partial t}&+&\left\langle V^k\right\rangle\nabla_k\left\langle v^iv^j\right\rangle+\left\langle v^iv^k\right\rangle\nabla_k\left\langle V^j\right\rangle+\left\langle v^jv^k\right\rangle\nabla_k\left\langle V^i\right\rangle+\nonumber\\
&+&\nabla_k\left\langle v^iv^jv^k\right\rangle=-\frac{1}{\rho}\left(\left\langle v^j\nabla^ip\right\rangle+\left\langle v^i\nabla^jp\right\rangle\right)+\label{eq:fluc_transport}\\
&+&\nu\left(\left\langle v^j\nabla^k\nabla_k v^i\right\rangle+\left\langle v^i\nabla^k\nabla_k v^j\right\rangle\right).\nonumber
\end{eqnarray}
The energy equations for the laminar-, mean-, and fluctuating parts of the velocity field are obtained by contracting \eqref{eq:laminar_transport}, \eqref{eq:mean_transport}, and \eqref{eq:fluc_transport} with, $z_{ij}$, and dividing by two. This operation yields the corresponding energy equations
\begin{eqnarray}
\frac{D K}{Dt}&=&-\frac{1}{\rho}\nabla_iV^iP +\nu\left(\nabla^j\nabla_jK-\left(\nabla^jV_i\right)\nabla_jV^i\right)\label{eq:laminar_energy},\\
\frac{D K_0}{D t}&+&\nabla_j\left\langle V^i\right\rangle\left\langle v_iv^j\right\rangle-\left\langle v_iv^j\right\rangle\nabla_j\left\langle V^i\right\rangle = -\frac{1}{\rho}\nabla_i\left\langle V^i\right\rangle P+\nonumber\\&+&\nu\left(\nabla^j\nabla_jK_0-\left(\nabla^j\left\langle V^i\right\rangle\right)\nabla_j\left\langle V_i\right\rangle\right)\label{eq:mean_energy},\\
\frac{D K_t}{D t}&+&\left\langle v_iv^j\right\rangle\nabla_j\left\langle V^i\right\rangle+\frac{1}{2}\nabla_j\left\langle v_iv^iv^j\right\rangle = -\frac{1}{\rho}\nabla_i\left\langle v^ip\right\rangle+\nonumber\\
&+&\nu\left(\nabla^j\nabla_j K_t-\left\langle \left(\nabla^jv^i\right)\nabla_jv_i\right\rangle\right),\label{eq:turb_energy}
\end{eqnarray}
where the material derivative is defined as
\begin{equation}
\frac{D}{Dt}=\frac{\partial }{\partial t}+\left\langle V^j\right\rangle\nabla_j.
\end{equation}
and the density-normalized kinetic energy for the laminar, mean, and fluctuating part are defined as 
\begin{eqnarray}
K&=&\frac{1}{2}V^iV_i\hspace{3pt},\hspace{3pt} K_0=\frac{1}{2}\left\langle V^i\right\rangle\left\langle V_i\right\rangle\hspace{3pt},\hspace{3pt} K_t=\frac{1}{2}\left\langle v^iv_i\right\rangle.
\end{eqnarray}
Note that the viscous term was separated into two parts in \eqref{eq:laminar_energy}-\eqref{eq:turb_energy}, using the following relation for an arbitrary tensor, $V^i$
\begin{equation}
V_i\nabla_j\nabla^jV^i=\frac{1}{2}\nabla^j\nabla_jV^iV_i-\left(\nabla^jV^i\right)\nabla_jV_i.
\end{equation}
In \eqref{eq:turb_energy} we note that the dissipation can be expressed explicitly by rewriting the viscous part of the equation as follows
\begin{equation}
\nu\left\langle v_i\nabla_j\nabla^jv^i\right\rangle=\left\langle v_i\nabla_j\tau^{ij}\right\rangle=\nabla_j\left\langle v_i\tau^{ij}\right\rangle-\left\langle \tau^{ij}\nabla_j v_i\right\rangle,\label{eq:viscous_dissipation}
\end{equation}
where the last term is the viscous dissipation. The deviatoric stress tensor is defined as
\begin{equation}
\tau^{ij}=2\nu s^{ij},
\end{equation}
and the strain tensor is
\begin{equation}
s^{ij}=\frac{1}{2}\left(\nabla^jv^i+\nabla^iv^j\right).
\end{equation}
In order to formulate the energy equation, \eqref{eq:turb_energy}, in terms of the viscous dissipation term we express the viscous terms by \eqref{eq:viscous_dissipation} in order to obtain the desired form
\begin{eqnarray}
\frac{D K_t}{D t}&+&\left\langle v_iv^j\right\rangle\nabla_j\left\langle V^i\right\rangle+\frac{1}{2}\nabla_j\left\langle v_iv^iv^j\right\rangle = -\frac{1}{\rho}\nabla_i\left\langle v^ip\right\rangle+\nonumber\\
&+&\nabla_j\left\langle v_i\tau^{ij}\right\rangle-\left\langle \tau^{ij}\nabla_jv_i\right\rangle,
\end{eqnarray}
which represents the ensemble averaged turbulence kinetic energy equation with an explicit viscous dissipation term.
\section{Derivation of SADFM in SSC\label{app:Derivation_of_POD_scaled}}
The LD in tensor form reads
\begin{equation}
\int_\Omega R^i_{\cdot\hat{j}}\varphi^{\hat{j}}
\widehat{w}\sqrt{\widehat{Z}}d\mu^{\widehat{4}}=\lambda\varphi^{i}.\label{eq:expression1}
\end{equation}
We know that we can scale the instantaneous physical velocity field by the centerline velocity, which is given by \cite{Hussein1994}
\begin{equation}
U_c=\frac{BM_0^{1/2}}{Ce^{\xi}}.
\end{equation}
Since the covariant bases are proportional to $Ce^\xi$, it means that the contravariant velocities at two different points, $V^i$, and $V^{\hat{i}}$ can be decomposed as $V^i=\widetilde{V}^i\widetilde{U}_c$ and $V^{\hat{i}}=\widetilde{V}^{\hat{i}}\widehat{\widetilde{U}}_c$, where $\widetilde{U}_c$ and $\widehat{\widetilde{U}}_c$ are the centerline velocities of the contravariant components defined as
\begin{eqnarray}
\widetilde{U}_c &=& \frac{BM_0^{1/2}}{{C^2e^{2\xi}}}\\
\widehat{\widetilde{U}}_c &=& \frac{BM_0^{1/2}}{{C^2e^{2\hat{\xi}}}}.
\end{eqnarray}
We see that this is true since we can reconstruct the original velocity field by noting that the physical field is expanded by the covariant basis in the following manner
\begin{equation}
\overline{V} = \underbrace{V^i}_{\widetilde{V}^i\widetilde{U}_c}\overline{z}_i =  \underbrace{\widetilde{V}^i\widetilde{U}_c\overbrace{\overline{z}_i}^{\propto{Ce^{\xi}}}}_{\propto U_c}.
\end{equation}
Since we know that the statistics based on the scaled contravariant velocity, $\widetilde{V}^i$, are homogeneous along $\xi$ the expression in \eqref{eq:expression1} can be rewritten as
\begin{equation}
\int_\Omega \underbrace{\overbrace{z_{\widehat{j}\widehat{k}}}^{\tilde{z}_{\widehat{j}\widehat{k}}C^2e^{2\widehat{\xi}}}  \overbrace{R^{i{\widehat{k}}}}^{\widetilde{U}_c\widehat{\widetilde{U}}_c\widetilde{R}^{i\widehat{k}}}}
_{\widetilde{R}^i_{\widehat{j}}\widetilde{U}_c\widehat{\widetilde{U}}_cC^2e^{2\widehat{\xi}}}\varphi^{\widehat{j}}\underbrace{\sqrt{\widehat{Z}}}_{\left(Ce^{\widehat{\xi}}\right)^3\sin\widehat{\theta}}\widehat{w}d\mu^{\widehat{4}}=\lambda\varphi^{i},
\end{equation}
which means that 
\begin{equation}
\int_\Omega \widetilde{R}^i_{\widehat{j}}\underbrace{\widetilde{U}_c\widehat{\widetilde{U}}_c}_{\frac{B^2M_0}{C^4e^{2\xi}e^{2\widehat{\xi}}}}
C^2e^{2\widehat{\xi}}\varphi^{\widehat{j}}C^3e^{3\widehat{\xi}}\sin\widehat{\theta}\widehat{w} d\mu^{\widehat{4}}=\lambda\varphi^{i}.
\end{equation}
from which is obtained
\begin{equation}
CB^2M_0\int_\Omega \widetilde{R}^i_{\cdot\widehat{j}}\varphi^{\hat{j}}\frac{e^{3\widehat{\xi}}}{e^{2\xi}}\sin\widehat{\theta}\widehat{w}d\mu^{\widehat{4}}=\lambda\varphi^i\label{eq:ss}
\end{equation}
where 
\begin{equation}
\widetilde{R}^i_{\cdot\widehat{j}}=\widetilde{z}_{\widehat{j}\widehat{k}}\widetilde{R}^{i\widehat{k}}.
\end{equation}
After canceling out terms \eqref{eq:ss} takes the form
\begin{equation}
\int_\Omega \widetilde{R}^i_{\widehat{j}}\varphi^{\widehat{j}}\frac{e^{3\widehat{\xi}}}{e^{2\xi}}\sin\widehat{\theta}\widehat{w}d\mu^{\widehat{4}}=\widetilde{\lambda}\varphi^{i},
\end{equation}
where 
\begin{equation}
\widetilde{\lambda} = \frac{\lambda}{CB^2M_0}.
\end{equation}
By setting $\widehat{w}=e^{-\widehat{\xi}}$ and defining $\widetilde{\varphi}^i=e^{2\xi}\varphi^i$ and $\widetilde{\varphi}^{\widehat{j}}=e^{2\widehat{\xi}}\varphi^{\widehat{j}}$, the final non-dimensionalized expression for the LD integral in SSC is obtained
\begin{equation}
\int_\Omega \widetilde{R}^i_{\widehat{j}}\widetilde{\varphi}^{\widehat{j}}\sin\widehat{\theta}d\mu^{\widehat{4}}=\widetilde{\lambda}\widetilde{\varphi}^{i},
\end{equation}
where $\widetilde{R}^i_{\widehat{j}}$ represents the two-point two-½time correlation tensor for the scaled field. This leads to the conclusion that the optimal $\xi$-dependent modal components for the scaled field, $\tilde{v}^i$, are stretched Fourier modes. From this the SADFM can be deduced for the physical, $\xi$-decaying velocity field. We note that this derivation of the modes was made possible by the introduction of the specific weight function, $\widehat{w}=e^{-\widehat{\xi}}$. 
\FloatBarrier
\section{Orthogonality of eigenfunctions \label{app:Orthogonality_of_eigenfunctions}}
In the following it is demonstrated that the spatially decaying stretched eigenfunctions are orthogonal with respect to $(\cdot,\cdot)_w:\Omega\times\Omega\rightarrow\mathbb{C}$, where $\Omega:=\Omega_t\times\Omega_\xi\times\Omega_\theta\times\Omega_\phi$. For different choices of $\Omega$, however, the weight function, $w$, must be adapted accordingly in order to achieve orthogonality between eigenfunctions. For a finite domain where $\Omega_t=[0:T]$, $\Omega_\xi=[0:L_\xi]$, $\Omega_\theta = [0:\theta_{\text{max}}]$ and $\Omega_\phi = [0:2\pi[$, we let~${\overline{\Phi}^\alpha,\overline{\Phi}_\beta\in L^2_w(\Omega,\mathbb{C})}$. The $L^2_w$-inner product of the two vectors is then
\begin{eqnarray}
\left(\overline{\Phi}^\alpha,\overline{\Phi}_\beta\right)_w&=&\int_\Omega z_{jk}\varphi^{j\alpha}\varphi_\beta^{k*}wd\mu,\nonumber\\
&=& \int_\Omega z_{jk}\varphi^{j\alpha}\varphi_\beta^{k*}w\sqrt{Z}d\mu^4,\nonumber\\
&=& \frac{1}{\Vol}\int_\Omega z_{jk}\frac{\widetilde{\psi}^{j\alpha}\widetilde{\psi}^{k*}_\beta}{C^5e^{4\xi}\sin\theta}e^{i\left(t\left(\omega'-\omega\right)+\xi\left(\kappa'-\kappa\right)+\phi\left(m-n\right)\right)}w\sqrt{Z}d\mu^4,\\
\end{eqnarray}
where $\Vol = 2\pi TL_\xi$, $d\mu^4=dt\,d\xi\,d\theta\,d\phi$. Since $z_{ij}=\tilde{z}_{ij}\left(Ce^\xi\right)^2$, $\sqrt{Z}=\sin\theta\left(Ce^\xi\right)^3$ and $w=e^{-\xi}$ this yields
\begin{eqnarray}
\left(\overline{\Phi}^\alpha,\overline{\Phi}_\beta\right)_w &=& \frac{1}{\Vol}\int_\Omega \tilde{z}_{jk}\widetilde{\psi}^{j\alpha}\widetilde{\psi}^{k*}_{\beta}%
e^{i\left(t\frac{2\pi\left(r-s\right)}{T}+\frac{2\pi\left(p-q\right)}{L_\xi}\xi+\left(m-n\right)\phi\right)}d\mu^4,\nonumber\\
&=& \underbrace{\int_{\Omega_\theta}\tilde{z}_{jk}\widetilde{\psi}^{j\alpha}\widetilde{\psi}^{k*}_{\beta}d\theta}_{\delta^\alpha_\beta}%
\underbrace{\frac{1}{\Vol}\int_{\Omega_t\times\Omega_\xi\times\Omega_\phi}e^{i\left(t\frac{2\pi\left(r-s\right)}{T}+\frac{2\pi\left(p-q\right)}{L_\xi}\xi+\left(m-n\right)\phi\right)}d\mu^3}_{\delta_{rs}\delta_{pq}\delta_{mn}}.
\end{eqnarray}
We can therefore write
\begin{equation}
\left(\overline{\Phi}^\alpha,\overline{\Phi}_\beta\right)_w = \delta^\alpha_\beta\delta_{rs}\delta_{pq}\delta_{mn},\label{eq:orthogonality_a_b}
\end{equation}
where $\delta^\alpha_\beta$, $\delta_{rs}$, $\delta_{pq}$, and $\delta_{mn}$ are Kronecker deltas. 
\FloatBarrier
\section{Component SADFM spectra\label{app:Formulation_energy_spectra}}
Having confirmed the form of the LD modes in Appendix \ref{app:Derivation_of_POD_scaled} as well as the orthogonality of the LD modes in Appendix \ref{app:Orthogonality_of_eigenfunctions} the current section depicts how the streamwise SADFM energy spectra are obtained for the far-field region of the jet. The aim is thus to obtain an analytical set of orthogonal basis functions from which an energy spectrum can be obtained. The spectrum should depict the true velocity field, and not the scaled one, such that the basis functions optimally represent the physical flow field. There exist multiple ways to deduce the optimal basis. We note that an alternative to the current approach is based on the LD integral for scalar functions with the introduction of a weight. This approach yields the same result as the following.
\\
\\
Consider a complex function $f=f(\kappa)=e^{i\kappa}$, where $\kappa\in\mathbb{R}$. Introducing another function, $g = f^{\ln \tilde{x}}$, where $\tilde{x} = (x-x_0)/C$ we obtain
\begin{equation}
g(\kappa)=f^{\ln \tilde{x}} = e^{i\kappa\ln \tilde{x}} = e^{\ln\left(\tilde{x}^{i\kappa}\right)} = \tilde{x}^{i\kappa}\label{eq:g_k},
\end{equation}
which shows that \eqref{eq:g_k} is a complex power series with a real base. Transforming \eqref{eq:g_k} to SSC by $\tilde{x}=e^\xi$, we obtain the trigonometric polynomials 
\begin{equation}
g = \tilde{x}^{i\kappa} = e^{i\kappa\xi},\label{eq:g_k_SSC}
\end{equation}
which appear in the kernel of \eqref{eq:LD_SSC}. In order to obtain energy spectra along the streamwise direction of the jet far-field with respect to \eqref{eq:g_k_SSC} the orthogonality of the basis used to expand the field must be ensured. This is achieved by choosing an appropriate inner product - analogous to the procedure in Appendix \ref{app:Derivation_of_POD_scaled}. Here the subspace of \eqref{eq:L2-function_space} is chosen, and is defined as
\begin{equation}
L^2_w\left(\Omega,\mathbb{C}\right):=\left\lbrace f:\Omega\rightarrow\mathbb{C} \rvert \int_\Omega \lvert f \rvert^2wd\mu <\infty\right\rbrace,\label{eq:L2-function_space_2}
\end{equation}
where the corresponding weighted inner product is defined as
\begin{equation}
\left(f,g\right)_w = \int_\Omega fg^*wd\mu.\label{eq:wieghted_inner_product}
\end{equation}
In the one-dimensional case, the domain is chosen to be an interval, $\Omega = \left[a,b\right]$, where $0<a<b$. The function \eqref{eq:g_k_SSC} can be evaluated in terms of orthogonality by taking the inner product of the functions $g=g\left(\kappa\right)$ and $g'=g\left(\kappa'\right)$
\begin{equation}
\left(g,g'\right)_w = \int_a^b g g'^{*}wd\mu = \int_a^b \tilde{x}^{i\kappa} \tilde{x}^{-i\kappa '}wd\tilde{x} = \int_a^b \tilde{x}^{i\left(\kappa-\kappa'\right)}wd\tilde{x}.
\end{equation}
In SSC this yields
\begin{equation}
\left(g,g'\right)_w = \int_{\xi\left(a\right)}^{\xi\left(b\right)} e^{i\left(\kappa-\kappa'\right)\xi}e^\xi wd\xi.\label{eq:g_g_orthogonality}
\end{equation}
Choosing $w=e^{-\xi}$ these are orthogonal for appropriately chosen integration limits (see Appendix \ref{app:Orthogonality_of_eigenfunctions}). Defining the function $h$ by $h=ge^{-\xi/2}$ the following is obtained from \eqref{eq:g_k_SSC}
\begin{equation}
h=ge^{-\xi/2} = e^{i\kappa\xi-\xi/2} = \tilde{x}^{i\kappa-\frac{1}{2}},
\end{equation}
which is orthogonal for $w=1$ on and appropriately chosen limits. This means that $h$ can be used as an orthonormal basis to expand the flow field. Since the same basis is obtained from the one-dimensional LD integral, it means that $h$ represent the optimal one-dimensional basis of the flow in the streamwise direction.

The projection coefficients of the velocity components onto the orthonormal eigenfunctions $g$ can then be obtained as follows
\begin{equation}
c\left(\kappa\right) = \left(v,h\right) = \int_a^b v\tilde{x}^{-i\kappa-\frac{1}{2}}d\tilde{x}= \int_{\xi\left(a\right)}^{\xi\left(b\right)} ve^{-i\kappa\xi+\frac{1}{2}\xi} d\xi.
\end{equation}
Since $v=\tilde{v}U_c$, where $U_c=BM_0^{1/2}/\left(Ce^\xi\right)$ the coefficients can be written as
\begin{equation}
c\left(\kappa\right) =  \frac{BM_0^{\frac{1}{2}}}{C}\int_{\xi\left(a\right)}^{\xi\left(b\right)} u e^{-i\kappa\xi} d\xi,
\end{equation}
where $u=\tilde{v}e^{-\frac{1}{2}\xi}$. Thereby the regular FFT algorithm can be used to obtain the projection coefficients, as long as the scaled velocities, $\tilde{v}$, are multiplied by $e^{-\frac{1}{2}\xi}$ prior to the FFT. The orthonormality of the bases, $g$, with respect to $\left(\cdot,\cdot\right)$ means that the field can be expanded as follows
\begin{equation}
v=c\left(\kappa\right)h\left(\kappa\right).
\end{equation}
The one-dimensional energy spectra are then obtained from 
\begin{equation}
E\left(\kappa\right)=\left\lvert c\left(\kappa\right)\right\rvert^2.
\end{equation}
%
%
\section{Effect of $w$ on the filtering of TKE}\label{app:effect_of_w}
Defining the LD on a weighted inner product space means that TKE is filtered away from the jet far-field. The filtering is directly related to the choice of weight. Since the weight is known it is possible to derive the filtering of the TKE of the flow based on the choice of $w$ and choice of domain $\Omega$. The weighted mean TKE is then 
\begin{equation}
K_w=\frac{\rho}{2}\left\langle\left(\overline{v},\overline{v}\right)_w\right\rangle = \frac{\rho}{2}\int_\Omega \left\langle\overline{v}\cdot\overline{v}\right\rangle wd\mu,\label{eq:app_WTKE_equation}
\end{equation}
where $\overline{v}$ represents the fluctuating part of the velocity field. Choosing to represent the domain in SSC, we have $\Omega:=\Omega_\xi\times\Omega_\theta\times\Omega_\phi$, where $\Omega_\xi:=[0;L_\xi]$, $\Omega_\theta:=[0;\theta']$, and $\Omega_\phi:=[0;2\pi[$. Converting \eqref{eq:app_WTKE_equation} into arithmetic form we obtain 
\begin{equation}
K_w = \frac{\rho}{2}\int_{\Omega_\xi}\int_{\Omega_\theta}\int_{\Omega_\phi}\left\langle\overline{v}\cdot\overline{v}\right\rangle w\sqrt{Z}d\xi d\theta d\phi\label{eq:app_TKE_arithmetic}.
\end{equation}
We can then split the averaged velocity dot-product into a $\theta$-dependent part, $f\left(\theta\right)$, and a $\xi$-dependent part in the following way due to axi-symmetry
\begin{equation}
\left\langle\overline{v}\cdot\overline{v}\right\rangle = f\left(\theta\right)e^{-2\xi}.
\end{equation}
By inserting the volume element from \eqref{eq:volume_element_SSC} together with the integration limits \eqref{eq:app_TKE_arithmetic} takes the following form
\begin{equation}
K_w = \rho C^3\pi\int_0^{\theta'}f\left(\theta\right)\sin\theta d\theta\int_0^{L_\xi}e^\xi wd\xi.
\end{equation}
\begin{center}
\begin{figure}[t]
\subfloat[]{\includegraphics[scale=0.45]{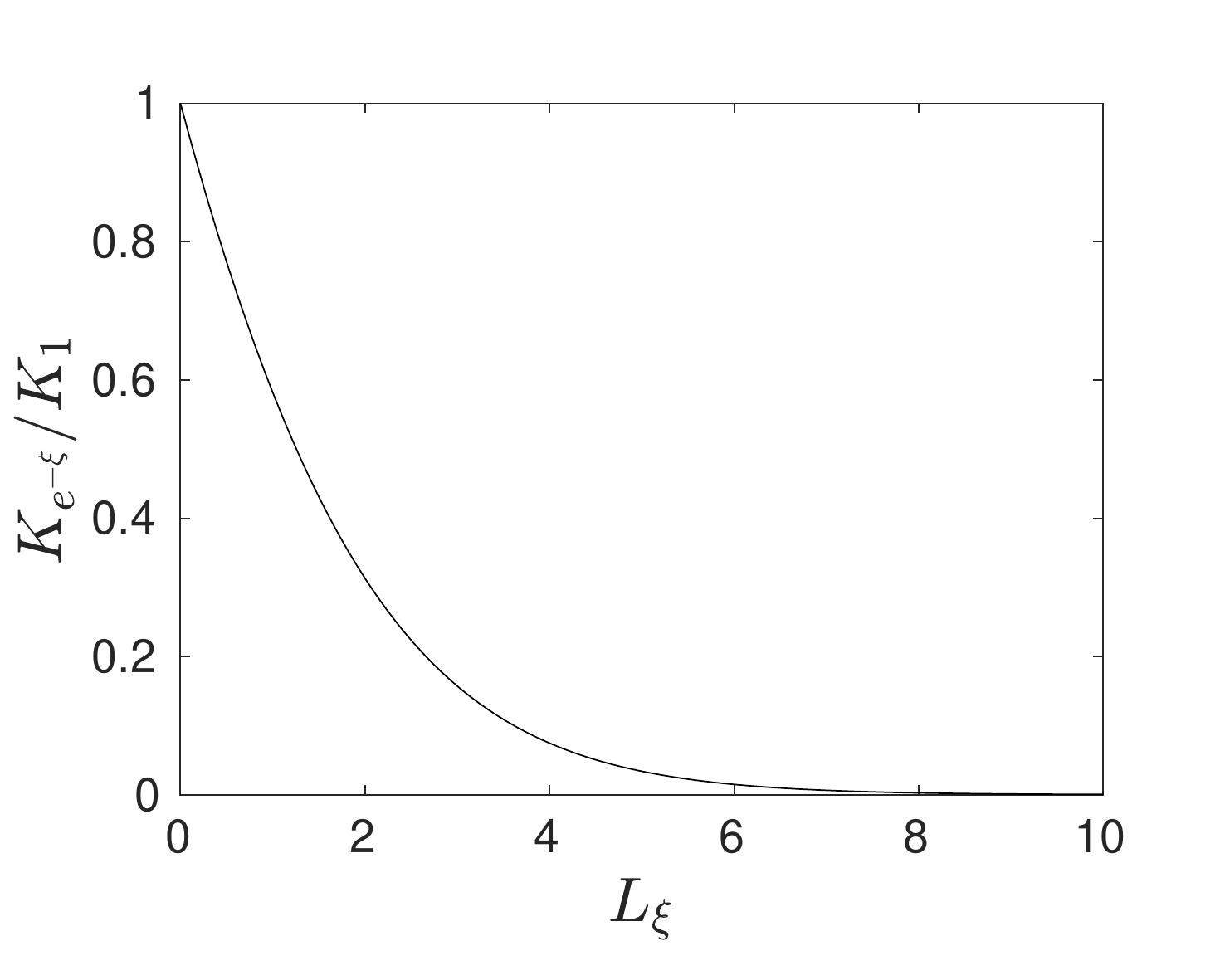}\label{fig:TKE_loss}}
\subfloat[]{\includegraphics[scale=0.45]{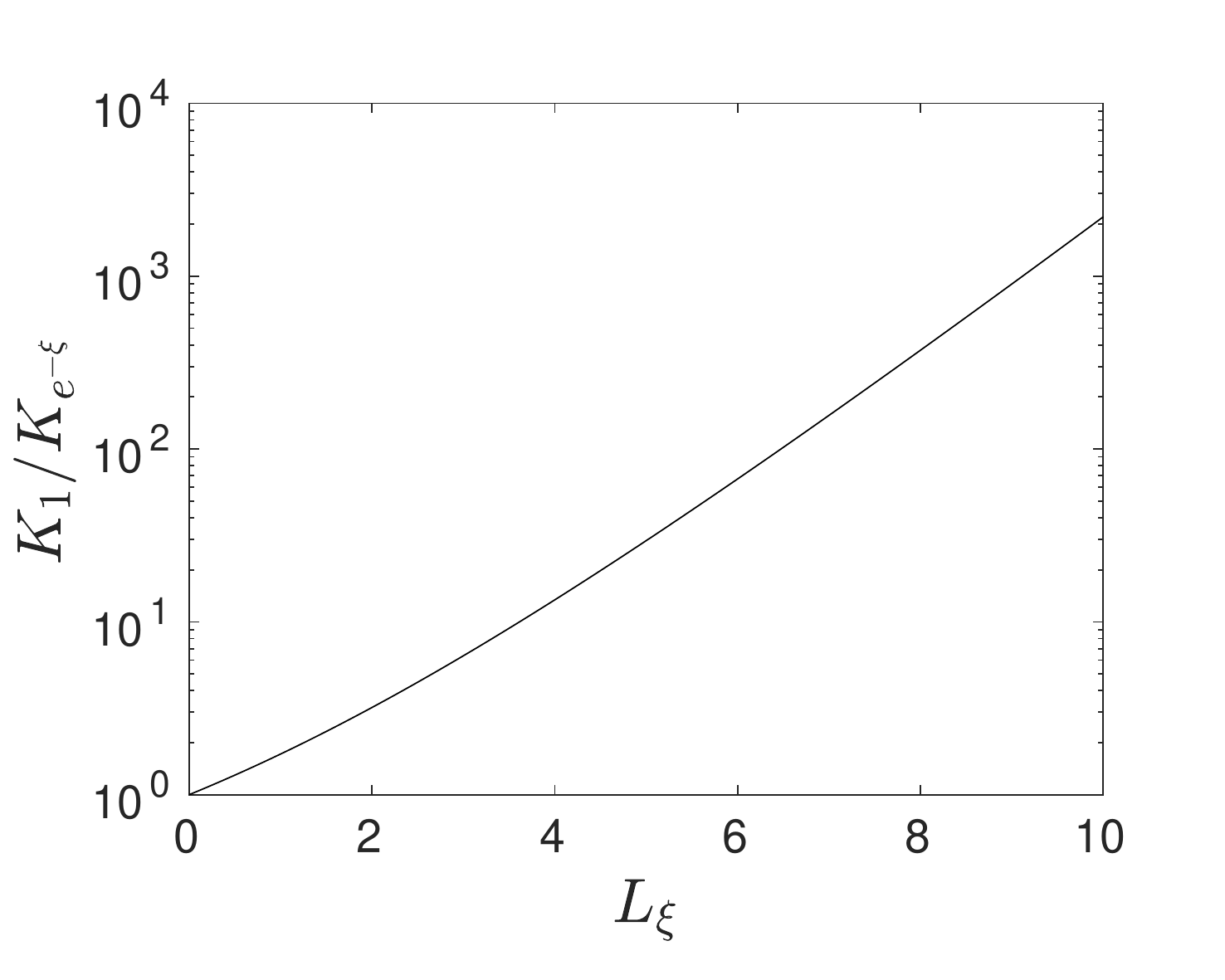}\label{fig:TKE_scaling_factor}}
\caption{(a): The loss of TKE, $K_{e^{-\xi}}/K_1 = L_\xi/\left(e^{L_\xi}-1\right)$, in the jet far-field by applying $w=e^{-\xi}$ in the weighted inner-product definition. (b): $K_1/K_{e^{-\xi}}$ in a semi-logarithmic plot.}
\end{figure}
\end{center}
\noindent
Now taking the ratio between the weighted ($w=e^{-\xi}$) and the unweighted ($w=1$) TKE integrals yields 
\begin{equation}
\frac{K_{e^{-\xi}}}{K_1} = \frac{L_\xi}{e^{L_\xi}-1},\label{eq:ratio}
\end{equation}
which is illustrated in figure \ref{fig:TKE_loss}. For the current case of $L_\xi = 1.21$, $K_{e^{-\xi}}/K_1 = 0.51$, or the energy filtering is approximately $49\%$ compared to the case of $w=1$. From this follows that the energy represented by the eigenvalues obtained from the weighted inner product are likewise weighted, as they are obtained from the same inner product space. This is seen directly from 
\begin{equation}
E_w=\frac{\rho}{2}\left\langle\left(\overline{v},\overline{v}\right)_w\right\rangle = \frac{\rho}{2}\sum_\alpha\lambda^\alpha_w,\label{eq:weighted_e_value_sum}
\end{equation}
where the $w$-subscript on the eigenvalue designates the weight used for the computation. From \eqref{eq:ratio} and \eqref{eq:weighted_e_value_sum} we obtain
\begin{equation}
\sum_\alpha\lambda^\alpha_1 =\frac{e^{L_{\xi}}-1}{L_\xi}\sum_\alpha\lambda^\alpha_{e^{-\xi}},
\end{equation}
where the multiplier is shown in figure \ref{fig:TKE_scaling_factor}. Note, however, that this does not necessarily imply that $\lambda^\alpha_1=\left(e^{L_{\xi}}-1\right)\lambda^\alpha_{e^{-\xi}}/L_\xi$ for each individual $\alpha$.
\FloatBarrier
\newpage
\section{LD modes as functions of wavenumber\label{app:LD_modes_as_function_of_wavenumber}}
\begin{figure}[h]
\centering   
\subfloat[$|\Re\left\lbrace\psi^\xi_1\left(\kappa,\theta\right)\right\rbrace|$]{\includegraphics[width=0.30\linewidth]{figs/POD/SSC/mode_u_1real}\label{fig:app_mode_u_1real}}
\subfloat[$|\Re\left\lbrace\psi^\xi_2\left(\kappa,\theta\right)\right\rbrace|$]{\includegraphics[width=0.30\linewidth]{figs/POD/SSC/mode_u_2real}\label{fig:app_mode_u_2real}}
\subfloat[$|\Re\left\lbrace\psi^\xi_3\left(\kappa,\theta\right)\right\rbrace|$]{\includegraphics[width=0.30\linewidth]{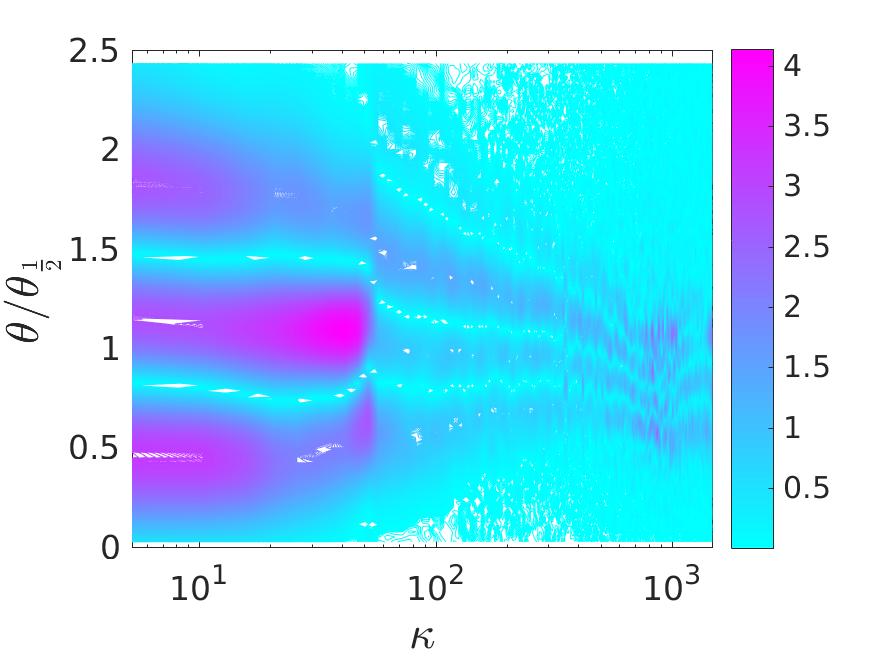}\label{fig:app_mode_u_3real}}\\
\subfloat[$|\Re\left\lbrace\psi^\xi_4\left(\kappa,\theta\right)\right\rbrace|$]{\includegraphics[width=0.30\linewidth]{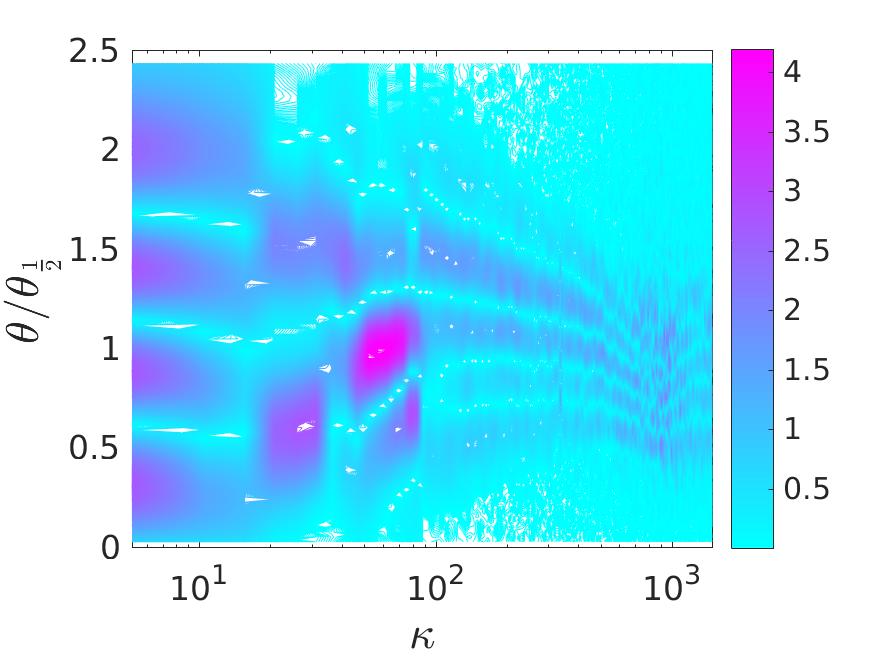}\label{fig:app_mode_u_4real}}
\subfloat[$|\Re\left\lbrace\psi^\xi_5\left(\kappa,\theta\right)\right\rbrace|$]{\includegraphics[width=0.30\linewidth]{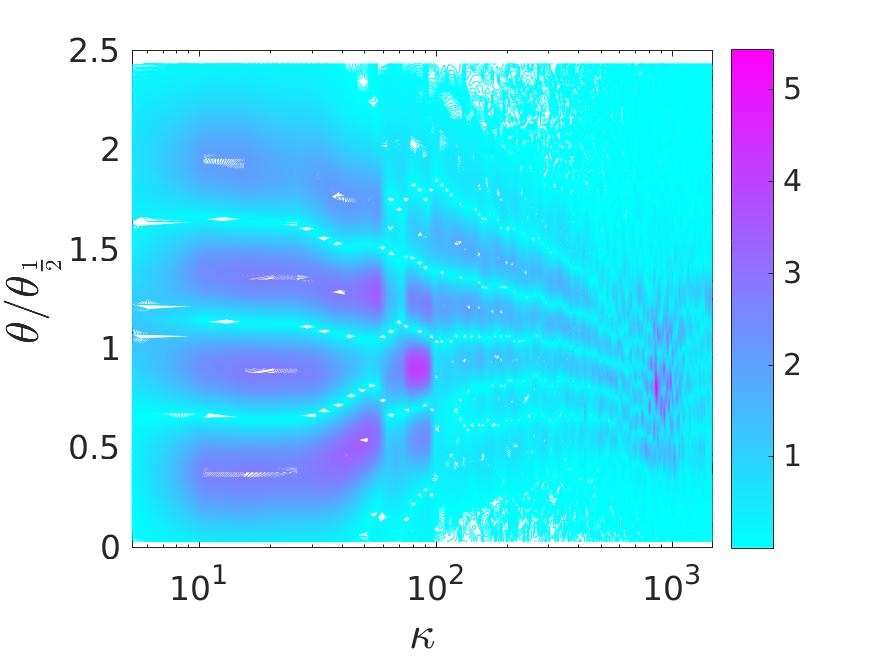}\label{fig:app_mode_u_5real}}
\subfloat[$|\Re\left\lbrace\psi^\xi_6\left(\kappa,\theta\right)\right\rbrace|$]{\includegraphics[width=0.30\linewidth]{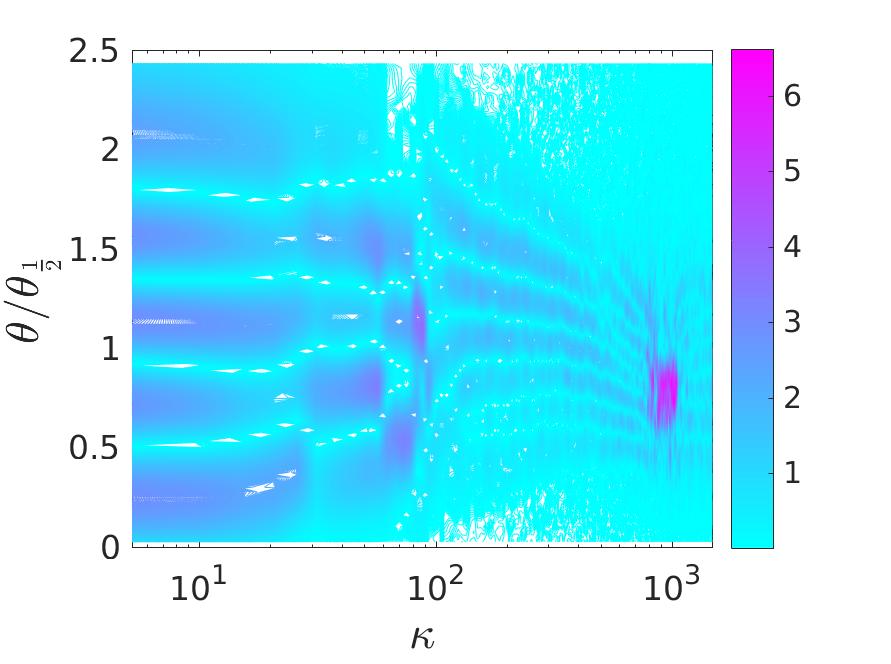}\label{fig:app_mode_u_6real}}\\
\subfloat[$|\Re\left\lbrace\psi^\xi_7\left(\kappa,\theta\right)\right\rbrace|$]{\includegraphics[width=0.30\linewidth]{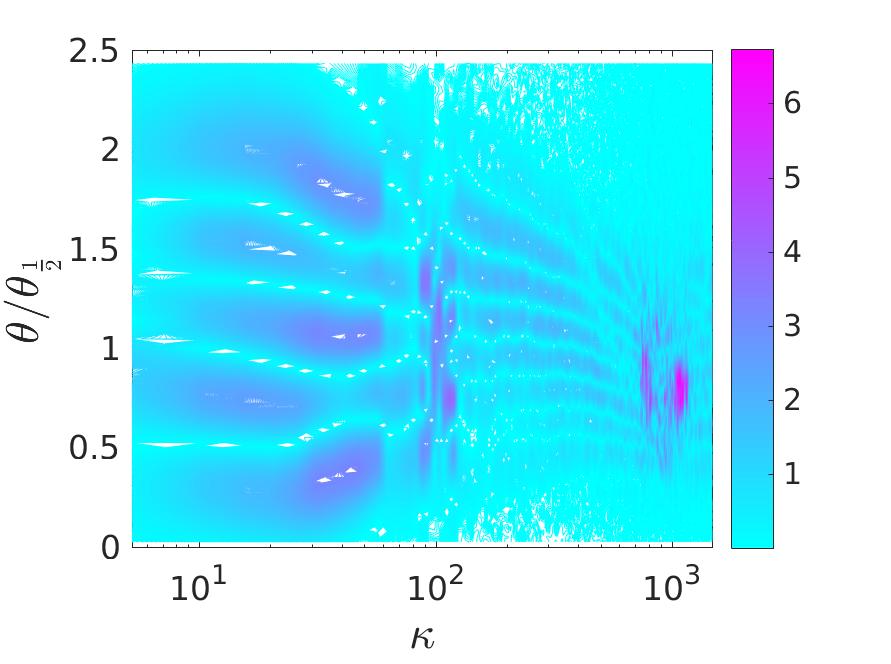}\label{fig:app_mode_u_7real}}
\subfloat[$|\Re\left\lbrace\psi^\xi_8\left(\kappa,\theta\right)\right\rbrace|$]{\includegraphics[width=0.30\linewidth]{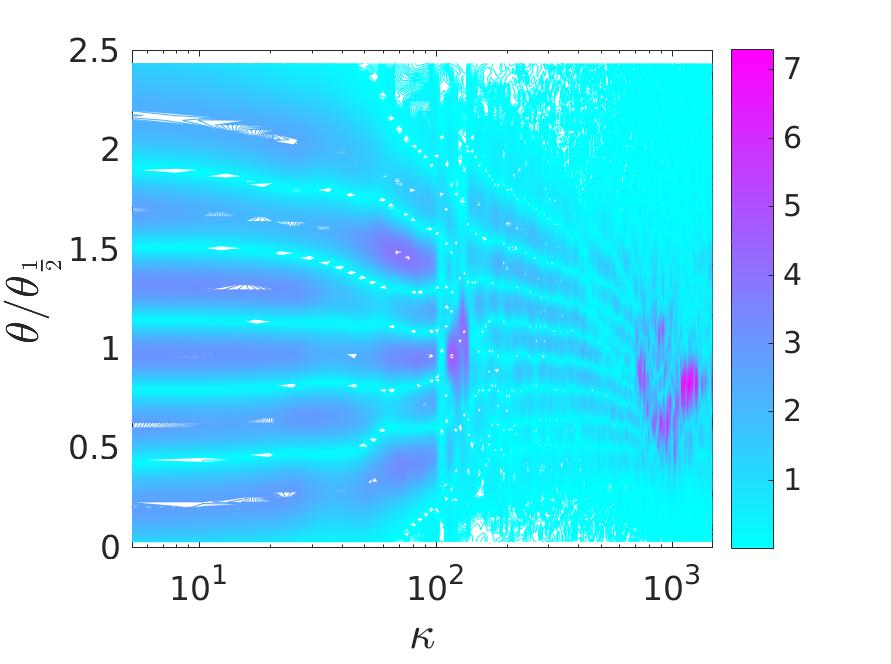}\label{fig:app_mode_u_8real}}
\subfloat[$|\Re\left\lbrace\psi^\xi_9\left(\kappa,\theta\right)\right\rbrace|$]{\includegraphics[width=0.30\linewidth]{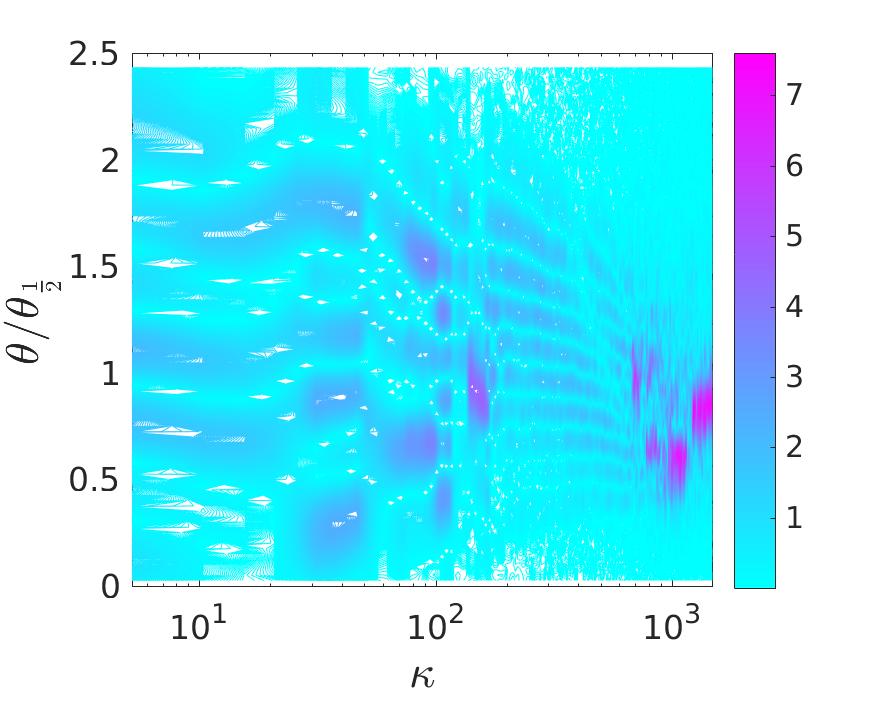}\label{fig:app_mode_u_9real}}
\caption{The absolute real parts of $\xi$-components of the LD modes $\alpha=1-9$, as a function of dimensionless wavenumber, $|\Re\left\lbrace\psi^\xi_\alpha\left(\kappa,\theta\right)\right\rbrace|$. The superscript indicates the directional component $\xi$ or $\theta$ and the subscript denotes the LD mode number.\label{fig:app_LD_modes_real_u_1}}
\end{figure}

\begin{figure}[h]
\centering   
\subfloat[$|\Re\left\lbrace\psi^\theta_1\left(\kappa,\theta\right)\right\rbrace|$]{\includegraphics[width=0.30\linewidth]{figs/POD/SSC/mode_v_1real}\label{fig:app_mode_v_1real}}
\subfloat[$|\Re\left\lbrace\psi^\theta_2\left(\kappa,\theta\right)\right\rbrace|$]{\includegraphics[width=0.30\linewidth]{figs/POD/SSC/mode_v_2real}\label{fig:app_mode_v_2real}}
\subfloat[$|\Re\left\lbrace\psi^\theta_3\left(\kappa,\theta\right)\right\rbrace|$]{\includegraphics[width=0.30\linewidth]{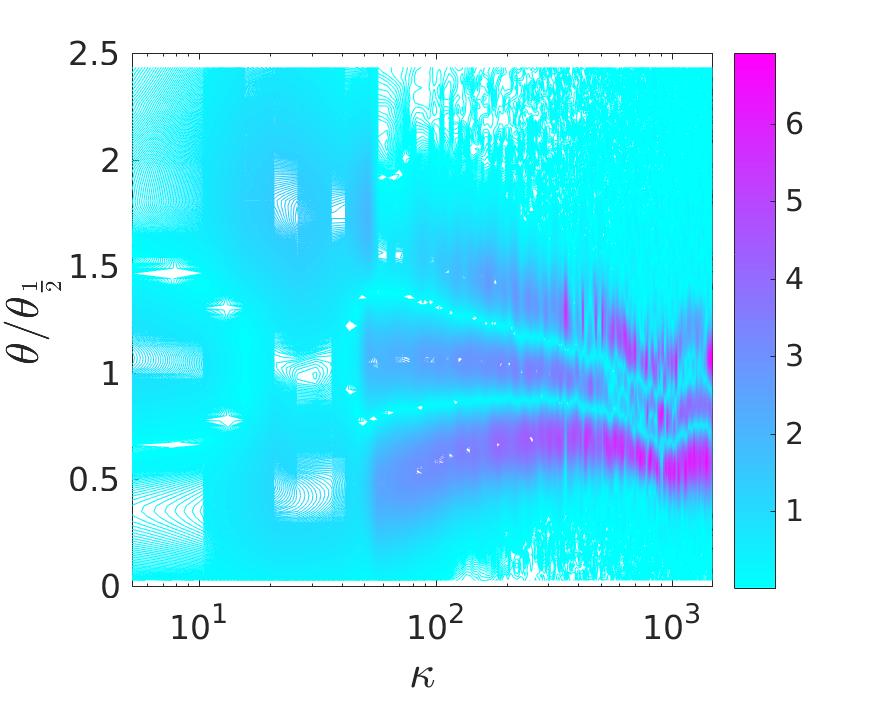}\label{fig:app_mode_v_3real}}\\
\subfloat[$|\Re\left\lbrace\psi^\theta_4\left(\kappa,\theta\right)\right\rbrace|$]{\includegraphics[width=0.30\linewidth]{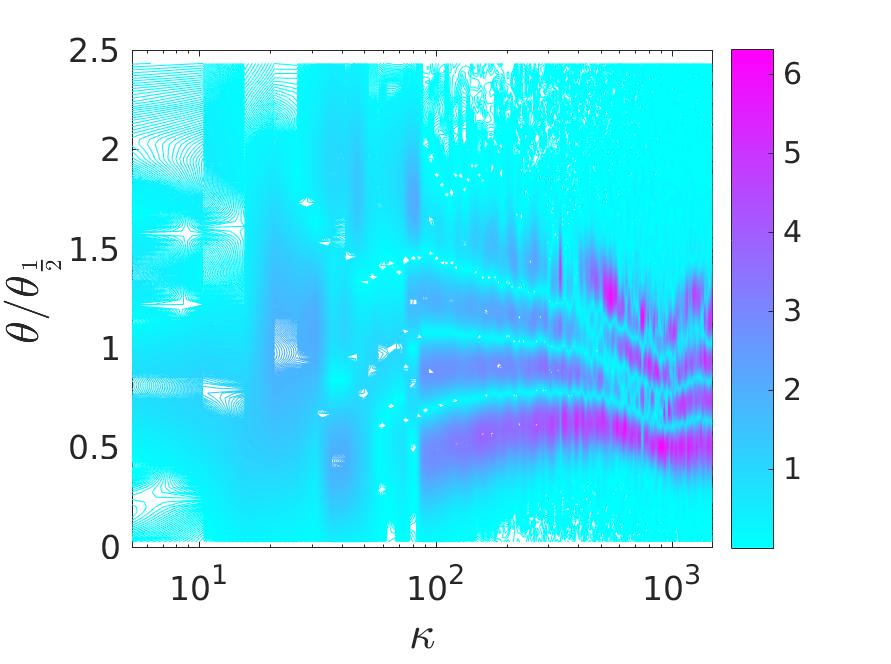}\label{fig:app_mode_v_4real}}
\subfloat[$|\Re\left\lbrace\psi^\theta_5\left(\kappa,\theta\right)\right\rbrace|$]{\includegraphics[width=0.30\linewidth]{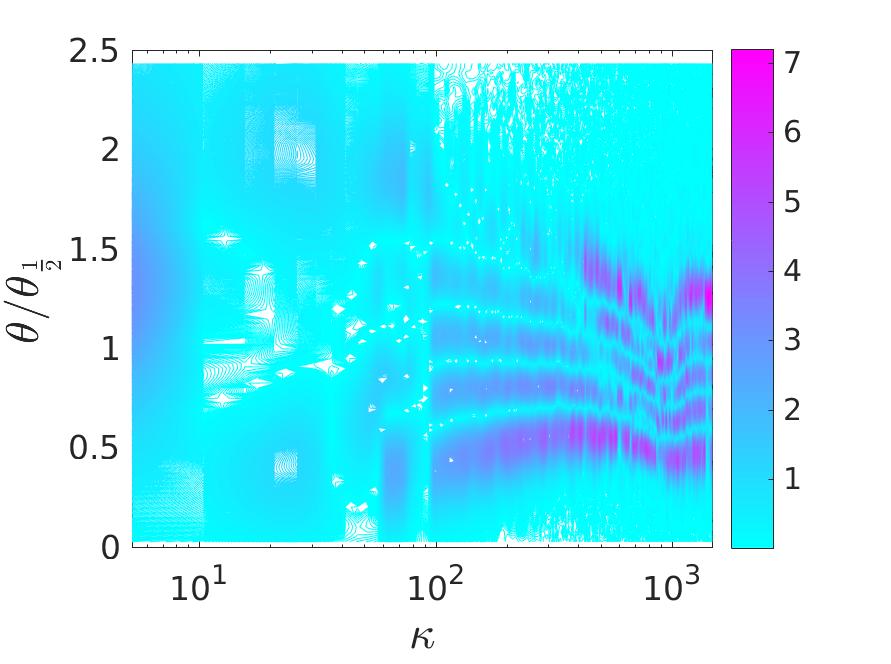}\label{fig:app_mode_v_5real}}
\subfloat[$|\Re\left\lbrace\psi^\theta_6\left(\kappa,\theta\right)\right\rbrace|$]{\includegraphics[width=0.30\linewidth]{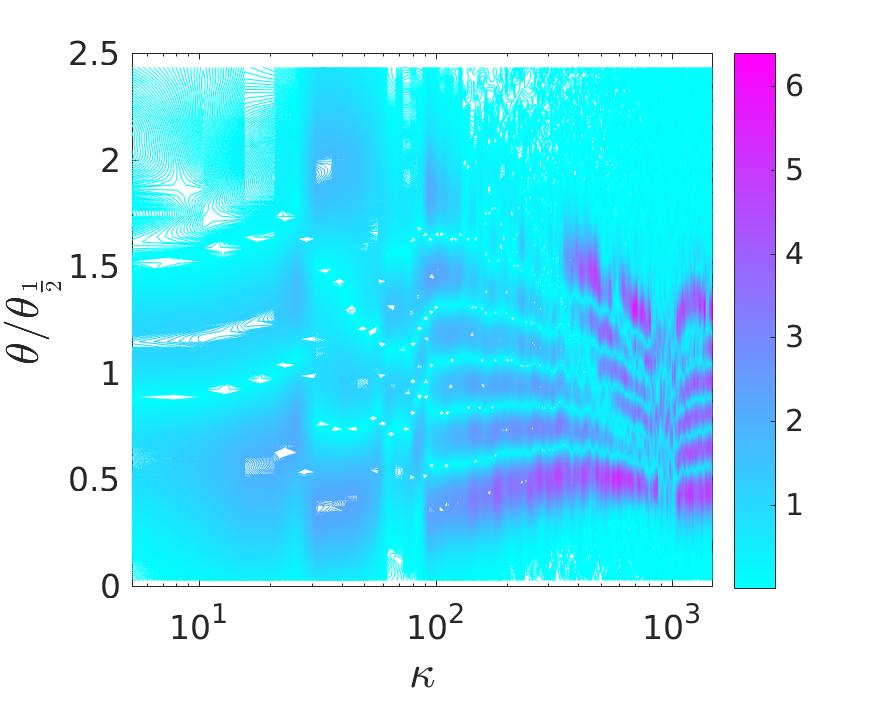}\label{fig:app_mode_v_6real}}\\
\subfloat[$|\Re\left\lbrace\psi^\theta_7\left(\kappa,\theta\right)\right\rbrace|$]{\includegraphics[width=0.30\linewidth]{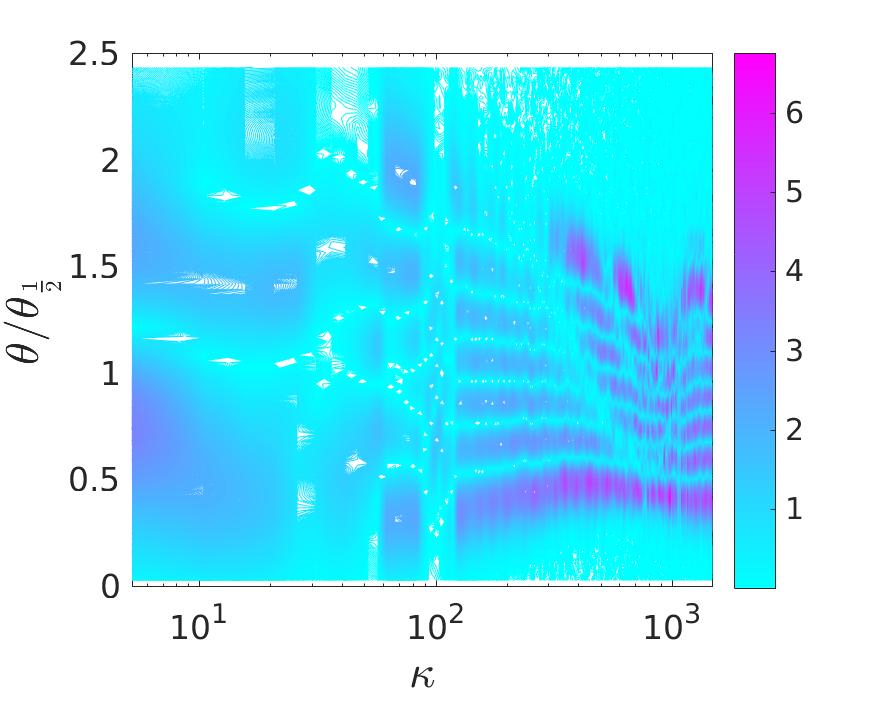}\label{fig:app_mode_v_7real}}
\subfloat[$|\Re\left\lbrace\psi^\theta_8\left(\kappa,\theta\right)\right\rbrace|$]{\includegraphics[width=0.30\linewidth]{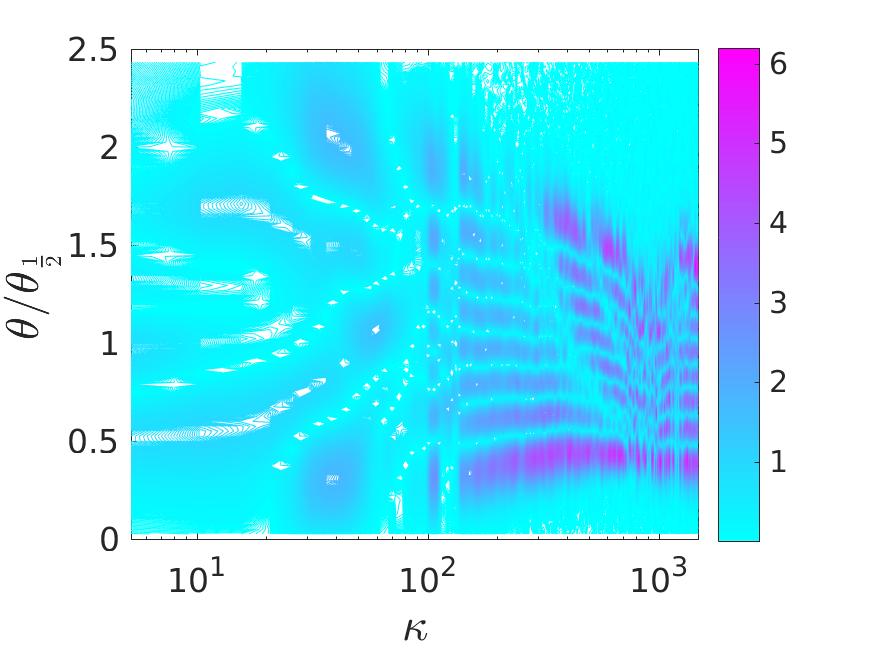}\label{fig:app_mode_v_8real}}
\subfloat[$|\Re\left\lbrace\psi^\theta_9\left(\kappa,\theta\right)\right\rbrace|$]{\includegraphics[width=0.30\linewidth]{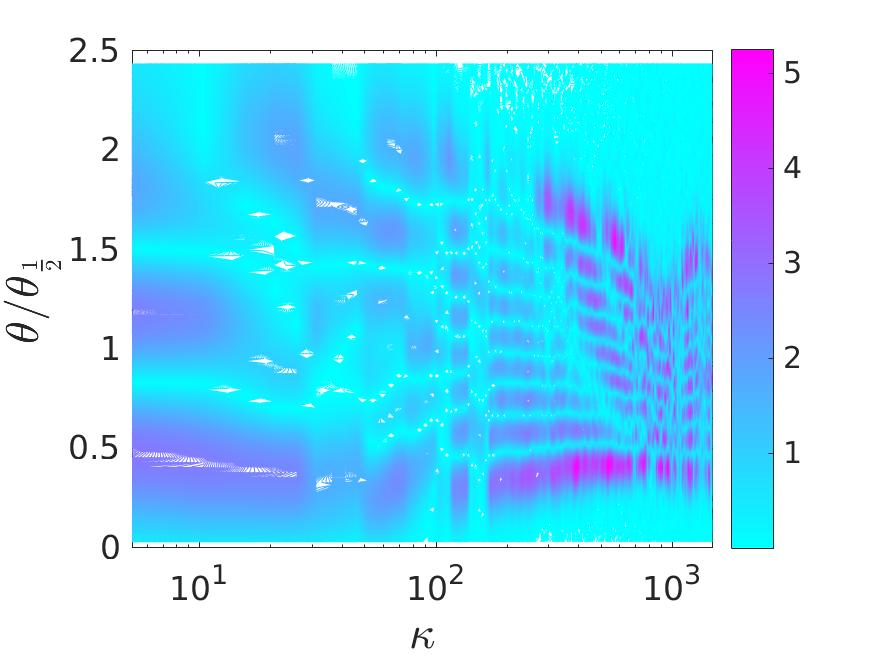}\label{fig:app_mode_v_9real}}
\caption{The absolute real parts of $\theta$-components of the LD modes $\alpha=1-9$, as a function of dimensionless wavenumber, $|\Re\left\lbrace\psi^\theta_\alpha\left(\kappa,\theta\right)\right\rbrace|$. The superscript indicates the directional component $\xi$ or $\theta$ and the subscript denotes the LD mode number.\label{fig:app_LD_modes_real_v_1}}
\end{figure}

\begin{figure}[h]
\centering   
\subfloat[$|\Im\left\lbrace\psi^\xi_1\left(\kappa,\theta\right)\right\rbrace|$]{\includegraphics[width=0.30\linewidth]{figs/POD/SSC/mode_u_1imag}\label{fig:app_mode_u_1imag}}
\subfloat[$|\Im\left\lbrace\psi^\xi_2\left(\kappa,\theta\right)\right\rbrace|$]{\includegraphics[width=0.30\linewidth]{figs/POD/SSC/mode_u_2imag}\label{fig:app_mode_u_2imag}}
\subfloat[$|\Im\left\lbrace\psi^\xi_3\left(\kappa,\theta\right)\right\rbrace|$]{\includegraphics[width=0.30\linewidth]{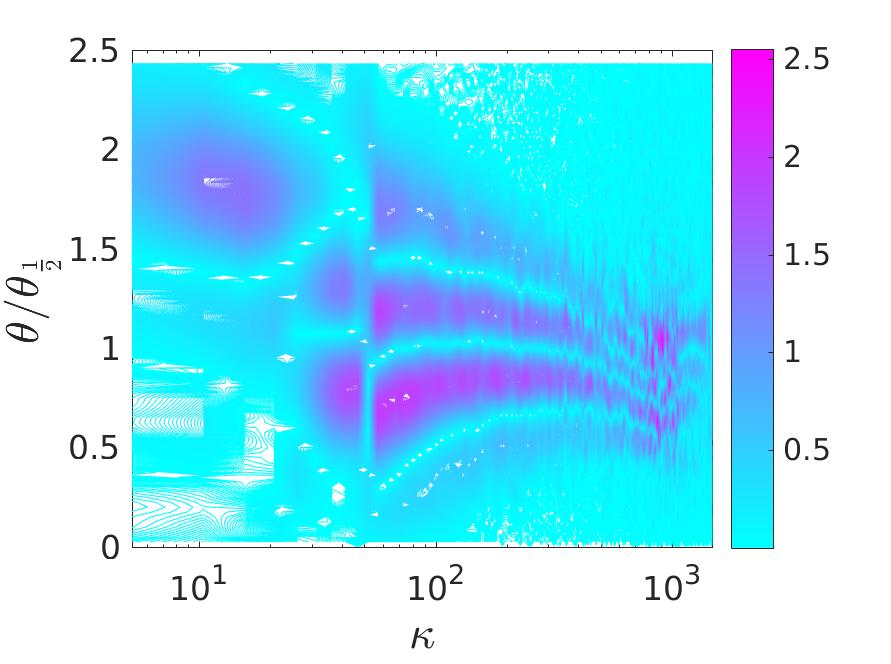}\label{fig:app_mode_u_3imag}}\\
\subfloat[$|\Im\left\lbrace\psi^\xi_4\left(\kappa,\theta\right)\right\rbrace|$]{\includegraphics[width=0.30\linewidth]{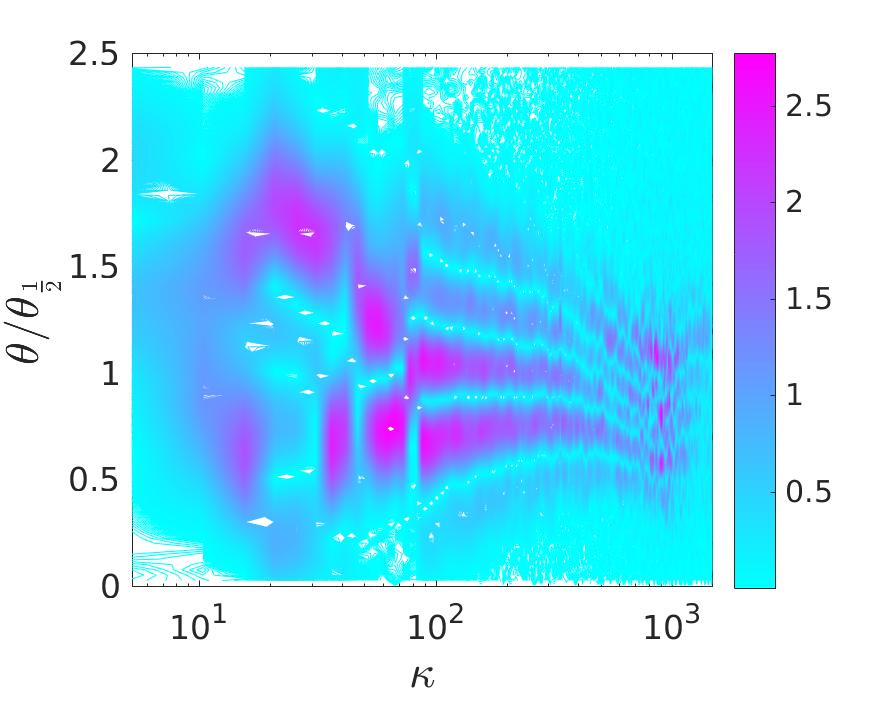}\label{fig:app_mode_u_4imag}}
\subfloat[$|\Im\left\lbrace\psi^\xi_5\left(\kappa,\theta\right)\right\rbrace|$]{\includegraphics[width=0.30\linewidth]{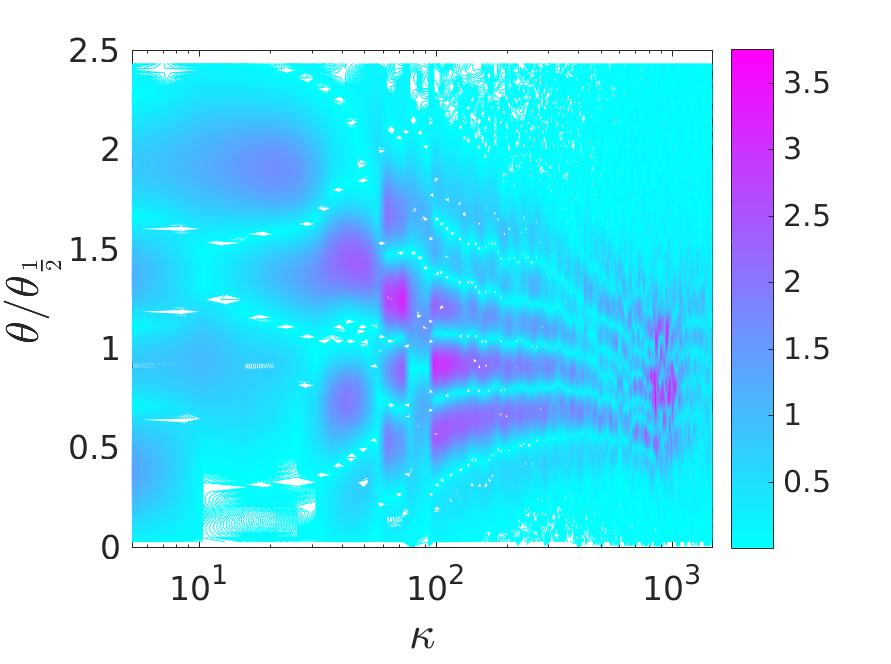}\label{fig:app_mode_u_5mag}}
\subfloat[$|\Im\left\lbrace\psi^\xi_6\left(\kappa,\theta\right)\right\rbrace|$]{\includegraphics[width=0.30\linewidth]{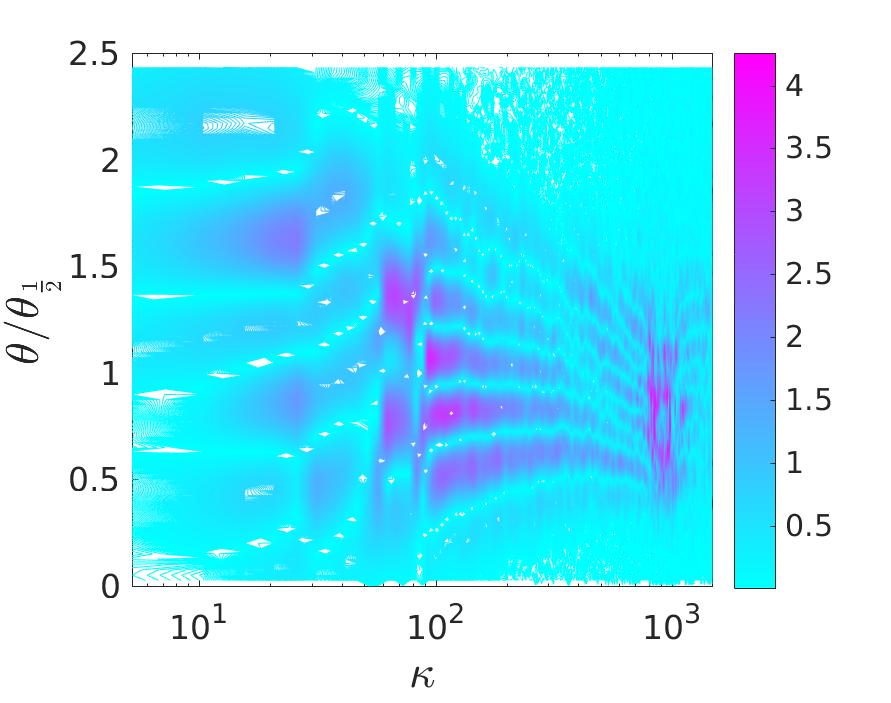}\label{fig:app_mode_u_6imag}}\\
\subfloat[$|\Im\left\lbrace\psi^\xi_7\left(\kappa,\theta\right)\right\rbrace|$]{\includegraphics[width=0.30\linewidth]{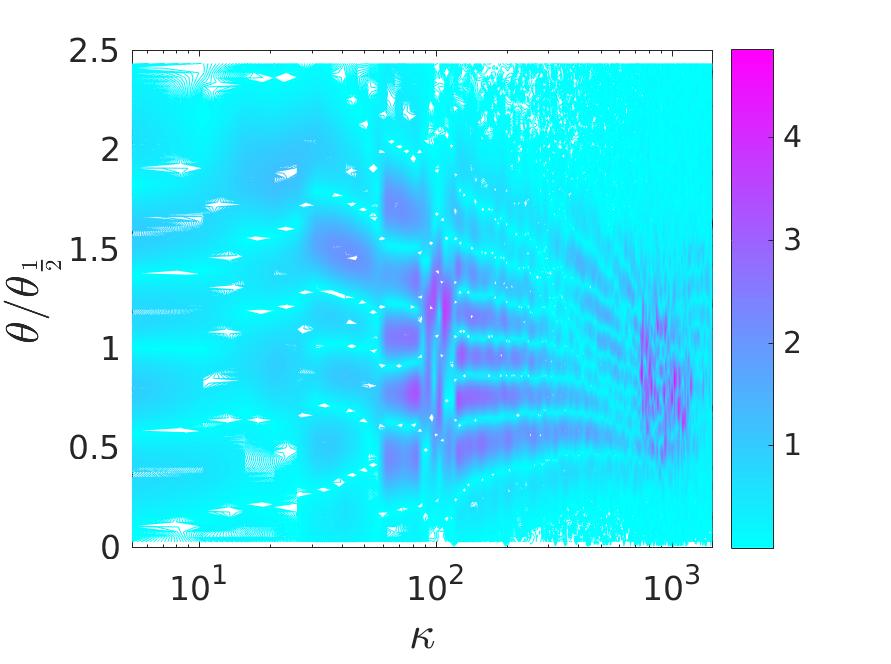}\label{fig:app_mode_u_7imag}}
\subfloat[$|\Im\left\lbrace\psi^\xi_8\left(\kappa,\theta\right)\right\rbrace|$]{\includegraphics[width=0.30\linewidth]{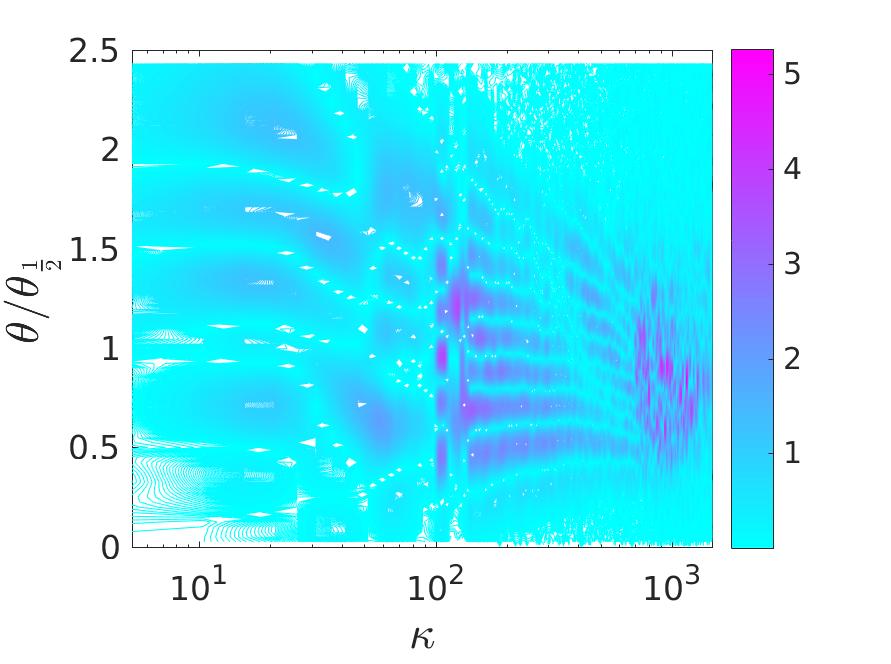}\label{fig:app_mode_u_8imag}}
\subfloat[$|\Im\left\lbrace\psi^\xi_9\left(\kappa,\theta\right)\right\rbrace|$]{\includegraphics[width=0.30\linewidth]{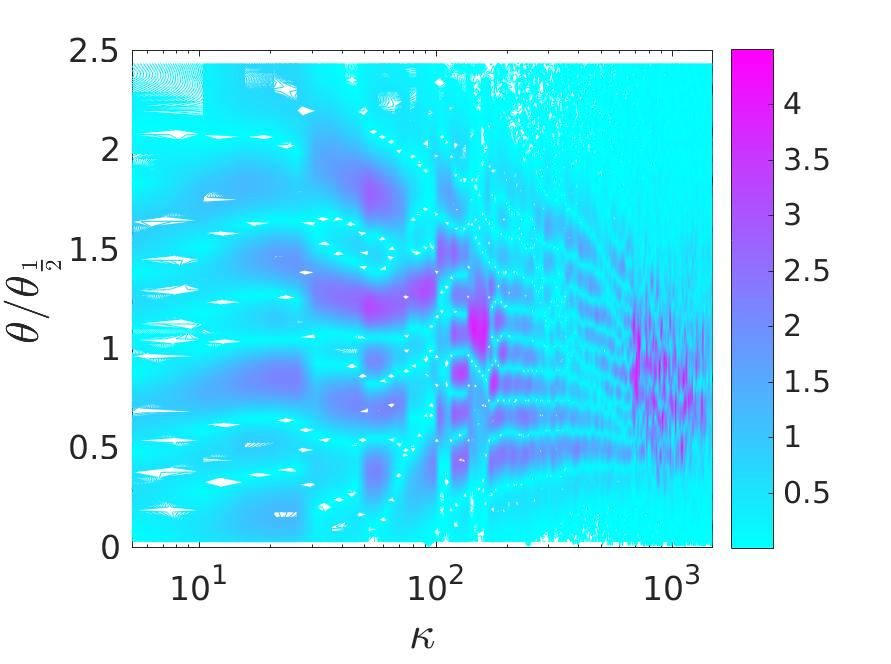}\label{fig:app_mode_u_9imag}}
\caption{The absolute imaginary parts of $\xi$-components of the LD modes $\alpha=1-9$, as a function of dimensionless wavenumber, $|\Im\left\lbrace\psi^\xi_\alpha\left(\kappa,\theta\right)\right\rbrace|$. The superscript indicates the directional component $\xi$ or $\theta$ and the subscript denotes the LD mode number.\label{fig:app_LD_modes_imag_u_1}}
\end{figure}

\begin{figure}[t]
\centering   
\subfloat[$|\Im\left\lbrace\psi^\theta_1\left(\kappa,\theta\right)\right\rbrace|$]{\includegraphics[width=0.30\linewidth]{figs/POD/SSC/mode_v_1imag}\label{fig:app_mode_v_1imag}}
\subfloat[$|\Im\left\lbrace\psi^\theta_2\left(\kappa,\theta\right)\right\rbrace|$]{\includegraphics[width=0.30\linewidth]{figs/POD/SSC/mode_v_2imag}\label{fig:app_mode_v_2imag}}
\subfloat[$|\Im\left\lbrace\psi^\theta_3\left(\kappa,\theta\right)\right\rbrace|$]{\includegraphics[width=0.30\linewidth]{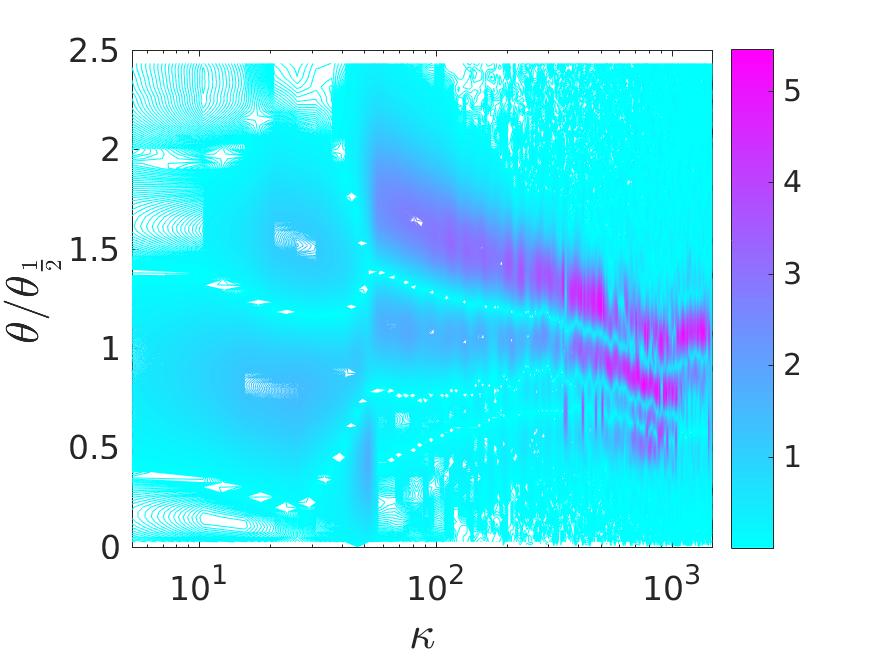}\label{fig:app_mode_v_3imag}}\\
\subfloat[$|\Im\left\lbrace\psi^\theta_4\left(\kappa,\theta\right)\right\rbrace|$]{\includegraphics[width=0.30\linewidth]{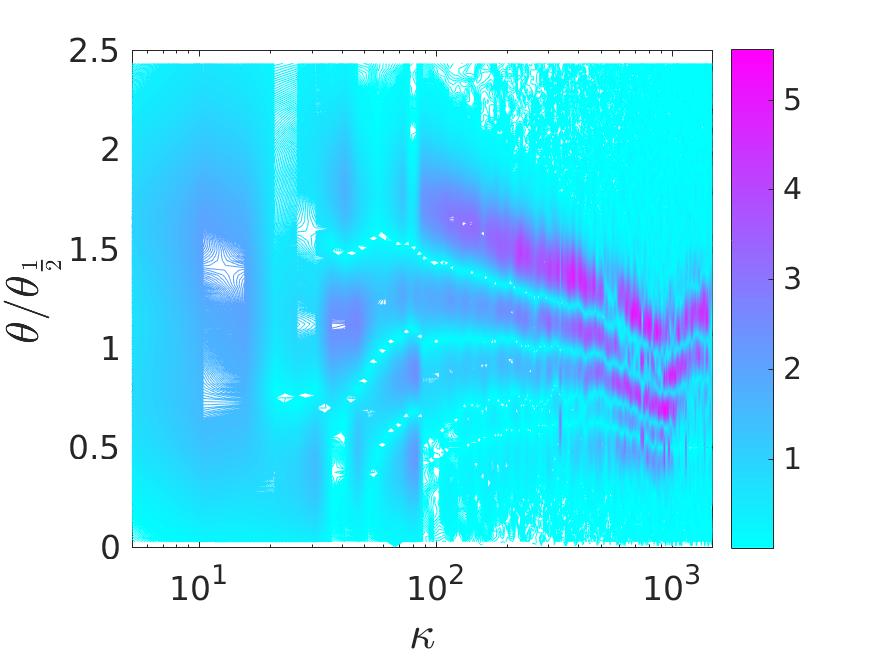}\label{fig:app_mode_v_4imag}}
\subfloat[$|\Im\left\lbrace\psi^\theta_5\left(\kappa,\theta\right)\right\rbrace|$]{\includegraphics[width=0.30\linewidth]{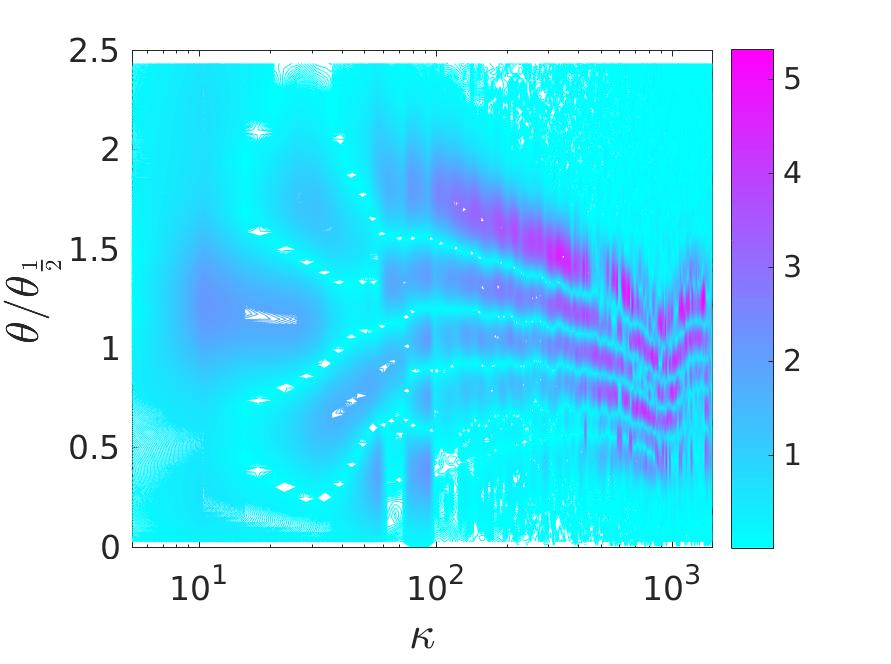}\label{fig:app_mode_v_5mag}}
\subfloat[$|\Im\left\lbrace\psi^\theta_6\left(\kappa,\theta\right)\right\rbrace|$]{\includegraphics[width=0.30\linewidth]{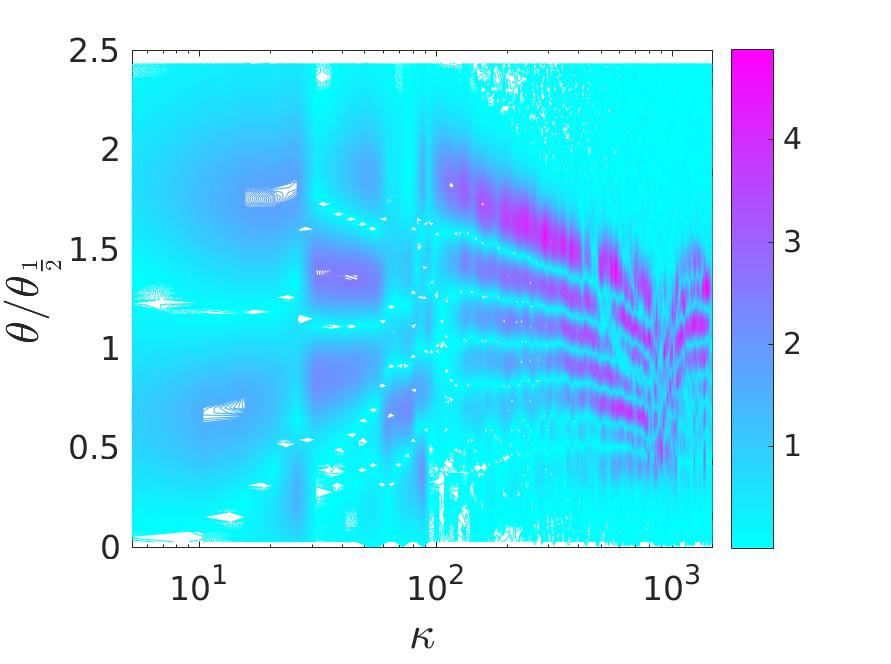}\label{fig:app_mode_v_6imag}}\\
\subfloat[$|\Im\left\lbrace\psi^\theta_7\left(\kappa,\theta\right)\right\rbrace|$]{\includegraphics[width=0.30\linewidth]{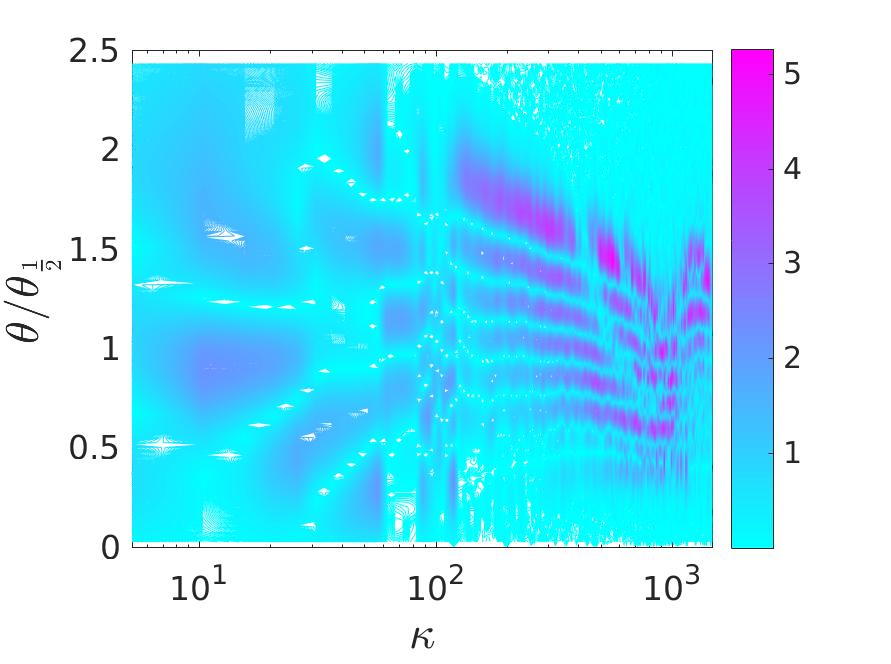}\label{fig:app_mode_v_7imag}}
\subfloat[$|\Im\left\lbrace\psi^\theta_8\left(\kappa,\theta\right)\right\rbrace|$]{\includegraphics[width=0.30\linewidth]{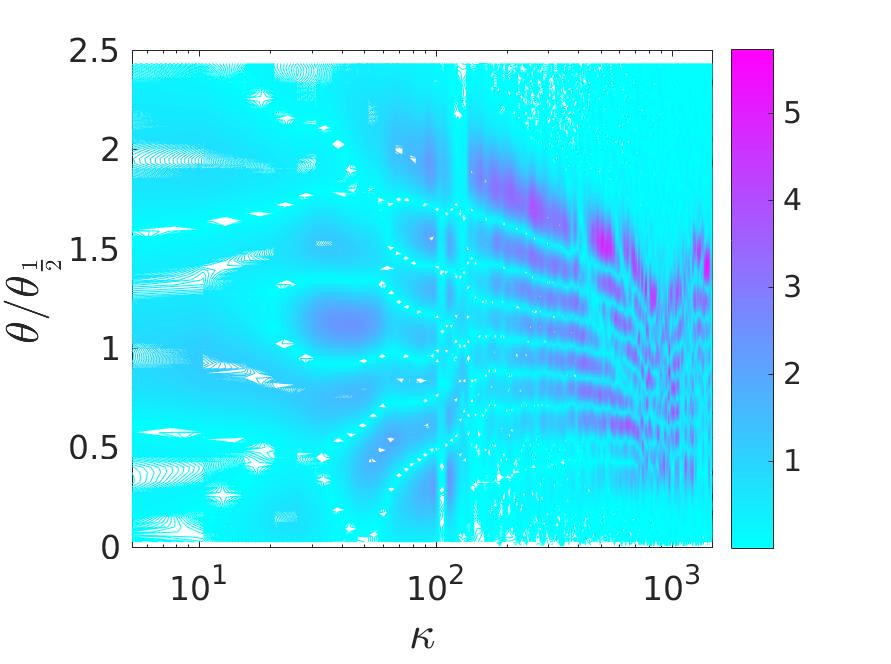}\label{fig:app_mode_v_8imag}}
\subfloat[$|\Im\left\lbrace\psi^\theta_9\left(\kappa,\theta\right)\right\rbrace|$]{\includegraphics[width=0.30\linewidth]{figs/POD/SSC/mode_v_8imag}\label{fig:app_mode_v_9imag}}
\caption{The absolute imaginary parts of $\theta$-components of the LD modes $\alpha=1-9$, as a function of dimensionless wavenumber, $|\Im\left\lbrace\psi^\theta_\alpha\left(\kappa,\theta\right)\right\rbrace|$. The superscript indicates the directional component $\xi$ or $\theta$ and the subscript denotes the LD mode number.\label{fig:app_LD_modes_imag_v_1}}
\end{figure}

Figures \ref{fig:app_LD_modes_real_u_1}, \ref{fig:app_LD_modes_real_v_1}, are the absolute real parts of the $\xi$- and $\theta$-components, respectively, of modes $\alpha=1-9$ as a function of dimensionless wavenumber, $\kappa$, and figures \ref{fig:app_LD_modes_imag_u_1}, and \ref{fig:app_LD_modes_imag_v_1} are the corresponding imaginary parts of the $\xi$- and $\theta$-components, respectively. 
\end{document}

%% file: experimentalsetup_pdf.tex
\begingroup%
  \makeatletter%
  \providecommand\color[2][]{%
    \errmessage{(Inkscape) Color is used for the text in Inkscape, but the package 'color.sty' is not loaded}%
    \renewcommand\color[2][]{}%
  }%
  \providecommand\transparent[1]{%
    \errmessage{(Inkscape) Transparency is used (non-zero) for the text in Inkscape, but the package 'transparent.sty' is not loaded}%
    \renewcommand\transparent[1]{}%
  }%
  \providecommand\rotatebox[2]{#2}%
  \ifx\svgwidth\undefined%
    \setlength{\unitlength}{229.5369752bp}%
    \ifx\svgscale\undefined%
      \relax%
    \else%
      \setlength{\unitlength}{\unitlength * \real{\svgscale}}%
    \fi%
  \else%
    \setlength{\unitlength}{\svgwidth}%
  \fi%
  \global\let\svgwidth\undefined%
  \global\let\svgscale\undefined%
  \makeatother%
  \begin{picture}(1,0.69449255)%
    \put(0,0){\includegraphics[width=\unitlength]{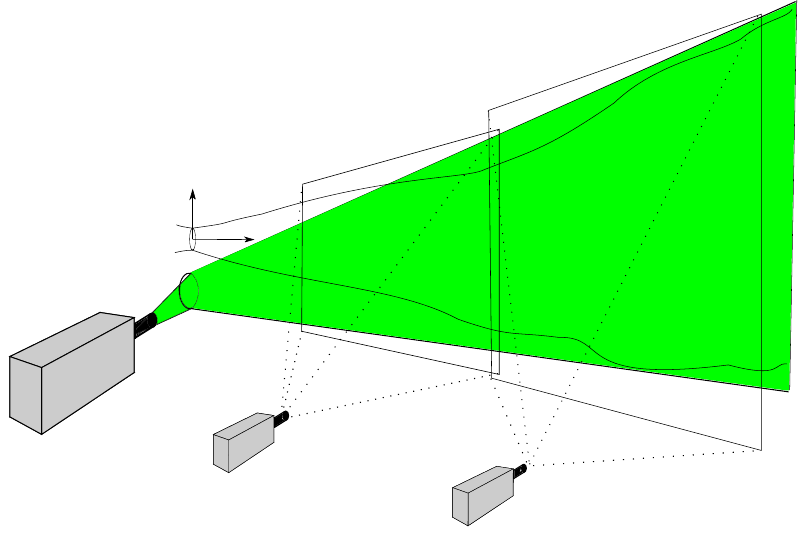}}%
    \put(0.23832228,0.46449366){\makebox(0,0)[lb]{\smash{r}}}%
    \put(0.30843299,0.4039435){\makebox(0,0)[lb]{\smash{x}}}%
    \put(0.21763879,0.05863158){\makebox(0,0)[lb]{\smash{Camera 1
}}}%
    \put(0.51890291,0.00059222){\makebox(0,0)[lb]{\smash{Camera 2
}}}%
    \put(0.37391587,0.52942329){\makebox(0,0)[lb]{\smash{Field of view 1
}}}%
    \put(0.64325872,0.65975358){\makebox(0,0)[lb]{\smash{Field of view 2
}}}%
    \put(0.10739734,0.33181077){\makebox(0,0)[lb]{\smash{Mirror}}}%
    \put(-0.00369037,0.11125806){\makebox(0,0)[lb]{\smash{Laser}}}%
    \put(0.12134394,0.38801091){\makebox(0,0)[lb]{\smash{Nozzle}}}%
  \end{picture}%
\endgroup%